\input harvmac

\input amssym
\input epsf


\newfam\frakfam
\font\teneufm=eufm10
\font\seveneufm=eufm7
\font\fiveeufm=eufm5
\textfont\frakfam=\teneufm
\scriptfont\frakfam=\seveneufm
\scriptscriptfont\frakfam=\fiveeufm


\def\bb{
\font\tenmsb=msbm10
\font\sevenmsb=msbm7
\font\fivemsb=msbm5
\textfont1=\tenmsb
\scriptfont1=\sevenmsb
\scriptscriptfont1=\fivemsb
}


\newfam\dsromfam
\font\tendsrom=dsrom10
\textfont\dsromfam=\tendsrom
\def\ds{\fam\dsromfam \tendsrom}


\newfam\mbffam
\font\tenmbf=cmmib10
\font\sevenmbf=cmmib7
\font\fivembf=cmmib5
\textfont\mbffam=\tenmbf
\scriptfont\mbffam=\sevenmbf
\scriptscriptfont\mbffam=\fivembf


\newfam\mbfcalfam
\font\tenmbfcal=cmbsy10
\font\sevenmbfcal=cmbsy7
\font\fivembfcal=cmbsy5
\textfont\mbfcalfam=\tenmbfcal
\scriptfont\mbfcalfam=\sevenmbfcal
\scriptscriptfont\mbfcalfam=\fivembfcal


\newfam\mscrfam
\font\tenmscr=rsfs10
\font\sevenmscr=rsfs7
\font\fivemscr=rsfs5
\textfont\mscrfam=\tenmscr
\scriptfont\mscrfam=\sevenmscr
\scriptscriptfont\mscrfam=\fivemscr
\def\scr{\fam\mscrfam \tenmscr}




\def\tilde{\widetilde}
\def\bw{{\bar w}}
\def\bz{{\bar z}}

\def\t{\tilde}
\def\hat{\widehat}

\def\bar{\overline}
\def\b{\bar}
\def\bsq#1{{{\b{#1}}^{\lower 2.5pt\hbox{$\scriptstyle 2$}}}}
\def\bexp#1#2{{{\b{#1}}^{\lower 2.5pt\hbox{$\scriptstyle #2$}}}}
\def\dotexp#1#2{{{#1}^{\lower 2.5pt\hbox{$\scriptstyle #2$}}}}


\def\rt2{\sqrt{2}}
\def\half {{1 \over 2}}
\def\Re{\mathop{\rm Re}}
\def\Im{\mathop{\rm Im}}
\def\d{\partial}

\def\grad{\nabla}

\def\det{\mathop{\rm det}}

\def\Tr{\mathop{\rm Tr}}


\font\tenbifull=cmmib10
\font\tenbimed=cmmib7
\font\tenbismall=cmmib5
\textfont9=\tenbifull \scriptfont9=\tenbimed
\scriptscriptfont9=\tenbismall

\mathchardef\bbGamma="7000
\mathchardef\bbDelta="7001
\mathchardef\bbPhi="7002
\mathchardef\bbAlpha="7003
\mathchardef\bbXi="7004
\mathchardef\bbPi="7005
\mathchardef\bbSigma="7006
\mathchardef\bbUpsilon="7007
\mathchardef\bbTheta="7008
\mathchardef\bbPsi="7009
\mathchardef\bbOmega="700A
\mathchardef\bbalpha="710B
\mathchardef\bbbeta="710C
\mathchardef\bbgamma="710D
\mathchardef\bbdelta="710E
\mathchardef\bbepsilon="710F
\mathchardef\bbzeta="7110
\mathchardef\bbeta="7111
\mathchardef\bbtheta="7112
\mathchardef\bbiota="7113
\mathchardef\bbkappa="7114
\mathchardef\bblambda="7115
\mathchardef\bbmu="7116
\mathchardef\bbnu="7117
\mathchardef\bbxi="7118
\mathchardef\bbpi="7119
\mathchardef\bbrho="711A
\mathchardef\bbsigma="711B
\mathchardef\bbtau="711C
\mathchardef\bbupsilon="711D
\mathchardef\bbphi="711E
\mathchardef\bbchi="711F
\mathchardef\bbpsi="7120
\mathchardef\bbomega="7121
\mathchardef\bbvarepsilon="7122
\mathchardef\bbvartheta="7123
\mathchardef\bbvarpi="7124
\mathchardef\bbvarrho="7125
\mathchardef\bbvarsigma="7126
\mathchardef\bbvarphi="7127


\def\alphadot{{\dot\alpha}}
\def\betadot{{\dot\beta}}



\def\thetasq{\theta^2}


\def\SA{{\scr A}}
\def\SF{{\scr F}}
\def\SL{{\scr L}}
\def\CA{{\cal A}}

\def\CD{{\cal D}}

\def\CF{{\cal F}}

\def\CI{{\cal I}}
\def\CJ{{\cal J}}
\def\CK{{\cal K}}
\def\CL{{\cal L}}
\def\CM{{\cal M}}
\def\CN{{\cal N}}
\def\CO{{\cal O}}

\def\CQ{{\cal Q}}
\def\CR{{\cal R}}
\def\CS{{\cal S}}
\def\CT{{\cal T}}


\def\1{{\ds 1}}
\def\R{\hbox{$\bb R$}}
\def\C{\hbox{$\bb C$}}

\def\Z{\hbox{$\bb Z$}}
\def\N{\hbox{$\bb N$}}
\def\P{\hbox{$\bb P$}}


\def\ep{\varepsilon}

\noblackbox

\def\unit{\relax{\rm 1\kern-.26em I}}
\def\nada{\relax{\rm 0\kern-.30em l}}
\def\tilde{\widetilde}
\def\t{\tilde}
\def\alphadot{{\dot \alpha}}
\def\betadot{{\dot\beta}}

\def\det{{\rm det}}

\noblackbox
\def\IL{\relax{\rm I\kern-.18em L}}
\def\IH{\relax{\rm I\kern-.18em H}}
\def\IR{\relax{\rm I\kern-.18em R}}
\def\IC{\relax\hbox{$\inbar\kern-.3em{\rm C}$}}
\def\IZ{\relax\ifmmode\mathchoice
{\hbox{\cmss Z\kern-.4em Z}}{\hbox{\cmss Z\kern-.4em Z}} {\lower.9pt\hbox{\cmsss Z\kern-.4em Z}}
{\lower1.2pt\hbox{\cmsss Z\kern-.4em Z}}\else{\cmss Z\kern-.4em Z}\fi}
\def\CM {{\cal M}}
\def\CN {{\cal N}}
\def\CR {{\cal R}}
\def\CD {{\cal D}}
\def\CF {{\cal F}}
\def\CJ {{\cal J}}
\def\partialslash{\not{\hbox{\kern-2pt $\partial$}}}

\def\CL {{\cal L}}

\def\CO {{\cal O}}

\def\CS {{\cal S}}
\def\CA{{\cal A}}
\def\CK{{\cal K}}
\def\CM {{\cal M}}
\def\CN {{\cal N}}

\def\CO {{\cal O}}

\def\CQ {{\cal Q }}

\def\CS {{\cal S }}

\def\det{{\rm det}}
\def\Tr{{\rm Tr}}

\font\manual=manfnt \def\dbend{\lower3.5pt\hbox{\manual\char127}}

\def\IZ{\relax\ifmmode\mathchoice
{\hbox{\cmss Z\kern-.4em Z}}{\hbox{\cmss Z\kern-.4em Z}} {\lower.9pt\hbox{\cmsss Z\kern-.4em Z}}
{\lower1.2pt\hbox{\cmsss Z\kern-.4em Z}}\else{\cmss Z\kern-.4em Z}\fi}
\def\half {{1\over 2}}

\def\bar{\overline}
\def\CS{{\cal S}}

\def\rt2{\sqrt{2}}
\def\irt2{{1\over\sqrt{2}}}

\def\t{\tilde}
\def\hat{\widehat}
\def\slashchar#1{\setbox0=\hbox{$#1$}           
   \dimen0=\wd0                                 
   \setbox1=\hbox{/} \dimen1=\wd1               
   \ifdim\dimen0>\dimen1                        
      \rlap{\hbox to \dimen0{\hfil/\hfil}}      
      #1                                        
   \else                                        
      \rlap{\hbox to \dimen1{\hfil$#1$\hfil}}   
      /                                         
   \fi}

\def\foursqr#1#2{{\vcenter{\vbox{
    \hrule height.#2pt
    \hbox{\vrule width.#2pt height#1pt \kern#1pt
    \vrule width.#2pt}
    \hrule height.#2pt
    \hrule height.#2pt
    \hbox{\vrule width.#2pt height#1pt \kern#1pt
    \vrule width.#2pt}
    \hrule height.#2pt
        \hrule height.#2pt
    \hbox{\vrule width.#2pt height#1pt \kern#1pt
    \vrule width.#2pt}
    \hrule height.#2pt
        \hrule height.#2pt
    \hbox{\vrule width.#2pt height#1pt \kern#1pt
    \vrule width.#2pt}
    \hrule height.#2pt}}}}
\def\psqr#1#2{{\vcenter{\vbox{\hrule height.#2pt
    \hbox{\vrule width.#2pt height#1pt \kern#1pt
    \vrule width.#2pt}
    \hrule height.#2pt \hrule height.#2pt
    \hbox{\vrule width.#2pt height#1pt \kern#1pt
    \vrule width.#2pt}
    \hrule height.#2pt}}}}
\def\sqr#1#2{{\vcenter{\vbox{\hrule height.#2pt
    \hbox{\vrule width.#2pt height#1pt \kern#1pt
    \vrule width.#2pt}
    \hrule height.#2pt}}}}

\def\figin{\epsfcheck\figin}\def\figins{\epsfcheck\figins}
\def\epsfcheck{\ifx\epsfbox\UnDeFiNeD
\message{(NO epsf.tex, FIGURES WILL BE IGNORED)}
\gdef\figin##1{\vskip2in}\gdef\figins##1{\hskip.5in}
\else\message{(FIGURES WILL BE INCLUDED)}%
\gdef\figin##1{##1}\gdef\figins##1{##1}\fi}
\def\DefWarn#1{}
\def\figinsert{\goodbreak\midinsert}
\def\ifig#1#2#3{\DefWarn#1\xdef#1{fig.~\the\figno}
\writedef{#1\leftbracket fig.\noexpand~\the\figno}%
\figinsert\figin{\centerline{#3}}\medskip\centerline{\vbox{\baselineskip12pt \advance\hsize by
-1truein\noindent\footnotefont{\bf Fig.~\the\figno:\ } \it#2}}
\bigskip\endinsert\global\advance\figno by1}


\lref\FestucciaWS{
  G.~Festuccia and N.~Seiberg,
  ``Rigid Supersymmetric Theories in Curved Superspace,''
JHEP {\bf 1106}, 114 (2011).
[arXiv:1105.0689 [hep-th]].
}

\lref\MartelliFU{
  D.~Martelli, A.~Passias and J.~Sparks,
  ``The Gravity dual of supersymmetric gauge theories on a squashed three-sphere,''
Nucl.\ Phys.\ B {\bf 864}, 840 (2012).
[arXiv:1110.6400 [hep-th]].
}

\lref\ClossetVG{
  C.~Closset, T.~T.~Dumitrescu, G.~Festuccia, Z.~Komargodski and N.~Seiberg,
  ``Contact Terms, Unitarity, and F-Maximization in Three-Dimensional Superconformal Theories,''
JHEP {\bf 1210}, 053 (2012).
[arXiv:1205.4142 [hep-th]].
}

\lref\ClossetVP{
  C.~Closset, T.~T.~Dumitrescu, G.~Festuccia, Z.~Komargodski and N.~Seiberg,
  ``Comments on Chern-Simons Contact Terms in Three Dimensions,''
JHEP {\bf 1209}, 091 (2012).
[arXiv:1206.5218 [hep-th]].
}

\lref\KomargodskiRB{
  Z.~Komargodski and N.~Seiberg,
  ``Comments on Supercurrent Multiplets, Supersymmetric Field Theories and Supergravity,''
JHEP {\bf 1007}, 017 (2010).
[arXiv:1002.2228 [hep-th]].
}

\lref\SohniusTP{
  M.~F.~Sohnius and P.~C.~West,
  ``An Alternative Minimal Off-Shell Version of N=1 Supergravity,''
Phys.\ Lett.\ B {\bf 105}, 353 (1981).
}

\lref\DolanRP{
  F.~A.~H.~Dolan, V.~P.~Spiridonov and G.~S.~Vartanov,
  ``From 4d superconformal indices to 3d partition functions,''
Phys.\ Lett.\ B {\bf 704}, 234 (2011).
[arXiv:1104.1787 [hep-th]].
}

\lref\GaddeIA{
  A.~Gadde and W.~Yan,
  ``Reducing the 4d Index to the $S^3$ Partition Function,''
JHEP {\bf 1212}, 003 (2012).
[arXiv:1104.2592 [hep-th]].
}

\lref\ImamuraUW{
  Y.~Imamura,
  ``Relation between the 4d superconformal index and the $S^3$ partition function,''
JHEP {\bf 1109}, 133 (2011).
[arXiv:1104.4482 [hep-th]].
}

\lref\DumitrescuHA{
  T.~T.~Dumitrescu, G.~Festuccia and N.~Seiberg,
  ``Exploring Curved Superspace,''
JHEP {\bf 1208}, 141 (2012).
[arXiv:1205.1115 [hep-th]].
}

\lref\DumitrescuAT{
  T.~T.~Dumitrescu and G.~Festuccia,
  ``Exploring Curved Superspace (II),''
JHEP {\bf 1301}, 072 (2013).
[arXiv:1209.5408 [hep-th]].
}

\lref\DumitrescuIU{
  T.~T.~Dumitrescu and N.~Seiberg,
  ``Supercurrents and Brane Currents in Diverse Dimensions,''
JHEP {\bf 1107}, 095 (2011).
[arXiv:1106.0031 [hep-th]].
}

\lref\SohniusFW{
  M.~Sohnius and P.~C.~West,
  ``The Tensor Calculus And Matter Coupling Of The Alternative Minimal Auxiliary Field Formulation Of N=1 Supergravity,''
Nucl.\ Phys.\ B {\bf 198}, 493 (1982).
}

\lref\KuzenkoXG{
  S.~M.~Kuzenko, U.~Lindstrom and G.~Tartaglino-Mazzucchelli,
  ``Off-shell supergravity-matter couplings in three dimensions,''
JHEP {\bf 1103}, 120 (2011).
[arXiv:1101.4013 [hep-th]].
}

\lref\KuzenkoBC{
  S.~M.~Kuzenko, U.~Lindstrom and G.~Tartaglino-Mazzucchelli,
  ``Three-dimensional (p,q) AdS superspaces and matter couplings,''
JHEP {\bf 1208}, 024 (2012).
[arXiv:1205.4622 [hep-th]].
}

\lref\WittenEV{
  E.~Witten,
  ``Supersymmetric Yang-Mills theory on a four manifold,''
J.\ Math.\ Phys.\  {\bf 35}, 5101 (1994).
[hep-th/9403195].
}

\lref\KlareGN{
  C.~Klare, A.~Tomasiello and A.~Zaffaroni,
  ``Supersymmetry on Curved Spaces and Holography,''
JHEP {\bf 1208}, 061 (2012).
[arXiv:1205.1062 [hep-th]].
}

\lref\KomargodskiPC{
  Z.~Komargodski and N.~Seiberg,
  ``Comments on the Fayet-Iliopoulos Term in Field Theory and Supergravity,''
JHEP {\bf 0906}, 007 (2009).
[arXiv:0904.1159 [hep-th]].
}

\lref\ImamuraSU{
  Y.~Imamura and S.~Yokoyama,
  ``Index for three dimensional superconformal field theories with general R-charge assignments,''
JHEP {\bf 1104}, 007 (2011).
[arXiv:1101.0557 [hep-th]].
}

\lref\HoweZM{
  P.~S.~Howe, J.~M.~Izquierdo, G.~Papadopoulos and P.~K.~Townsend,
  ``New supergravities with central charges and Killing spinors in (2+1)-dimensions,''
Nucl.\ Phys.\ B {\bf 467}, 183 (1996).
[hep-th/9505032].
}

\lref\KuzenkoRD{
  S.~M.~Kuzenko and G.~Tartaglino-Mazzucchelli,
  ``Three-dimensional N=2 (AdS) supergravity and associated supercurrents,''
JHEP {\bf 1112}, 052 (2011).
[arXiv:1109.0496 [hep-th]].
}

\lref\RomelsbergerEC{
  C.~Romelsberger,
  ``Calculating the Superconformal Index and Seiberg Duality,''
[arXiv:0707.3702 [hep-th]].
}

\lref\RomelsbergerEG{
  C.~Romelsberger,
  ``Counting chiral primaries in N = 1, d=4 superconformal field theories,''
Nucl.\ Phys.\ B {\bf 747}, 329 (2006).
[hep-th/0510060].
}

\lref\DolanQI{
  F.~A.~Dolan and H.~Osborn,
  ``Applications of the Superconformal Index for Protected Operators and q-Hypergeometric Identities to N=1 Dual Theories,''
Nucl.\ Phys.\ B {\bf 818}, 137 (2009).
[arXiv:0801.4947 [hep-th]].
}

\lref\ClossetRU{
  C.~Closset, T.~T.~Dumitrescu, G.~Festuccia and Z.~Komargodski,
  ``Supersymmetric Field Theories on Three-Manifolds,''
JHEP {\bf 1305}, 017 (2013).
[arXiv:1212.3388].
}

\lref\KSii{
  K.~Kodaira and D.C.~Spencer,
  ``On Deformations of Complex Analytic Structures II,''
Ann. Math. {\bf 67}, 403 (1958).
}

\lref\Kodairabook{
K.~Kodaira, ``Complex Manifolds and Deformation of Complex Structures,'' Springer (1986).
}

\lref\Kobayashi{
S.~Kobayashi, ``Differential Geometry of Complex Vector Bundles,'' Princeton University Press (1987).
}

\lref\WittenDF{
  E.~Witten,
  ``Constraints on Supersymmetry Breaking,''
Nucl.\ Phys.\ B {\bf 202}, 253 (1982).
}

\lref\spivak{
M.~Spivak, ``Calculus on Manifolds,'' Perseus (1965).
}

\lref\ImamuraWG{
  Y.~Imamura and D.~Yokoyama,
  ``N=2 supersymmetric theories on squashed three-sphere,''
Phys.\ Rev.\ D {\bf 85}, 025015 (2012).
[arXiv:1109.4734 [hep-th]].
}

\lref\MartelliAQA{
  D.~Martelli and A.~Passias,
  ``The gravity dual of supersymmetric gauge theories on a two-parameter deformed three-sphere,''
[arXiv:1306.3893 [hep-th]].
}

\lref\AharonyDHA{
  O.~Aharony, S.~S.~Razamat, N.~Seiberg and B.~Willett,
  ``3d dualities from 4d dualities,''
JHEP {\bf 1307}, 149 (2013).
[arXiv:1305.3924 [hep-th]].
}

\lref\SpiridonovZA{
  V.~P.~Spiridonov and G.~S.~Vartanov,
  ``Elliptic Hypergeometry of Supersymmetric Dualities,''
Commun.\ Math.\ Phys.\  {\bf 304}, 797 (2011).
[arXiv:0910.5944 [hep-th]].
}

\lref\KinneyEJ{
  J.~Kinney, J.~M.~Maldacena, S.~Minwalla and S.~Raju,
  ``An Index for 4 dimensional super conformal theories,''
Commun.\ Math.\ Phys.\  {\bf 275}, 209 (2007).
[hep-th/0510251].
}

\lref\SenPH{
  D.~Sen,
  ``Supersymmetry In The Space-time $\R \times S^3$,''
Nucl.\ Phys.\ B {\bf 284}, 201 (1987).
}

\lref\Kodairasone{
  K.~Kodaira, ``Complex structures on $S^{1}\times S^{3}$,''
Proceedings of the National Academy of Sciences of the United States of America {\bf 55}, 240 (1966).
}

\lref\cs{
C.~Closset and I.~Shamir, to appear.
}

\lref\Belgun{
F.~A.~Belgun, ``On the metric structure of non-K\"ahler complex surfaces,''
Math. Ann. {\bf 317}, 1 (2000).
}

\lref\gaudorn{
P.~Gauduchon and~L.~Ornea, ``Locally conformally K\"ahler metrics on Hopf surfaces,'' Ann. Inst. Fourier {\bf 48}, 4 (1998).
}

\lref\AharonyHDA{
  O.~Aharony, N.~Seiberg and Y.~Tachikawa,
  ``Reading between the lines of four-dimensional gauge theories,''
JHEP {\bf 1308}, 115 (2013).
[arXiv:1305.0318 [hep-th]].
}

\lref\CassaniDBA{
  D.~Cassani and D.~Martelli,
  ``Supersymmetry on curved spaces and superconformal anomalies,''
[arXiv:1307.6567 [hep-th]].
}

\lref\JohansenAW{
  A.~Johansen,
  ``Twisting of $N=1$ SUSY gauge theories and heterotic topological theories,''
Int.\ J.\ Mod.\ Phys.\ A {\bf 10}, 4325 (1995).
[hep-th/9403017].
}

\lref\Spiridonov{
  V.~Spiridonov,
  ``Elliptic Hypergeometric Functions,''
[arXiv:0704.3099].
}

\lref\Rains{
  E.~M.~Rains,
  ``Transformations of Elliptic Hypergeometric Integrals,''
[math/0309252].
}

\lref\Bult{
  F.~van de Bult,
  ``Hyperbolic Hypergeometric Functions,''
University of Amsterdam, Ph.D. Thesis, [http://www.its.caltech.edu/$\sim$vdbult/Thesis.pdf].
}

\lref\KodairaCCASii{
K.~Kodaira, ``On the structure of compact complex analytic surfaces, II,'' American Journal of Mathematics {\bf 88}, 682 (1966).
}

\lref\SpiridonovHF{
  V.~P.~Spiridonov and G.~S.~Vartanov,
  ``Elliptic hypergeometry of supersymmetric dualities II. Orthogonal groups, knots, and vortices,''
[arXiv:1107.5788 [hep-th]].
}

\lref\BeniniNC{
  F.~Benini, T.~Nishioka and M.~Yamazaki,
  ``4d Index to 3d Index and 2d TQFT,''
Phys.\ Rev.\ D {\bf 86}, 065015 (2012).
[arXiv:1109.0283 [hep-th]].
}

\lref\RazamatOPA{
  S.~S.~Razamat and B.~Willett,
  ``Global Properties of Supersymmetric Theories and the Lens Space,''
[arXiv:1307.4381 [hep-th]].
}

\lref\Nakagawa{
N.~Nakagawa, ``Complex structures on $L (p, q)\times S^1$,'' Hiroshima Mathematical Journal {\bf 25} 423 (1995).
}

\lref\GaddeDDA{
  A.~Gadde and S.~Gukov,
  ``2d Index and Surface operators,''
[arXiv:1305.0266 [hep-th]].
}

\lref\BeniniNDA{
  F.~Benini, R.~Eager, K.~Hori and Y.~Tachikawa,
  ``Elliptic genera of two-dimensional N=2 gauge theories with rank-one gauge groups,''
[arXiv:1305.0533 [hep-th]].
}

\lref\BeniniXPA{
  F.~Benini, R.~Eager, K.~Hori and Y.~Tachikawa,
  ``Elliptic genera of 2d N=2 gauge theories,''
[arXiv:1308.4896 [hep-th]].
}

\lref\KapustinKZ{
  A.~Kapustin, B.~Willett and I.~Yaakov,
  ``Exact Results for Wilson Loops in Superconformal Chern-Simons Theories with
  Matter,''
  JHEP {\bf 1003}, 089 (2010)
  [arXiv:0909.4559 [hep-th]].
}

\lref\JafferisUN{
  D.~L.~Jafferis,
  ``The Exact Superconformal R-Symmetry Extremizes Z,''
JHEP {\bf 1205}, 159 (2012).
[arXiv:1012.3210 [hep-th]].
}

\lref\HamaAV{
  N.~Hama, K.~Hosomichi and S.~Lee,
  ``Notes on SUSY Gauge Theories on Three-Sphere,''
JHEP {\bf 1103}, 127 (2011).
[arXiv:1012.3512 [hep-th]].
}

\lref\HamaEA{
  N.~Hama, K.~Hosomichi and S.~Lee,
  ``SUSY Gauge Theories on Squashed Three-Spheres,''
JHEP {\bf 1105}, 014 (2011).
[arXiv:1102.4716 [hep-th]].
}

\lref\AldayLBA{
  L.~F.~Alday, D.~Martelli, P.~Richmond and J.~Sparks,
  ``Localization on Three-Manifolds,''
[arXiv:1307.6848 [hep-th]].
}

\lref\NianQWA{
  J.~Nian,
  ``Localization of Supersymmetric Chern-Simons-Matter Theory on a Squashed $S^3$ with $SU(2)\times U(1)$ Isometry,''
[arXiv:1309.3266 [hep-th]].
}

\lref\BhattacharyaZY{
  J.~Bhattacharya, S.~Bhattacharyya, S.~Minwalla and S.~Raju,
  ``Indices for Superconformal Field Theories in 3,5 and 6 Dimensions,''
JHEP {\bf 0802}, 064 (2008).
[arXiv:0801.1435 [hep-th]].
}

\lref\KapustinJM{
  A.~Kapustin and B.~Willett,
  ``Generalized Superconformal Index for Three Dimensional Field Theories,''
[arXiv:1106.2484 [hep-th]].
}

\lref\DimoftePY{
  T.~Dimofte, D.~Gaiotto and S.~Gukov,
  ``3-Manifolds and 3d Indices,''
[arXiv:1112.5179 [hep-th]].
}

\lref\NishiokaHAA{
  T.~Nishioka and I.~Yaakov,
  ``Supersymmetric Renyi Entropy,''
[arXiv:1306.2958 [hep-th]].
}

\lref\Huybrechts{
D.~Huybrechts, ``Complex Geometry: An Introduction,'' Springer (2006).
}

\lref\TanakaDCA{
  A.~Tanaka,
  ``Localization on round sphere revisited,''
[arXiv:1309.4992 [hep-th]].
}

\lref\SamtlebenGY{
  H.~Samtleben and D.~Tsimpis,
  ``Rigid supersymmetric theories in 4d Riemannian space,''
JHEP {\bf 1205}, 132 (2012).
[arXiv:1203.3420 [hep-th]].
}

\lref\LiuBI{
  J.~T.~Liu, L.~A.~Pando Zayas and D.~Reichmann,
  ``Rigid Supersymmetric Backgrounds of Minimal Off-Shell Supergravity,''
JHEP {\bf 1210}, 034 (2012).
[arXiv:1207.2785 [hep-th]].
}

\lref\AldayAU{
  L.~F.~Alday, M.~Fluder and J.~Sparks,
  ``The Large N limit of M2-branes on Lens spaces,''
JHEP {\bf 1210}, 057 (2012).
[arXiv:1204.1280 [hep-th]].
}

\lref\BG{
M.~Brunella and~E.~Ghys, ``Umbilical Foliations and Transversely Holomorphic Flows,'' J.~Differential Geometry {\bf 41} 1 (1995).
}

\lref\DKi{
T.~Duchamp and~M.~Kalka, ``Deformation Theory for Holomorphic Foliations,'' J.~Differential Geometry {\bf 14} 317 (1979).
}

\lref\GM{
X.~Gomez-Mont, ``Transversal Holomorphic Structures,'' J.~Differential Geometry {\bf 15} 161 (1980).
}

\lref\Haef{
J.~Girbau, A.~Haefliger, and~D.~Sundararaman, ``On deformations of transversely holomorphic foliations,'' Journal f\"ur die reine und angewandte Mathematik {\bf 345} 122 (1983).
}

\lref\Ghys{
E.~Ghys, ``On transversely holomorphic flows II,'' Inventiones mathematicae {\bf 126} 281 (1996).
}

\lref\DKiii{
T.~Duchamp and~M.~Kalka, ``Holomorphic Foliations and Deformations of the Hopf Foliation,'' Pacific Journal of Mathematics {\bf 112} 69 (1984).
}

\lref\Haefii{
A.~Haefliger, ``Deformations of transversely holomorphic flows on spheres and deformations of Hopf manifolds,'' Compositio Mathematica {\bf 55} 241 (1985)
}

\lref\Brunella{
M.~Brunella, ``On transversely holomorphic flows I,'' Inventiones mathematicae {\bf 126} 265 (1996).
}

\lref\WittenHU{
  E.~Witten and J.~Bagger,
  ``Quantization of Newton's Constant in Certain Supergravity Theories,''
Phys.\ Lett.\ B {\bf 115}, 202 (1982)..
}

\lref\SeibergQD{
  N.~Seiberg,
  ``Modifying the Sum Over Topological Sectors and Constraints on Supergravity,''
JHEP {\bf 1007}, 070 (2010).
[arXiv:1005.0002 [hep-th]].
}

\lref\DistlerZG{
  J.~Distler and E.~Sharpe,
  ``Quantization of Fayet-Iliopoulos Parameters in Supergravity,''
Phys.\ Rev.\ D {\bf 83}, 085010 (2011).
[arXiv:1008.0419 [hep-th]].
}

\lref\BanksZN{
  T.~Banks and N.~Seiberg,
  ``Symmetries and Strings in Field Theory and Gravity,''
Phys.\ Rev.\ D {\bf 83}, 084019 (2011).
[arXiv:1011.5120 [hep-th]].
}

\lref\HellermanFV{
  S.~Hellerman and E.~Sharpe,
  ``Sums over topological sectors and quantization of Fayet-Iliopoulos parameters,''
Adv.\ Theor.\ Math.\ Phys.\  {\bf 15}, 1141 (2011).
[arXiv:1012.5999 [hep-th]].
}

\lref\KapustinIJ{
  A.~Kapustin and D.~Orlov,
  ``Remarks on A branes, mirror symmetry, and the Fukaya category,''
J.\ Geom.\ Phys.\  {\bf 48}, 84 (2003).
[hep-th/0109098].
}



\rightline{WIS/09/13 SEP-DPPA}
\vskip-20pt
\Title{
} {\vbox{\centerline{The Geometry of Supersymmetric}
\centerline{Partition Functions}}}
\vskip-15pt
\centerline{Cyril Closset,$^1$ Thomas T. Dumitrescu,$^{2}$ Guido Festuccia,$^3$ Zohar Komargodski$^{1}$}
\vskip15pt
\centerline{ $^{1}$ {\it Weizmann Institute of Science, Rehovot
76100, Israel}}
 \centerline{$^{2}$ {\it Department of Physics, Harvard University, Cambridge, MA 02138, USA}}
 \centerline{$^{3}${\it Niels Bohr International Academy and Discovery Center, Niels Bohr Institute,}} 
  \vskip-5pt
  \centerline{\it University of Copenhagen, Blegdamsvej 17, 2100 Copenhagen~\O, Denmark}

\vskip35pt

\noindent We consider supersymmetric field theories on compact manifolds~$\CM$ and obtain constraints on the parameter dependence of their partition functions~$Z_{\CM}$. Our primary focus is the dependence of~$Z_{\CM}$ on the geometry of~$\CM$, as well as background gauge fields that couple to continuous flavor symmetries. For~$\CN=1$ theories with a~$U(1)_R$ symmetry in four dimensions, $\CM$ must be a complex manifold with a Hermitian metric. We find that~$Z_{\CM}$ is independent of the metric and depends holomorphically on the complex structure moduli. Background gauge fields define holomorphic vector bundles over~$\CM$ and~$Z_{\CM}$ is a holomorphic function of the corresponding bundle moduli. We also carry out a parallel analysis for three-dimensional~$\CN=2$ theories with a~$U(1)_R$ symmetry, where the necessary geometric structure on~$\CM$ is a transversely holomorphic foliation (THF) with a transversely Hermitian metric. Again, we find that~$Z_{\CM}$ is independent of the metric and depends holomorphically on the moduli of the THF. We discuss several applications, including manifolds diffeomorphic to~$S^3 \times S^1$ or~$S^2 \times S^1$, which are related to supersymmetric indices, and manifolds diffeomorphic to~$S^3$ (squashed spheres). In examples where~$Z_{\CM}$ has been calculated explicitly, our results explain many of its observed properties.

\Date{September 2013}

\listtoc\writetoc

\newsec{Introduction}

We consider supersymmetric field theories on compact manifolds~$\CM$, focusing on theories with four supercharges and a~$U(1)_R$ symmetry in four dimensions ($\CN=1$) and three dimensions~($\CN=2$). In addition to a Riemannian metric~$g_{\mu\nu}$, the existence of one or several supercharges on~$\CM$ generally requires the presence of additional geometric structures. The supersymmetric Lagrangian~${\scr L}_\CM$ of the theory on~$\CM$ depends on these structures, as well as other data, such as the couplings in the original flat-space Lagrangian or background gauge fields that couple to continuous flavor symmetries. In some examples, the partition function~$Z_\CM$ of the supersymmetric field theory on~$\CM$ has been calculated exactly and was found to only depend on a finite number of continuous parameters, rather than all the data used to define~$\SL_\CM$. In this paper, we derive {\it a priori} constraints on the parameter dependence of~$Z_\CM$. Our primary focus is the dependence on the geometry of the supersymmetric background. The constraints follow from general properties of supersymmetric field theories with a~$U(1)_R$ symmetry and do not rely on a weakly coupled Lagrangian description or detailed computations in specific models.  

Following~\FestucciaWS, we describe supersymmetric field theories on~$\CM$ by embedding the metric~$g_{\mu\nu}$ into a non-dynamical supergravity multiplet, which also includes a gravitino~$\Psi_\mu$ and various other background fields. A configuration of the bosonic supergravity background fields on~$\CM$ preserves a rigid supercharge~$Q$ if and only if the corresponding variation of the gravitino vanishes, $\delta_Q \Psi_\mu =0$. Once a supersymmetric background on~$\CM$ has been found, supersymmetric Lagrangians for the dynamical fields, as well as their supersymmetry transformations, follow from the corresponding supergravity formulas. 

Around flat space, a supersymmetric field theory couples to supergravity via its supercurrent multiplet, which contains the energy-momentum tensor and the supersymmetry current. In principle, these linearized couplings can be non-linearly completed via the Noether procedure.\foot{In general, the non-linear completion is not unique, e.g.~due to higher-order curvature couplings. We will restrict ourselves to a minimal completion of the linearized theory, which reduces to the original flat-space theory at short distances.} This highlights the importance of the supercurrent multiplet for elucidating the structure of the the supersymmetric theory on~$\CM$. We will analyze how~$Z_{\CM}$ depends on the geometry of~$\CM$ by studying the case when~$\CM$ is a small deformation around flat space. The Lagrangian~$\SL_\CM$ then differs from the original flat-space Lagrangian by a deformation~$\Delta \SL$, which is constructed from the operators in the supercurrent multiplet. The transformation properties of the supercurrent under the preserved supercharge~$Q$ imply that certain deformations of~$\CM$ lead to a~$Q$-exact~$\Delta \SL$. Hence, they cannot affect the partition function~$Z_\CM$. Similarly, the dependence of~$Z_\CM$ on background vector fields can be studied using the linearized coupling to a flavor current multiplet. 

It may seem surprising that a linearized analysis around flat space leads to general conclusions about the parameter dependence of~$Z_{\CM}$ for arbitrary supersymmetric backgrounds. This logic is standard in the context of topologically twisted theories (see for instance~\WittenEV). There one defines a scalar supercharge~$Q$ on~$\CM$ so that a suitably defined energy-momentum tensor in the original flat-space theory is~$Q$-exact. Since~$Q$ is a scalar, this is sufficient to conclude that the partition function~$Z_{\CM}$ is independent of the metric for arbitrary~$\CM$, not just small deformations around flat space, and hence it is a topological invariant. Below, we will study manifolds~$\CM$ with additional geometric structure and the supercharge~$Q$ will transform as a scalar under certain adapted coordinate transformations. This allows us to make an analogous argument for the independence of~$Z_{\CM}$ on deformations that couple to~$Q$-exact operators in the supercurrent multiplet.\foot{More precisely, this statement holds for a particular choice of local counterterms in the background fields, which must be consistent with other physical requirements. The non-existence of such a choice signals the presence of an anomaly. (See~\refs{\ClossetVG,\ClossetVP} for an example in three dimensions.) Throughout, we will ignore possible anomalies and assume that the counterterms have been suitably adjusted. (In this context, four-dimensional superconformal anomalies were recently discussed in~\CassaniDBA.)} Even though~$Z_{\CM}$ is no longer a topological invariant, we will see that supersymmetry restricts its dependence on the geometry of~$\CM$ to a finite-dimensional parameter space. We have also analyzed the parameter dependence of~$Z_{\CM}$ using the full non-linear background supergravity formalism. This requires additional technical machinery, which obscures the conceptual simplicity of the results. In this paper we only present the linearized analysis; the non-linear analysis will be discussed elsewhere.

Our approach hinges on a detailed understanding of the relevant supercurrent multiplets. The different supercurrents that arise in theories with four supercharges were analyzed in~\refs{\KomargodskiRB, \DumitrescuIU}. In the presence of a~$U(1)_R$ symmetry, the appropriate supercurrent is the~$\CR$-multiplet.\foot{Some field theories admit more than one kind of supercurrent multiplet. They can therefore be coupled to different background supergravity fields, which give rise to different classes of supersymmetric manifolds~$\CM$. See~\refs{\FestucciaWS,\SamtlebenGY\DumitrescuHA\LiuBI-\DumitrescuAT} for a discussion in four dimensions.} Below, we will review the~$\CR$-multiplet, the associated background supergravity fields, and the geometric structures that are needed to preserve supersymmetry. We also identify supersymmetric configurations for background gauge fields. This will enable us to summarize our results on the parameter dependence of supersymmetric partition functions. We first consider~$\CN=1$ theories in four dimensions (sections~1.1-1.3), and then give a parallel discussion for~$\CN=2$ theories in three dimensions (sections~1.4-1.6). In section~1.7 we sketch the implications of our results for two examples of recent interest. Section~1.8 contains an outline of the paper.

\subsec{Supersymmetric Backgrounds in Four Dimensions}

In four-dimensional~$\CN=1$ theories, the~$\CR$-multiplet contains the operators\foot{In the presence of continuous Abelian flavor symmetries that can mix with the~$R$-symmetry, the~$\CR$-multiplet is not unique. We will assume throughout that a fixed~$\CR$-multiplet has been chosen. As we will see below, some supersymmetric backgrounds impose restrictions on the allowed~$R$-charges.}
\eqn\rmultops{j_\mu^{( R )}~, \quad  S_{\alpha\mu}~, \quad {\, \t S^\alphadot}_{\mu}~,\quad T_{\mu\nu}~,\quad \CF_{\mu\nu}~,}
where~$j_\mu^{(R )}$ is the~$R$-current, $S_{\mu\alpha}$ and~${\, \t S^\alphadot}_{\mu}$ are the supersymmetry currents,\foot{In Euclidean signature, all spinors are complex. Moreover, left- and right-handed spinors in four dimensions are not related by complex conjugation. Our conventions are summarized in appendix~A.} $T_{\mu\nu}$ is the energy-momentum tensor, and~$\CF_{\mu\nu}$ is a closed two-form, which gives rise to a two-form current~$i \ep_{\mu\nu\rho\lambda} \CF^{\rho\lambda}$. The supergravity background fields that couple to these conserved currents are given by\foot{Our~$A_\mu^{( R)}$ is related to the~$R$-symmetry gauge field~$A_\mu$ used in~\DumitrescuHA\ by~$A^{(R )}_\mu = A_\mu - {3 \over 2} V_\mu$, where~$V_\mu$ is defined in~(1.3).}
\eqn\fdsgra{A^{( R)}_\mu~, \quad \Psi_{\alpha\mu}~,\quad {\, \t \Psi^\alphadot}_\mu~,\quad \Delta g_{\mu\nu}~,\quad B_{\mu\nu}~,}
where~$A^{(R )}_\mu$ is an Abelian gauge field, $\Psi_{\alpha\mu}$ and~${\, \t \Psi^\alphadot}_\mu$ are gravitinos, $\Delta g_{\mu\nu}$ is the linearized metric (so that~$g_{\mu\nu} = \delta_{\mu\nu} + \Delta g_{\mu\nu}$), and~$B_{\mu\nu}$ is a two-form gauge field. We will also use the three-form field strength~$H_{\mu\nu\rho}$ of~$B_{\mu\nu}$, and its dual~$V^\mu$, which is a conserved vector,
\eqn\vdef{H = dB~, \qquad V^\mu = {i\over 6}\ep^{\mu\nu\rho\lambda} H_{\nu\rho\lambda}~, \qquad \grad_\mu V^\mu = 0~.}
The linearized supergravity couplings around flat space are given by
\eqn\fdlinlag{\Delta {\scr L} = - \half \Delta g^{\mu\nu} T_{\mu\nu} + A^{(R )\mu} j_\mu^{( R )} + {i \over 4} \ep^{\mu\nu\rho\lambda} \CF_{\mu\nu} B_{\rho\lambda} + ({\rm fermions})~.
}
Their non-linear completion is the new minimal supergravity theory of~\refs{\SohniusTP, \SohniusFW}.

A configuration of~$g_{\mu\nu}$, $A^{( R)}_\mu$, $B_{\mu\nu}$ on a four-manifold~$\CM_4$ preserves a supercharge~$Q$ or~$\t Q$ of~$R$-charge~$-1$ or~$+1$, if the supergravity variations~$\delta_Q \Psi_{\mu}$ or~$\delta_{\t Q} {\t \Psi}_\mu$ vanish~\FestucciaWS. (The other variations~$\delta_{ Q} \t \Psi_\mu$ and~$\delta_{\t Q} \Psi_\mu$ vanish automatically, due to the~$R$-symmetry.) This happens when the corresponding spinor parameters~$\zeta_\alpha$ or~$\t \zeta^\alphadot$, which have~$R$-charge~$+1$ and~$-1$, satisfy the following (generalized) Killing spinor equations,
\eqn\speq{\eqalign{& \big(\grad_\mu - i A^{( R)}_\mu\big) \zeta = { i\over 2} V_\mu \zeta - i V^\nu \sigma_{\mu\nu} \zeta~,\cr
&\big(\grad_\mu + i A^{( R)}_\mu\big) \t \zeta =  -{i \over 2} V_\mu \t \zeta  + i V^\nu \t \sigma_{\mu\nu} \t \zeta~.}}
It was shown in~\refs{\DumitrescuHA, \KlareGN} that a solution of the first equation exists if and only if~$\CM_4$ admits an integrable complex structure~${J^\mu}_\nu$ and~$g_{\mu\nu}$ is a compatible Hermitian metric.\foot{A solution with~$V^\mu = 0$ exists if and only if~$\CM_4$ is K\"ahler. This corresponds to the twisted theories studied in~\refs{\JohansenAW,\WittenEV}.} (See appendix~B for a review of complex manifolds.) In this case the Killing spinor~$\zeta_\alpha$ is everywhere non-zero and determines the complex structure,\foot{This differs from the conventions of~\DumitrescuHA\ by a sign, so that the holomorphic coordinates used there correspond our anti-holomorphic coordinates.}
\eqn\ccbil{{J^\mu}_\nu = -{2 i \over |\zeta|^2} \zeta^\dagger {\sigma^\mu}_\nu \zeta~.}
The spinor also defines a nowhere vanishing~$(2,0)$-form~$\zeta \sigma_{\mu\nu}\zeta$. The supercharge~$Q$ corresponding to~$\zeta_\alpha$ transforms as a scalar under holomorphic coordinate transformations twisted by suitable~$R$-symmetry transformations~\DumitrescuHA. As was discussed above, this is essential for the validity of our linearized analysis. 

The complex structure and the metric constrain, but do not completely determine, the other supergravity background fields. The field strength of~$B_{\mu\nu}$ takes the form
\eqn\fdthreform{H= {i \over 2}\, dJ + W~, \qquad W \in \Lambda^{2,1}~, \qquad \b \d W = 0~.}
Here~$(dJ)_{\mu\nu\rho} = \d_\mu J_{\nu\rho} + \d_\nu J_{\rho\mu} + \d_\rho J_{\mu\nu}$ and~$W_{i j \b k}$ is a~$\b\d$-closed~$(2,1)$-form. Note that the~$B$-field is only determined up to a flat piece. The~$R$-symmetry gauge field is given by 
\eqn\fdbgfields{\eqalign{& A^{( R )}_\mu = \hat A_\mu - {1 \over 4} \left(2{\delta_\mu}^\nu - i {J_\mu}^\nu\right) \grad_\rho {J^\rho}_\nu~,\cr
&\hat A_i={i\over 8} \d_i \log g ~, \qquad \hat A_{\b i}= -{i\over 8}\partial_{\b i} \log g~,}}
up to a globally defined complex gauge transformation. The formula for~$\hat A_\mu$ is only valid in holomorphic coordinates adapted to~${J^\mu}_\nu$. However, its field strength can be written in fully covariant form using the Riemann curvature tensor and the complex structure.

We will also consider backgrounds that admit another Killing spinor~$\t\zeta^\alphadot$, which solves the second equation in \speq. (See~\DumitrescuHA\ for additional details.)  In this case we obtain a nowhere vanishing vector~$K^\mu = \zeta \sigma^\mu \t \zeta$, which is Killing and anti-holomorphic with respect to the complex structure~${J^\mu}_\nu$ in~\ccbil. As long as~$K$ commutes with its complex conjugate, $[K, \b K] = 0$, we can choose holomorphic coordinates~$w, z$ adapted to~${J^\mu}_\nu$ and~$K^\mu$ such that~$K = \d_{\b w}$.\foot{The spinor~$\t \zeta^\alphadot$ defines another integrable complex structure
\eqn\tjdef{{\, \t J^\mu}_\nu = - {2i \over |\t \zeta|^2} \t \zeta^\dagger {\, \t \sigma^\mu}_\nu \t \zeta~,}
whose holomorphic coordinates are~$w$ and~$\b z$. The bilinear~$\t \zeta\,\t \sigma_{\mu\nu} \t \zeta$ is a nowhere vanishing~$(2,0)$-form with respect to~${\, \t J^\mu}_\nu$, so that it only has a~$w\b z$-component. We will also need the fact that~$J^{\mu\nu} + {\t J}^{\mu\nu}$ is proportional to~$K^\mu \b K^\nu - K^\nu \b K^\mu$.} The metric then takes the form
\eqn\metricTtwofiber{ds^2 = \Omega(z, \b z)^2 \left((dw + h (z,\b z)  d z)(d\b w + \b h (z,\b z)  d \b z) + c(z,\b z)^2  dz d\b z \right)~,}
which describes a~$T^2$ fibration over a Riemann surface~$\Sigma$.\foot{In general, the orbits of~$\Re K$ and~$\Im K$ do not close, which implies the existence of additional Killing vectors. See section~5.1 of~\ClossetRU\ for a related discussion in three dimensions.} Now both supercharges~$Q$ and~$\t Q$ are scalars under holomorphic coordinate transformations of the form~$w' = w + F(z)$, $z' = G(z)$, which preserve~$K$ and the form of the metric in~\metricTtwofiber. The other background fields are still given by~\fdthreform\ and~\fdbgfields, except that only~$W_{wz \b z}$ may be non-zero.  

Manifolds that preserve four supercharges are very restricted. In the compact case, they include the flat torus~$T^4$ and~$S^3 \times S^1$ with the usual round metric~\refs{\FestucciaWS,\DumitrescuHA}.

\subsec{Background Vector Fields in Four Dimensions}

If the field theory possesses a continuous flavor symmetry, which commutes with the supercharges, we can couple it to a background gauge field. For simplicity, we will focus on the Abelian case throughout this paper. (The generalization to the non-Abelian case is straightforward, and we will refer to it occasionally.) In flat space, the conserved flavor current~$j_\mu$ is embedded in a real linear multiplet~$\CJ$, along with a scalar~$J$ and fermions~$j_\alpha$ and~$\t j^\alphadot$. These operators couple to a background gauge field~$A_\mu$, a scalar~$D$, and gauginos~$\lambda_\alpha$ and~$\t \lambda^\alphadot$. 

In order to determine which configurations of the bosonic fields~$A_\mu$ and~$D$ are consistent with the preserved supercharge~$Q$ on a complex manifold~$\CM_4$, we follow the same logic as for the gravity multiplet and set the variation of the gaugino to zero,~$\delta_Q \lambda = 0$. From the corresponding formula in new minimal supergravity, 
\eqn\gauginotrans{\delta_Q \lambda = i \zeta D + \sigma^{\mu\nu} \zeta F_{\mu\nu}~,}
where~$F_{\mu\nu} = \d_\mu A_\nu - \d_\nu A_\mu$ is the field strength of~$A_\mu$. Using the fact that~$\zeta_\alpha$ determines the complex structure through~\ccbil\ and the nowhere vanishing~$(2,0)$-form~$\zeta \sigma_{\mu\nu} \zeta$, we find that setting~\gauginotrans\ to zero implies the following constraints,
\eqn\gaugepreserve{
F_{\b i \b j}=0~, \qquad D= -\half J^{\mu\nu} F_{\mu\nu}~.
}
As we will review in section~2.3, the first condition implies that~$A_\mu$ defines a holomorphic line bundle over the complex manifold~$\CM_4$. The second condition determines~$D$ in terms of the~$(1,1)$ part of the field strength. (These formulas generalize to the non-Abelian case, where the background gauge field defines a holomorphic vector bundle over~$\CM_4$.) As for the background fields in the supergravity multiplet, we do not assume that~$A_\mu$, $D$ are real.

For the case of two supercharges~$Q$ and~$\t Q$ considered around~\metricTtwofiber, we must also set~$\delta_{\t Q} \t \lambda = 0$, so that we find additional constraints,\foot{The variation of~$\t \lambda^\alphadot$ is given by
\eqn\varlambtil{\delta_{\t Q} \t \lambda = - i \t \zeta D +\t \sigma^{\mu\nu} \t \zeta F_{\mu\nu}~.}
To obtain the constraints that follow from setting it to zero, we use the bilinears determined by~$\t \zeta^\alphadot$, which are summarized in footnote~9.}
\eqn\twoscfdbundle{F_{w \b w} = F_{z \b w} = 0~.}
If we choose to restrict ourselves to real~$A_\mu$, this implies that the only non-vanishing component of the curvature is~$F_{z\b z}$. 

In order to preserve four supercharges, $A_\mu$ must be a flat connection, i.e.~$F_{\mu\nu} = 0$.

\subsec{Summary of Results in Four Dimensions}

We can now summarize the data that enters the supersymmetric Lagrangian on the complex manifold~$\CM_4$, in the presence of one supercharge~$Q$:
\medskip
\item{$\bullet$} The integrable complex structure~${J^\mu}_\nu$.
\item{$\bullet$} A compatible Hermitian metric~$g_{i \b j}$.
\item{$\bullet$} The~$(2,1)$-form~$W$ in~\fdthreform, which satisfies~$\b \d W = 0$.
\item{$\bullet$} Abelian background gauge fields, which satisfy~\gaugepreserve\ and define holomorphic line bundles. (In the non-Abelian case, holomorphic vector bundles.) 
\item{$\bullet$} Coupling constants, including the parameters in the original flat-space Lagrangian.\foot{On some backgrounds, it is possible to promote certain coupling constants to background fields, without breaking the preserved supercharges. See section~1.5 for an example.}
\medskip
\noindent This data can often be varied continuously. For complex structures, Hermitian metrics, and holomorphic line bundles, this requires basic aspects of deformation theory, which are reviewed in section~2. A key fact is that moduli spaces of complex structures and holomorphic line bundles are parametrized by finitely many complex parameters, as long as~$\CM_4$ is compact. The Lagrangian also depends on various discrete choices, such as the topology of~$\CM_4$ and the topology of the gauge bundle for~$A_\mu$, or discrete coupling constants.\foot{See~\AharonyHDA\ for a recent example. Sometimes, couplings that are continuous in flat space must be quantized in non-trivial backgrounds (see below).}

We would like to understand how the partition function~$Z_{\CM_4}$ behaves as we vary the continuous data summarized above. For the geometric data, this is analyzed in section~3, where we study small deformations around flat space using the~$Q$-cohomology of the operators in the~$\CR$-multiplet or the flavor current multiplet~$\CJ$. This leads to the following constraints on the partition function~$Z_{\CM_4}$:
\medskip
\item{$\bullet$} For a fixed complex structure, $Z_{\CM_4}$ does not depend on the Hermitian metric.
\item{$\bullet$} $Z_{\CM_4}$ is a locally holomorphic function of the complex structure moduli.
\item{$\bullet$} $Z_{\CM_4}$ only depends on the~$(2,1)$-form~$W$ through its Dolbeault cohomology class in~$H^{2,1}(\CM_4)$. Moreover, $Z_{\CM_4}$ does not depend on~$W$ at all, unless the theory has Fayet-Iliopoulos (FI) terms or a target space whose K\"ahler form is not exact.\foot{Under these conditions the flat-space theory does not admit a Ferrara-Zumino (FZ) supercurrent multiplet~\refs{\KomargodskiPC,\KomargodskiRB}.}
\item{$\bullet$} $Z_{\CM_4}$ only depends on background gauge fields through the corresponding holomorphic line (or vector) bundles. It is a locally holomorphic function of the bundle moduli.
\medskip
\noindent We have also obtained the following results about the dependence of the partition function on continuous couplings:
\medskip
\item{$\bullet$} $Z_{\CM_4}$ does not depend on~$D$-terms.\foot{By a~$D$-term we mean a term in the Lagrangian that is constructed from the top component of a well-defined superfield. This does not apply to FI-terms, which reside in the top components of gauge non-invariant vector superfields.}

\item{$\bullet$} $Z_{\CM_4}$ does not depend on chiral~$F$-terms, but it may be a locally holomorphic function of anti-chiral~$F$-terms. 
\medskip
\noindent We will not present the proof of these two statements here, since they do not readily follow from a linearized analysis around flat space. However, it is clear that the corresponding terms in the flat-space Lagrangian are~$Q$-exact. 

In the presence of additional supercharges, the partition function is further constrained. For the case of two supercharges~$Q$ and~$\t Q$ of opposite~$R$-charge, the dependence of~$Z_{\CM_4}$ on some complex structure and holomorphic line bundle moduli, as well as on anti-chiral~$F$-terms, disappears.

\subsec{Supersymmetric Backgrounds in Three Dimensions}

The~$\CR$-multiplet in three-dimensional~$\CN=2$ theories was studied in~\refs{\DumitrescuIU,\ClossetVG,\ClossetVP}. It contains the following operators:
\eqn\tdrmultops{j_\mu^{( R)}~, \quad S_{\alpha\mu}~, \quad \t S_{\alpha\mu}~, \quad T_{\mu\nu}~, \quad j_\mu^{( Z)}~, \quad J^{(Z)}~.}
Here~$j_\mu^{( R)}$ is the~$R$-current, $S_{\alpha\mu}$ and~$\t S_{\alpha\mu}$ are the supersymmetry currents, $T_{\mu\nu}$ is the energy-momentum tensor, $j_\mu^{(Z)}$ is the current that gives rise to the central charge~$Z$ in the flat-space supersymmetry algebra, and~$J^{(Z)}$ is a well-defined scalar, which gives rise to a two-form current~$i \ep_{\mu\nu\rho} \d^\nu J^{(Z)}$. The corresponding supergravity theory was recently studied in superspace~\refs{\KuzenkoXG\KuzenkoRD-\KuzenkoBC}, as well as in components~\refs{\ClossetVG,\ClossetVP,\ClossetRU}. The supergravity background fields that couple to the operators in~\tdrmultops\ are given by\foot{Again, our~$A_\mu^{( R)}$ is related to the~$R$-symmetry gauge field~$A_\mu$ used in~\ClossetRU\ by~$A^{(R )}_\mu = A_\mu - {3 \over 2} V_\mu$, where~$V_\mu$ is defined in~(1.17).} 
\eqn\tdsgra{A_\mu^{( R)}~, \quad \Psi_{\alpha\mu}~, \quad \t \Psi_{\alpha\mu}~, \quad \Delta g_{\mu\nu}~, \quad C_\mu~, \quad H~,}
where~$A^{( R)}_\mu$ and~$C_\mu$ are Abelian gauge fields, $\Psi_{\alpha\mu}$ and~$\t \Psi_{\alpha\mu}$ are the gravitinos, $\Delta g_{\mu\nu}$ is the linearized metric, and~$H$ is a scalar. Note that the graviphoton~$C_\mu$, which couples to the dimension three operator~$j_\mu^{( Z)}$, is dimensionless. We will also use the dual graviphoton field strength~$V^\mu$, which is a conserved vector,
\eqn\tddfsdef{V^\mu = -i \ep^{\mu\nu\rho} \d_\nu C_\rho~, \qquad \grad_\mu V^\mu = 0~.}
The linearized coupling of the~$R$-multiplet to the supergravity fields takes the form
\eqn\linearizedthreed{\Delta \SL=-\half \Delta g^{\mu\nu} T_{\mu\nu} + A^{( R)\mu} j_\mu^{( R)}+C^\mu j_\mu^{(Z)}+HJ^{(Z)} + ({\rm fermions})~.}
The formulas above are closely related to their four-dimensional counterparts in section~1.2. This was used in~\ClossetRU\ to obtain the non-linear completion of~\linearizedthreed\ from new minimal supergravity in four dimensions~\refs{\SohniusTP, \SohniusFW}. 

The variations of~$\Psi_\mu$ and~$\t \Psi_\mu$ lead to generalized Killing spinor equations in three dimensions. A supercharge~$Q$ or~$\t Q$ corresponds to a spinor~$\zeta_\alpha$ or~$\t \zeta_\alpha$ that satisfies
\eqn\ksei{ \big(\grad_\mu - i A^{( R)}_\mu\big) \zeta =  -\half H \gamma_\mu\zeta + {i\over 2} V_\mu\zeta -\half \ep_{\mu\nu\rho}V^\nu \gamma^\rho\zeta~,}
\eqn\kseii{\big(\grad_\mu + iA^{( R)}_\mu\big) \t\zeta =  -\half H \gamma_\mu\t\zeta - {i\over 2} V_\mu\t\zeta +\half\ep_{\mu\nu\rho}V^\nu \gamma^\rho\t\zeta~.}
These equations were analyzed in~\ClossetRU\ (see also~\refs{\HoweZM,\KlareGN}). A solution of the first equation exists if and only if~$\CM_3$ admits a certain mathematical structure -- a transversely holomorphic foliation (THF) with a compatible transversely Hermitian metric -- whose properties closely parallel those of integrable complex structures and Hermitian metrics.\foot{In~\ClossetRU, the conditions for supersymmetry were stated in terms of an almost contact metric structure that satisfies a certain integrability condition. This structure is equivalent to a THF with a transversely Hermitian metric. We are grateful to Maxim Kontsevich for pointing this out to us.} These structures have been studied in the mathematical literature, see for instance~\refs{\DKi\GM-\Haef}. Three-manifolds admitting a THF are very restricted and were classified in~\refs{\BG\Brunella-\Ghys}. Topologically, all of them are Seifert manifolds or~$T^2$ bundles over~$S^1$.

A review of THFs can be found in section~5.1. For now we will only need to know that they can be described in terms of a nowhere vanishing one-form~$\eta_\mu$ and a tensor~${\Phi^\mu}_\nu = - {\ep^\mu}_{\nu\rho}\eta^\rho$,\foot{This differs from the definition in~\ClossetRU\ by a sign.} which is defined using a compatible metric. The role of~${\Phi^\mu}_\nu$ is similar to that of the complex structure tensor~${J^\mu}_\nu$. The THF implies the existence of adapted coordinates~$\tau, z, \b z$, in which
\eqn\tdadcoordint{\eta = d\tau + h (\tau, z, \b z) dz + \b h(\tau, z, \b z) d\b z~,\qquad ds^2 = \eta^2 + c(\tau, z, \b z)^2 dz d\b z~.}
Adapted coordinate patches are related by transformations of the form $\tau' = \tau + t(z, \b z)$ and~$z' = f(z)$, where~$f(z)$ is holomorphic. As in four dimensions, the supercharge~$Q$ corresponding to~$\zeta_\alpha$ transforms as a scalar under such adapted coordinate changes after a suitable twist by the~$R$-symmetry, which allows us to study the dependence of the partition function on the background geometry by considering small deformations around flat space.  The one-form~$\eta_\mu$ can be expressed in terms of~$\zeta_\alpha$, which is everywhere non-zero,
\eqn\etazeta{\eta_\mu = {1 \over |\zeta|^2} \zeta^\dagger \gamma_\mu \zeta~.}
The spinor also determines a nowhere vanishing one-form~$\zeta \gamma_\mu \zeta$, which in adapted coordinates only has a~$z$-component. 

The dual graviphoton field strength~$V^\mu$ is determined in terms of~$\eta_\mu$ and~$g_{\mu\nu}$, up to an ambiguity parametrized by a vector~$U^\mu$ and a function~$\kappa$, both of which may be complex, 
\eqn\tdbgs{V^\mu = \ep^{\mu\nu\rho} \d_\nu \eta_\rho + U^\mu + \kappa \eta^\mu~,\qquad {\Phi^\mu}_\nu U^\nu = - i U^\mu, \qquad \grad_\mu \left(U^\mu + \kappa \eta^\mu\right) = 0~.}
This only determines the graviphoton up to a flat connection. The other supergravity background fields are given by
\eqn\tdrgf{\eqalign{& H = - \half \grad_\mu \eta^\mu + {i \over 2} \ep^{\mu\nu\rho} \eta_\mu \d_\nu \eta_\rho + i \kappa~,\cr
&A^{( R)}_\mu = \hat A_\mu +  \ep_{\mu\nu\rho} \d^\nu \eta^\rho + {i \over 4} \eta_\mu \grad_\nu \eta^\nu - {i \over 2} \eta^\nu \grad_\nu \eta_\mu~,\cr
& \hat A_\mu = {1 \over 8} {\Phi^\nu}_\mu \d_\nu \log g~,}}
up to a globally well-defined complex gauge transformation for~$A_\mu^{( R)}$. As in four dimensions, the expression for~$\hat A_\mu$ in~\tdrgf\ is only valid in adapted~$\tau, z,\b z$ coordinates, while its field strength can be written covariantly using the Riemann curvature tensor and~$\eta_\mu$.

Conditions for the presence of additional supercharges were analyzed in~\ClossetRU.

\subsec{Background Vector Fields in Three Dimensions}

As before, we can couple an Abelian flavor current~$j_\mu$, which resides in a real linear multiplet~$\CJ$ together with two other scalar operators~$J, K$ as well as the fermions~$j_\alpha, \t j_\alpha$, to a background vector multiplet containing a gauge field~$A_\mu$, two scalars~$D, \sigma$, and the gauginos~$\lambda_\alpha, \t \lambda_\alpha$. The configurations that are consistent with the preserved supercharge~$Q$ on a three-manifold with a THF must satisfy~$\delta_Q \lambda = 0$. It was shown in~\ClossetRU\ that
\eqn\tdgauginovar{\delta_Q \lambda = i \zeta \left(D + \sigma H\right) - {i \over 2}\gamma_\mu \zeta \ep^{\mu\nu\rho} F_{\nu\rho} -i\gamma^\mu \zeta \left( \d_\mu \sigma + i V_\mu \sigma\right)~.}
Unlike~\gauginotrans, this explicitly contains the background supergravity fields~$V^\mu$ and~$H$. To derive the constraints that follow from setting~\tdgauginovar\ to zero, we substitute~\tdbgs\ and~\tdrgf. We then use~\tdadcoordint\ and~\etazeta, as well as the fact that~$\zeta\gamma_\mu\zeta$ only has a~$z$-component in adapted coordinates. These constraints are conveniently expressed in terms of a (generally complex) gauge field~$\SA_\mu$ and its field strength,
\eqn\sadef{\SA_\mu = A_\mu + i \sigma \eta_\mu~,\qquad {\scr F}_{\mu\nu} = \d_\mu \SA_\nu - \d_\nu \SA_\mu~.}
In adapted~$\tau, z, \b z$ coordinates, the configurations consistent with the supercharge~$Q$ satisfy
\eqn\tdgfcons{{\scr F}_{\tau \b z} = 0~,\qquad D = - \half \Phi^{\mu\nu} \SF_{\mu\nu} + \eta^\mu \d_\mu \sigma + \sigma \left(\half \grad_\mu \eta^\mu - {i \over 2} \ep^{\mu\nu\rho} \eta_\mu \d_\nu \eta_\rho\right)~.}
Note that~$U^\mu$ and~$\kappa$ do not appear in these formulas. The first condition is reminiscent of the condition~$F_{\b i \b j} = 0$ for holomorphic line bundles over complex manifolds. As we will see in section~5, an analogous structure can be defined on three-manifolds that carry a THF. Note that we can always satisfy~\tdgfcons\ by choosing~$A_\mu = - i \sigma \eta_\mu$ for arbitrary complex~$\sigma$.

\subsec{Summary of Results in Three Dimensions}

Using the background fields discussed above, we can write supersymmetric Lagrangians on~$\CM_3$ that are invariant under the preserved supercharge~$Q$~\ClossetRU. These Lagrangians depend on the following continuously variable data:
\medskip
\item{$\bullet$} The transversely holomorphic foliation (THF).
\item{$\bullet$} The compatible transversely Hermitian metric in~\tdadcoordint.
\item{$\bullet$} The ambiguity in~$V^\mu$ and~$H$, parametrized by~$U^\mu$ and~$\kappa$ in~\tdbgs\ and~\tdrgf.
\item{$\bullet$} Abelian background gauge fields, which satisfy~\tdgfcons\ and define holomorphic line bundles over~$\CM_3$.
\item{$\bullet$} Continous coupling constants.
\medskip
\noindent Real masses correspond to real, constant values of~$\sigma$ in a background vector multiplet. The Lagrangian also depends on data that is not continuously variable, such as the topology of~$\CM_3$, and quantized coupling constants, e.g.~Chern-Simons levels. 

Our analysis of the partition function~$Z_{\CM_3}$ proceeds as in four dimensions. In section~5, we review the theory of infinitesimal deformations for THFs, transversely Hermitian metrics, and holomorphic line bundles, which closely parallels the deformation theory of complex manifolds reviewed in section~2. In particular, we introduce a notion of~$(p,q)$-forms and a~$\b \d$-like operator~$\t \d$, which are used to define various cohomology groups that describe infinitesimal deformations. As in four dimensions, moduli spaces of THFs and holomorphic line bundles are parametrized by a finite number of complex parameters, if~$\CM_3$ is compact.

In section~6, we use this machinery to study how the partition function~$Z_{\CM_3}$ depends on the geometry of~$\CM_3$ by considering small deformations around flat space and using the~$Q$-cohomology of the~$\CR$-multiplet or the flavor current multiplet~$\CJ$. We find:
\medskip
\item{$\bullet$} Given a fixed THF, $Z_{\CM_3}$ does not depend on the transversely Hermitian metric.
\item{$\bullet$} $Z_{\CM_3}$ is a locally holomorphic function of the complex moduli that parametrize deformations of the THF.
\item{$\bullet$} $Z_{\CM_3}$ only depends on~$U^\mu$ and~$\kappa$ through the~$\t \d$-cohomology class of the closed~$(1,1)$-form~$W_{\mu\nu} = \half \ep_{\mu\nu\rho} (U^\rho + \kappa \eta^\rho)$. Moreover, the dependence on~$W_{\mu\nu}$ drops out completely, unless the theory has real masses (including FI-terms) or a target space whose K\"ahler form is not exact.\foot{As in four dimensions, these are the conditions under which the flat-space theory does not admit an FZ-multiplet~\DumitrescuIU.}
\item{$\bullet$} $Z_{\CM_3}$ only depends on background gauge fields through the corresponding holomorphic line bundles. It is a locally holomorphic function of the bundle moduli. 
\medskip
\noindent As in four dimensions, we also state without proof:
\medskip
\item{$\bullet$} $Z_{\CM_3}$ does not depend on~$D$-terms.
\item{$\bullet$} $Z_{\CM_3}$ does not depend on chiral~$F$-terms, but it may be a locally holomorphic function of anti-chiral~$F$-terms.
\medskip
\noindent However, $Z_{\CM_3}$ can depend on real masses through background vector multiplets. 

\subsec{Applications}

In sections~4 and~7 we will explore the implications of our general results in various examples. Here we briefly summarize some results in two cases of recent interest.

Four-dimensional~$\CN=1$ theories with a~$U(1)_R$ symmetry can be placed on~$S^3 \times \R$ while preserving four supercharges~\refs{\SenPH, \RomelsbergerEG, \FestucciaWS}. The supersymmetric index~$\CI(p,q,u)$ with fugacities~$p,q,u$ is defined by a trace over states on~$S^3$ in Hamiltonian quantization~\refs{\RomelsbergerEG, \KinneyEJ,\FestucciaWS}. Alternatively, it corresponds to a partition function on~$S^3 \times S^1$. All complex manifolds diffeomorphic to~$S^3 \times S^1$ are primary Hopf surfaces~\Kodairasone. They contain a two-parameter family, whose complex structure moduli precisely coincide with~$p$ and~$q$. Similarly, $u$ is a holomorphic line bundle modulus. Explicit computations of the index (see for instance~\refs{\RomelsbergerEC\DolanQI \SpiridonovZA-\SpiridonovHF}) give rise to a class of special functions that has recently been studied in the mathematical literature (see for instance~\refs{\Rains\Spiridonov-\Bult,\DolanQI\SpiridonovZA-\SpiridonovHF} and references therein). The fact that~$\CI(p,q,u)$ can be viewed as a holomorphic function of complex structure and holomorphic line bundle moduli on certain primary Hopf surfaces may shed new light on some of their interesting properties. 

Following~\refs{\KapustinKZ\JafferisUN-\HamaAV}, who studied partition functions of three-dimensional $\CN=2$ theories with a~$U(1)_R$ symmetry on round spheres preserving four supercharges, much recent work has focused on squashed spheres~\refs{\HamaEA\ImamuraUW\ImamuraWG\MartelliFU\NishiokaHAA\MartelliAQA\AldayLBA\NianQWA-\TanakaDCA}, i.e.~three-manifolds that are diffeomorphic to~$S^3$ but carry a more general metric with less symmetry. Explicit computations in these examples lead to the following observations about the partition function on squashed spheres:
\medskip
\item{a)} $Z_{S^3}$ only depends on the squashing through a single complex parameter, usually called~$b$, where~$b=1$ corresponds to the round~$S^3$ of~\refs{\KapustinKZ\JafferisUN-\HamaAV}.
\item{b)} Some squashings do not affect $Z_{S^3}$, i.e.~they give~$b=1$.
\medskip
\noindent We will show how these observations follow from our general results. In order to explain~a), we demonstrate that all known examples of squashed spheres belong to a one-parameter family of THFs on~$S^3$. (An example that does not belong to this family and preserves a single supercharge is presented in appendix~D.) The trivial squashings in~b) do not deform the THF, although they change the compatible transversely Hermitian metric, and hence they do not affect~$Z_{S^3}$.

\subsec{Outline}

Section~2 reviews basic aspects of deformation theory for complex structures, Hermitian metrics, and holomorphic line bundles.

In section~3, we study the the partition function~$Z_{\CM_4}$ by considering small deformations around flat space. The main tool is the~$Q$-cohomology of the flat-space~$\CR$-multiplet, which is used to constrain the dependence of~$Z_{\CM_4}$ on the geometry of~$\CM_4$ and the ambiguity parametrized by the~$(2,1)$-form~$W$ in~\fdthreform. The dependence on background gauge fields is constrained by the~$Q$-cohomology of the flavor current~$\CJ$. We also comment on additional constraints due to multiple supercharges.

Section~4 describes several four-dimensional examples. Complex manifolds diffeomorphic to~$S^3 \times S^1$ and their relation to the supersymmetric index are discussed in detail. We also examine complex manifolds diffeomorphic to~$L(r,s) \times S^1$ and~$S^2 \times T^2$.

In section~5 we summarize basic properties of THFs. We define three-dimensional analogues of~$(p,q)$-forms, the~$\b \d$-operator, and holomorphic line bundles. These feature prominently in the deformation theory of THFs, which we also review.

In Section~6 we constrain the dependence of the partition function~$Z_{\CM_3}$ on the geometry of~$\CM_3$, the ambiguity parametrized by~$U^\mu$ and~$\kappa$ in~\tdbgs\ and~\tdrgf, as well as background gauge fields by considering small deformations around flat space and analyzing the~$Q$-cohomology of the corresponding current multiplets. 

Section~7 describes several examples in three dimensions. We discuss THFs and holomorphic line bundles on three-manifolds diffeomorphic to~$S^3$ and~$S^2 \times S^1$, and we comment on the implications for the corresponding partition functions.

Appendix~A summarizes our conventions. Basic aspects of complex manifolds are reviewed in appendix~B. Appendix~C contains some additional material related to section~3.4. In appendix~D we consider manifolds diffeomorphic to~$S^3 \times S^1$ or~$S^3$ that preserve only one supercharge. In appendix~E we review several examples of squashed spheres that have appeared in the literature. Appendix F contains additional material related to section~7.2.

\newsec{Deformation Theory in Four Dimensions}

Complex manifolds often belong to a continuous moduli space of manifolds that share the same underlying differentiable structure but possess different complex structures. Locally, this moduli space can be studied by considering small deformations around a given complex manifold~$\CM$. The theory of complex structure deformations is a well-established subject (see for instance~\Kodairabook). We review those aspects that are used in section~3 to study the dependence of the partition function~$Z_{\CM_4}$ on the geometry of the complex manifold~$\CM_4$. Analyzing the dependence of~$Z_{\CM_4}$ on background gauge fields requires the deformation theory of holomorphic vector bundles over~$\CM_4$ (see for instance~\Kobayashi). Here we outline the theory for the simple case of Abelian gauge fields and holomorphic line bundles. 

\subsec{Complex Structures}

Given a complex manifold~$\CM$, we consider infinitesimal deformations~$\Delta {J^\mu}_\nu$ of its complex structure, i.e.~we work to first order in~$\Delta {J^\mu}_\nu$. In holomorphic coordinates adapted to~${J^\mu}_\nu$, the requirement that~${J^\mu}_\nu + \Delta {J^\mu}_\nu$ is an almost complex structure implies
\eqn\acreq{\Delta{J^i}_j = \Delta {J^{\b i}}_{\b j} = 0~.}
Additionally requiring that~${J^\mu}_\nu + \Delta {J^\mu}_\nu$ is an integrable complex structure leads to 
\eqn\integrability{ \d_{\b j} \, \left(\Delta {J^i}_{\b k}\right) - \d_{\b k} \, \big(\Delta {J^i}_{\b j}\big)=0~,}
and its complex conjugate. Introducing~$\Theta^i = \Delta {J^i}_{\b j} d{\b z}^{\b j}$, viewed as a~$(0,1)$-form with coefficients in the holomorphic tangent bundle~$T^{1,0}\CM$, we can succinctly express~\integrability\ as
\eqn\intshort{\b \d \Theta^i = 0~.}
Not all choices of~$\Delta {J^\mu}_\nu$ lead to a new complex structure on~$\CM$. An infinitesimal diffeomorphism parametrized by a real vector field~$\ep^\mu$ gives rise to~$\Delta {J^i}_{\b j} = {\left(\CL_\ep J\right)^i}_{\b j} = 2 i \d_{\b j} \ep^i$ and its complex conjugate, where~$\CL_\ep$ is the Lie derivative along~$\ep^\mu$. A trivial complex structure deformation thus corresponds to
\eqn\exthet{\Theta^i =2 i \b  \d \ep^i~.}
Quotienting by such trivial deformations, we conclude that infinitesimal complex structure deformations are parametrized by the cohomology class of~$\Theta^i$ in the Dolbeault cohomology with coefficients in~$T^{1,0} \CM$,
\eqn\thcoho{\left[ \Theta^i \right] \in H^{0,1}\left(\CM, T^{1,0}\CM\right)~.}

If~$\CM$ admits a moduli space of complex structure deformations, each modulus is associated with an infinitesimal deformation, and hence an element of~$H^{0,1}(\CM, T^{1,0}\CM)$. However, not every cohomology class necessarily corresponds to a modulus, since there may be higher-order obstructions to integrating an infinitesimal deformation. This will not affect our discussion below. 

If~$\CM$ is compact, the elements of~$H^{0,1}(\CM, T^{1,0}\CM)$ can be represented by harmonic forms, which comprise a finite-dimensional complex vector space. Therefore, the number of complex structure moduli is finite.

\subsec{Hermitian Metrics}

Given a complex manifold~$\CM$, we can always choose a Hermitian metric~$g_{\mu\nu}$ compatible with the complex structure~${J^\mu}_\nu$,
\eqn\metcomp{g_{\mu\nu} {J^\mu}_\alpha {J^\nu}_\beta = g_{\alpha\beta}~.}
In holomorphic coordinates adapted to~${J^\mu}_\nu$, the only non-zero elements of~$g_{\mu\nu}$ are~$g_{i \bar j} = g_{\b j i}$. Thus, infinitesimal deformations~$\Delta g_{\mu\nu}$ of the metric that are compatible with~${J^\mu}_\nu$ only have~$\Delta g_{i \b j} = \Delta g_{\b j i}$ components. 

If we deform the complex structure, we must also deform the metric to ensure that~$g_{\mu\nu} + \Delta g_{\mu\nu}$ is compatible with~${J^\mu}_\nu + \Delta {J^\mu}_\nu$. At linear order, this leads to the constraint
\eqn\hermcons{g_{\mu\nu} \left(\Delta {J^\mu}_\alpha {J^\nu}_\beta + {J^\mu}_\alpha \Delta {J^\nu}_\beta \right) + \Delta g_{\mu\nu} {J^\mu}_\alpha {J^\nu}_\beta = \Delta g_{\alpha\beta}~.}
In holomorphic coordinates adapted to~${J^\mu}_\nu$ we find that~$\Delta g_{i \b j} = \Delta g_{\b j i}$ is unconstrained, while the other components of~$\Delta g_{\mu\nu}$ are determined in terms of~$\Delta {J^\mu}_\nu$,
\eqn\defmet{\Delta g_{i j}={i \over 2} \left( g_{i\bar k} \Delta {J^{\bar k}}_j+g_{j\bar k} \Delta {J^{\bar k}}_i\right)~,\qquad \Delta g_{\bar i\bar j}=-{i \over 2} \left(g_{ k\bar i} \Delta {J^{ k}}_{\bar j}+g_{ k\bar j} \Delta {J^{ k}}_{\bar i}\right)~.}

\subsec{Abelian Gauge Fields and Holomorphic Line Bundles}

Consider an Abelian gauge field~$A_\mu$ on a complex manifold~$\CM$, such that its field strength~$F_{\mu\nu} = \d_\mu A_\nu - \d_\nu A_\mu$ satisfies
\eqn\holcond{F_{\b i \b j} = 0~.}
Locally, we can then express~$A_{\b i} = \d_{\b i} \lambda$, where the function~$\lambda(z, \b z)$ is generally complex and only defined in a patch. If the gauge group is~$U(1)$, only the real part of~$\lambda$ can be removed by a gauge transformation. Since~$\CM$ is complex, it is natural to allow complex gauge transformations, so that the gauge group is~$GL(1,\C)$. In this case we can locally set~$A_{\b i} = 0$. The transition functions that preserve this gauge choice consist of holomorphic functions valued in~$GL(1, \C)$, which define a holomorphic line bundle over~$\CM$. 

The structure of the holomorphic line bundle only depends on~$A_{\b i}$. We can change this structure if we deform~$A_{\b i}$ by a well-defined~$(0,1)$-form~$\Delta A_{\b i}$ while preserving~\holcond,
\eqn\dacond{\b \d_{\b i} \,\big( \Delta A_{\b j}\big) - \b \d_{\b j}\, \left(\Delta A_{\b i}\right) = 0~.} 
As in the case of complex structure deformations, we must quotient by trivial deformations, which are induced by globally defined gauge transformations,
\eqn\globalgt{\Delta A_{\b i} = \d_{\b i} \ep~.}
Here~$\ep(z, \b z)$ is a well-defined complex function on~$\CM$. Deformations of the holomorphic line bundle defined by~$A_{\b i}$ are thus parametrized by the Dolbeault cohomology class of~$\Delta A_{\b i}$,
\eqn\acoho{\left[\Delta A_{\b i} \right] \in H^{0,1}(\CM)~.} 
In the Abelian case there are no higher obstructions, so that each element of~$H^{0,1}(\CM)$ gives rise to a finite deformation. If~$\CM$ is compact, $H^{0,1}(\CM)$ is a finite-dimensional complex vector space, whose dimension counts the number of holomorphic line bundle moduli.

\newsec{Parameter Dependence of~$Z_{\CM_4}$}

In this section we study the dependence of the partition function~$Z_{\CM_4}$ on the geometry of the complex manifold~$\CM_4$, as well as Abelian background gauge fields. We also discuss the dependence of~$Z_{\CM_4}$ on the~$(2,1)$-form~$W$ in~\fdthreform. As explained in the introduction, it is sufficient to study these questions around flat space. This amounts to analyzing the cohomology of the preserved supercharge~$Q$ on the bosonic operators in the~$\CR$-multiplet or a flavor current multiplet. We also briefly comment on situations with multiple supercharges. 

{\it Note: In this section, we raise and lower indices using the usual flat-space metric.}

\subsec{The~$Q$-Cohomology of the~$\CR$-Multiplet}

The operators~\rmultops\ that reside in the~$\CR$-multiplet can be embedded into superfields,\foot{These expressions contain several factors of~$i$ that do not appear in~\refs{\KomargodskiRB, \DumitrescuIU}, because we are working in Euclidean signature.}
\eqn\rmultcomp{\eqalign{& \CR_\mu = j^{(R)}_\mu - i \theta S_\mu + i \t \theta \t S_\mu + \theta \sigma^\nu \t \theta \, \Big(2 T_{\mu\nu} + {i \over 2} \ep_{\mu\nu\rho\lambda} \CF^{\rho\lambda} - {i \over 2} \ep_{\mu\nu\rho\lambda} \d^\rho j^{(R)\lambda}\Big) \cr
&\hskip25pt - \half \thetasq \t \theta \, \t \sigma^\nu \d_\nu S_\mu + \half {\t \theta\, }^2 \theta \sigma^\nu \d_\nu \t S_\mu - {1 \over 4} \thetasq {\t \theta\, }^2 \d^2 j^{(R)}_\mu~,\cr
& \chi_\alpha(y) = - 2i \big(\sigma^\mu {\t S}_\mu\big)_\alpha -4 \theta_\beta \left({\delta_\alpha}^\beta \, {T_\mu}^\mu - i {\left(\sigma^{\mu\nu}\right)_\alpha}^\beta \CF_{\mu\nu}\right) -4 \thetasq \left(\sigma^{\mu\nu} \d_\mu S_\nu\right)_\alpha~.}}
Here~$\chi_\alpha$ is chiral, $\t D_\alphadot \chi_\alpha = 0$, and~$y^\mu = x^\mu + i \theta \sigma^\mu \t \theta$ is the usual chiral superspace coordinate. This implies the following transformation rules under the preserved supercharge~$Q$ with spinor parameter~$\zeta_\alpha$,
\eqn\rmctrans{\eqalign{& \big\{Q, j_\mu^{( R )} \big\} = - i \zeta S_\mu~,\cr
& \big\{Q, S_{\alpha\mu}\big\} = 0~,\cr
&\big \{Q, {\, \t S^\alphadot}_\mu \big\} = 2 i \left(\t \sigma^\nu \zeta\right)^\alphadot \CT_{\mu\nu}~,\cr
& \big\{Q, T_{\mu\nu} \big\} = \half \zeta \sigma_{\mu\rho} \d^\rho S_\nu + \half \zeta \sigma_{\nu\rho} \d^\rho S_\mu~,\cr
& \big\{Q, \CF_{\mu\nu}\big\} = {i \over 2} \zeta \sigma_\nu \t \sigma_\rho \d_\mu S^\rho - {i \over 2} \zeta \sigma_\mu \t \sigma_\rho \d_\nu S^\rho~.
}}
Here we have defined the complex, non-symmetric, conserved tensor
\eqn\ctdef{\CT_{\mu\nu} = T_{\mu\nu} +{i \over 4} \ep_{\mu\nu\rho\lambda} \CF^{\rho\lambda} - {i \over 4} \ep_{\mu\nu\rho\lambda} \d^\rho j^{( R ) \lambda} - {i\over 2} \d_\nu j_\mu^{(R )}~,\qquad \d^\mu \CT_{\mu\nu} = 0~.}
There are eight bosonic~$Q$-exact operators,~$\{Q, {\, \t S^\alphadot}_\mu \}$. It is convenient to multiply this expression by~${1 \over  |\zeta|^2} \zeta^\dagger \sigma_\rho$. Since~${J^\mu}_\nu = -{2 i \over |\zeta|^2} \zeta^\dagger {\sigma^\mu}_\nu \zeta$, we find
\eqn\qst{\Big\{Q,  {1 \over  |\zeta|^2} \zeta^\dagger \sigma_\rho {\t S}_\mu \Big\} = - 2i \left({\delta^\nu}_\rho + i {J^\nu}_\rho\right) \CT_{\mu\nu}~.}
The projector onto anti-holomorphic indices shows that the eight~$Q$-exact bosonic operators are given by~$\CT_{\mu \b i}$. In holomorphic coordinates~$z^1 = w$ and~$z^2 = z$,
\eqn\detailqex{\eqalign{
& \CT_{w \b w} = T_{w \b w} +{i \over 2} \CF_{z\b z} -{i \over 2} \d_{\b w} j_w^{( R )} + {i \over 4} \d_{\b z} j_z^{( R )} -{i \over 4} \d_z j_{\b z}^{( R )}~,\cr
& \CT_{w \b z} = T_{w \b z} - {i \over 2} \CF_{w \b z} - {3 i \over 4} \d_{\b z} j_w^{( R )} +{i \over 4} \d_w j_{\b z}^{( R )}~,\cr
& \CT_{\b w \b w} = T_{\b w \b w} -{i \over 2} \d_{\b w} j_{\b w}^{( R )}~,\cr
& \CT_{\b w \b z} = T_{\b w \b z} +{i \over 2} \CF_{\b w \b z} -{i \over 4} \d_{\b w} j_{\b z}^{( R )} -{i \over 4} \d_{\b z} j_{\b w}^{( R )}~,
}} 
and four more with~$w \leftrightarrow z, \b w \leftrightarrow \b z$. (Here~$w,z$ are on equal footing and we only introduce them to simplify the notation.) Note that~$\CF_{\b w \b z} = i (\CT_{\b z \b w} - \CT_{\b w \b z})$ is by itself~$Q$-exact. Since~$Q^2 =0$, all~$Q$-exact operators are also~$Q$-closed. It can be checked that there are no other~$Q$-closed bosonic operators in the~$\CR$-multiplet. As we will see, this does not mean that every supersymmetric deformation of the Lagrangian is~$Q$-exact. 

\subsec{Deformations around Flat Space}

At the linearized level, the coupling of the~$\CR$-multiplet to the bosonic supergravity fields is given by~\fdlinlag, 
\eqn\linearizedlag{\Delta {\scr L}= -\half \Delta g^{\mu\nu} T_{\mu\nu} + A^{( R ) \mu} j^{( R )}_{ \mu} + {i \over 4} \ep^{\mu\nu\rho\lambda} B_{\mu\nu} \CF_{\rho\lambda}~.}
Following the discussion in section~2, we deform the complex structure and the Hermitian metric around their flat space values, while keeping the deformed metric compatible with the deformed complex structure as in~\defmet. By linearizing~\fdthreform\ and~\fdbgfields, we find that the background fields~$V^\mu$ and~$A^{( R )\mu}$ are given by
\eqn\linbgs{\eqalign{& V^{\b w} = \half \d_{\b z} \left(\Delta {J^{\b z}}_w - \Delta {J^{\b w}}_z\right) + 2 i \left(\d_w\, \Delta g_{z \b z} - \d_z \, \Delta g_{w \b z} \right) + 4 i W_{w z \b z}~,\cr
& V^{ w} = \half \d_{z} \left(\Delta {J^ z}_{\b w} - \Delta {J^ w}_{\b z}\right) + 2 i \left(\d_{\b z}\, \Delta g_{z \b w} - \d_{\b w} \, \Delta g_{z \b z} \right)~,\cr
& A^{( R ) w} = \half \d_w \, \Delta {J^w}_{\b w} -{1 \over 4} \d_z \, \Delta {J^z}_{\b w}  + {3 \over 4} \d_z \, \Delta {J^w}_{\b z} \cr
& \hskip37pt - 3 i \d_{\b z} \, \Delta g_{z \b w} + 2 i \d_{\b w} \, \Delta g_{z \b z} - i \d_{\b w} \, \Delta g_{w \b w}~,\cr
& A^{( R ) \b w} = \half \d_{\b w} \, \Delta {J^{\b w}}_w + {1 \over 4} \d_{\b z} \, \Delta {J^{\b z}}_w + {1 \over 4} \d_{\b z} \, \Delta {J^{\b w}}_z \cr
& \hskip37pt + i \d_w \, \Delta g_{w \b w} + i \d_z \, \Delta g_{w \b z}~,}}
and four more with~$w \leftrightarrow z, \b w \leftrightarrow \b z$. Locally, we may express the~$\b \d$-closed~$(2,1)$-form~$W_{i j \b k}$ in terms of a~$(2,0)$-form $\t B_{i j}$, so that~$W = \b \d \t B$. Substituting into~\linearizedlag, using the conservation of the~$R$-current, and dropping a total derivative, we find the following Lagrangian,
\eqn\deflag{\eqalign{\Delta {\scr L} = & - \Delta g^{i \b j} \CT_{i\b j} + i \t B_{wz} \CF_{\b w \b z}  - i \sum_{j = \b j} \Delta {J^{\b i}}_j \CT_{\b j \b i} \cr
&   + i \Delta {J^w}_{\b w} \left(T_{ww} + {i\over 2} \d_w j_w^{( R )} \right) + i \Delta {J^z}_{\b z} \left(T_{zz} + {i\over 2} \d_z j_z^{( R )} \right)\cr
& + i \Delta {J^w}_{\b z} \left(T_{wz} + {i \over2 } \CF_{wz} -{i \over 4} \d_w j_z^{( R )} + {3 i \over 4} \d_z j_w^{( R)}\right)\cr
& + i \Delta {J^z}_{\b w} \left(T_{wz} -{i \over 2} \CF_{wz} -{i \over 4} \d_z j_w^{( R )} +{3 i \over 4} \d_w j_z^{( R )}\right)~.}}
It is instructive to verify that this Lagrangian is supersymmetric, even though this must be the case on general grounds. Using~\rmctrans\ and the conservation of the supersymmetry current, we find
\eqn\checksusy{\big\{Q, \Delta {\scr L}\big\} = {i \over 4} \left(\d_{\b i} \, \Delta {J^k}_{\b j} - \d_{\b j} \,\Delta {J^k}_{\b i} \right) \zeta \sigma^{\b i \b j} S_k + \left({\rm total \; \,derivative}\right)~.}
Note that~$\t B_{w z}$ does not appear, because~$\{Q, \CF_{\b w \b z} \} = 0$. If we use the integrability condition~\integrability\ for the deformed complex structure, \checksusy\ reduces to a total derivative.

The fact that the operators~$\CT_{\mu\b i}$ are~$Q$-exact implies that the corresponding terms in~\deflag\ cannot affect the partition function~$Z_{\CM_4}$. We conclude:
\medskip
\item{$\bullet$} Given a fixed complex structure, $Z_{\CM_4}$ does not depend on the Hermitian metric.
\item{$\bullet$} $Z_{\CM_4}$ only depends on complex structure deformations through~$\Delta {J^{ i}}_{\b j}$. Therefore, it is a locally holomorphic function of the complex structure moduli.\foot{Our linearized analysis is not powerful enough to constrain global properties of the partition function. For instance, it may have singularities at certain loci in moduli space.}
\medskip
\noindent The fact that~$Z_{\CM_4}$ does not depend on trivial complex structure deformations of the form~$\Delta {J^i}_{\b j} = 2 i \d_{\b j} \ep^i$ follows from the underlying diffeomorphism invariance of our formalism. We can check this explicitly by substituting such a trivial deformation into~\deflag, using the conservation equation for the stress tensor, $\d^\mu T_{\mu\nu} = 0$, and dropping a total derivative. The terms involving~$\ep^i$,
\eqn\deflagcoho{\Delta {\scr L} = -2 \ep^w\left(\d_w \CT_{w \b w} + \d_z \CT_{w \b z}\right) -2  \ep^z \left(\d_z \CT_{z \b z} + \d_w \CT_{z \b w}\right) + \cdots~,}
only contain the~$Q$-exact operators~$\CT_{\mu\b i}$.

Since~$\CF_{\b w \b z}$ is also~$Q$-exact, we find that a well-defined~$(2,0)$-form~$\t B_{i j}$, which gives rise to a~$\b \d$-exact~$W$, cannot affect the partition function. Therefore:
\medskip
\item{$\bullet$} $Z_{\CM_4}$ only depends on~$W$ through its cohomology class in~$H^{2,1}(\CM_4)$. 
\medskip
\noindent We can obtain a stronger result if it is possible to improve the~$\CR$-multiplet to an~FZ-multiplet, i.e.~if the field theory does not possess any FI-terms and the K\"ahler form of the target space for chiral scalars is exact~\refs{\KomargodskiPC,\KomargodskiRB}. In this case one can find a well-defined operator~$\CA_\mu$, such that~$\CF_{\mu\nu} = \d_\mu \CA_\nu - \d_\nu \CA_\mu$. Substituting into~\deflag\ and integrating by parts, we find that~$\CA_{\b i}$ multiplies~$W_{i j \b k}$. Since~$\CA_{\b i}$ is~$Q$-exact up to a total~$\b \d_{\b i}$-derivative, which drops out because~$\b \d W = 0$, we find that the partition function does not depend on~$W$. Conversely, $Z_{\CM_4}$ can only depend on~$W$ in the presence of FI-terms or a non-trivial target-space K\"ahler class. Note that these may be quantized in the presence of topologically non-trivial flux for~$B_{\mu\nu}$, which necessarily arises if~$W$ is cohomologically non-trivial. For instance, this happens for the~$S^3 \times S^1$ background discussed in section~4. In this example, the quantization of FI-terms was recently discussed in~\AharonyDHA.\foot{This is reminiscent of the quantization of FI-terms or K\"ahler classes in supergravity~\refs{\WittenHU\SeibergQD\DistlerZG\BanksZN-\HellermanFV}.}

\subsec{Background Gauge Fields}

It is straightforward to repeat the preceding discussion to constrain the dependence of~$Z_{\CM_4}$ on background gauge fields. An Abelian flavor current~$j_\mu$ and its supersymmetric partners reside in a real linear multiplet,
\eqn\fdflavsf{\CJ = J + i \theta j - i \t \theta \t j + \big(\theta \sigma^\mu \t \theta\big) j_\mu + \half \theta^2 \t \theta \, \t \sigma^\mu \d_\mu j - \half {\t \theta}^2 \theta \sigma^\mu \d_\mu \t j - {1 \over 4} \theta^2 {\tilde \theta}^2 \d^2 J~.}
The~$Q$-transformations of the operators in~$\CJ$ are thus given by
\eqn\jvar{\eqalign{& \big\{ Q, J\big\} = i \zeta j~,\cr
& \big\{Q, j_\alpha\big\} = 0~,\cr
&\big \{Q,  \t j^\alphadot\big\} = - i (\t \sigma^\mu \zeta)^\alphadot \CJ_\mu~,\cr
&\big \{Q, j_\mu\big\} = - 2 \zeta \sigma_{\mu\nu} \d^\nu j~,}}
where we have defined the complex vector
\eqn\cjdef{\CJ_\mu = j_\mu - i \d_\mu J~.}
There are two bosonic~$Q$-exact operators, $\big\{Q, \t j^\alphadot\big\}$. As in the discussion around~\qst, we can multiply by~$\zeta^\dagger$ to conclude that these two operators are~$\CJ_{\b i}$. They are also the only~$Q$-closed operators. 

The linearized couplings of~$\CJ$ to the bosonic fields in a background vector multiplet take the form
\eqn\flavorcoup{\Delta {\scr L} = A^\mu j_\mu + D J~.}
The background field~$D$ is given by~\gaugepreserve,
\eqn\flatd{D = -2i \left(F_{w \b w} + F_{z \b z}\right)~.}
Substituting into~\flavorcoup\ and dropping a total derivative, we obtain
\eqn\fldlag{\eqalign{\Delta {\scr L} =~& 2 A_w \CJ_{\b w} + 2 A_z \CJ_{\b z} + 2 A_{\b w} \left(j_w + i \d_w J\right) + 2 A_{\b z} \left(j_z + i \d_z J\right)~.}}
As in the discussion around~\checksusy, the Lagrangian~\fldlag\ is supersymmetric provided~$A_{\b i}$ satisfies the integrability condition~$F_{\b w \b z} = 0$ for holomorphic line bundles. Since~$\CJ_{\b i}$ is~$Q$-exact, we conclude:
\medskip
\item{$\bullet$} $Z_{\CM_4}$ only depends on the anti-holomorphic part~$A_{\b i}$ of the background gauge field. Hence, it is a locally holomorphic function of the corresponding holomorphic line bundle moduli.
\medskip
\noindent As in the previous subsection, the reduction to the cohomology class of~$A_{\b i}$ in~$H^{0,1}(\CM_4)$ follows from background gauge invariance, as long as we allow complex gauge transformations.

\subsec{Additional Supercharges}

If the background fields on~$\CM_4$ preserve more than one supercharge, the parameter dependence of~$Z_{\CM_4}$ is further constrained. A dramatic example is the Witten index~\WittenDF, which is given by the partition function~$Z_{T^4}$ on a flat torus that preserves four supercharges. Under favorable conditions, this partition function does not depend on any continuous parameters. One way to show this is by noting that the entire deformation Lagrangian~$\Delta \SL$ in~\deflag\ is exact with respect to a suitable linear combination of the four supercharges on~$T^4$. Here we will briefly explore how the dependence of~$Z_{\CM_4}$ on complex structure moduli and background gauge fields is restricted if~$\CM_4$ admits two supercharges~$Q$ and~$\t Q$ of opposite~$R$-charge. 

The conditions for the presence of two such supercharges were analyzed in~\DumitrescuHA\ and reviewed in section~1.1, where we saw that~$\CM_4$ must be a~$T^2$ fibration over a Riemann surface~$\Sigma$. In this case we can choose holomorphic coordinates~$w, z$, such that the anti-holomorphic Killing vector~$K^\mu = \zeta \sigma^\mu \t \zeta$ is given by~$K = \d_{\b w}$ and the metric takes the form~\metricTtwofiber. The supercharges~$Q, \t Q$ transform as scalars under holomorphic coordinate changes that preserve~$K$ and the form of the metric, which allows us to rely on a linearized analysis around flat space to study the parameter dependence of the partition function. 

We consider deformations of the complex structure and the metric that are consistent with the presence of the anti-holomorphic Killing vector~$K$. This analysis is summarized in appendix~C. The operators multiplying~$\Delta {J^w}_{\b z}$ and~$\Delta {J^z}_{\b z}$ in the deformation Lagrangian~\deflag\ turn out to be~$\t Q$-exact. Together with the previous results, this implies:
\medskip
\item{$\bullet$} In the presence of~$Q$ and~$\t Q$, the partition function~$Z_{\CM_4}$ only depends on the complex structure moduli corresponding to~$\Delta {J^i}_\bw$. 
\medskip
\noindent In particular, $Z_{\CM_4}$ does not depend on the complex structure of the Riemann surface~$\Sigma$. As explained in~\DumitrescuHA, reducing on the~$T^2$ fiber leads to an~$A$-twisted~$\CN=(2,2)$ theory on~$\Sigma$ (more precisely, a deformation thereof). The supercharge~$Q_A = Q+ \t Q$ reduces to the two-dimensional BRST charge of the~$A$-model. The fact that~$Z_{\CM_4}$ does not depend on~$\Delta {J^z}_{\b z}$ reflects the fact that the theory on~$\Sigma$ is topological. 

We can also analyze the implications of~$\t Q$ for the dependence on holomorphic line bundle moduli. Since the operator multiplying~$A_{\b z}$ in~\fldlag\ is~$\t Q$-exact, we find:
\medskip
\item{$\bullet$} In the presence of~$Q$ and~$\t Q$, $Z_{\CM_4}$ only depends on~$A_{\bw}$.
\bigskip

Finally, we state without proof that the presence of~$\t Q$ also renders the partition function independent of anti-chiral~$F$-terms, so that~$Z_{\CM_4}$ does not depend on any~$D$-term or~$F$-term parameters.

\newsec{Examples in Four Dimensions}

In this section we explore complex manifolds~$\CM_4$ that are diffeomorphic to~$S^3 \times S^1$, $L(r,s) \times S^1$ with~$L(r,s)$ a Lens space, and~$S^2 \times T^2$. This leads to general statements about the partition function~$Z_{\CM_4}$ on these spaces. Complex manifolds diffeomorphic to~$S^3 \times S^1$ are closely related to the supersymmetric index. In each case, we discuss the amount of supersymmetry that can be preserved and discuss possible restrictions on the allowed~$R$-charges of the field theory. 

\subsec{$S^3\times S^1$ as a Complex Manifold}

It was shown in~\Kodairasone\ that every complex manifold diffeomorphic to~$S^3 \times S^1$ belongs to a family of complex manifolds known as primary Hopf surfaces (see also~\refs{\KSii,\KodairaCCASii,\Kodairabook}). They are quotients of~$\C^2 - (0,0)$, with coordinates~$(w,z)$, by an infinite cyclic group. There are two different types of primary Hopf surfaces:
\medskip
\item{1.)} If we identify 
\eqn\Hopfsurf{
\left(w,z\right) \sim \left(p w, q z\right)~, \qquad 0 < |p| \leq |q| <1~,
}
we obtain a complex manifold~$\CM^{p,q}_4$, whose complex structure depends on two complex parameters~$p, q$. 
\item{2.)} Identifying
\eqn\Hopfsurfii{
\left(w,z\right) \sim \left(q^n w + \lambda z^n, q z\right)~, \qquad 0 < |q| <1~, \quad \lambda \in \C^*~, \quad n \in \N~,
}
superficially leads to a complex manifold that depends on two continuous parameters~$q, \lambda$ and an integer~$n \geq 1$. However, $\lambda$ can be set to any non-zero value by a holomorphic coordinate change (simply rescale~$z$), and hence all non-zero values of~$\lambda$ give rise to the same complex structure. Taking~$\lambda \rightarrow 0$, we can make this complex structure arbitrarily close to a Hopf surface of the first type with~$p = q^n$. However, the complex structures with~$\lambda = 0$ and~$\lambda \neq 0$ are distinct. For instance, they do not admit the same number of holomorphic vector fields.
\bigskip

We will now discuss the Hopf surfaces of the first type, $\CM^{p,q}_4$, in more detail. It is convenient to express the parameters~$p, q$ in~\Hopfsurf\ as follows,
\eqn\deftausigma{
p= e^{- \beta_p + i \vartheta_p}~, \qquad q = e^{- \beta_q + i \vartheta_q}~, \qquad 0 < \beta_q \leq \beta_p~, \qquad \vartheta_{p,q} \sim \vartheta_{p,q} + 2\pi~.
}
We can explicitly see that~$\CM^{p,q}_4$ is diffeomorphic to~$S^3 \times S^1$ by introducing real variables~$x, \theta, \varphi, \chi$ subject to suitable identifications, 
\eqn\zoztworealangles{\eqalign{& w = e^{(- \beta_p + i \vartheta_p)x} \cos{\theta\over 2}\, e^{i \varphi}~, \qquad z = e^{(- \beta_q + i \vartheta_q)x} \sin{\theta\over 2} \,e^{i\chi}~,\cr
& x \sim x + 1~, \qquad 0 \leq \theta \leq \pi~, \qquad \varphi \sim \varphi + 2\pi~,\qquad \chi \sim \chi + 2\pi~.}}
Here~$x$ is a coordinate on~$S^1$, while~$\theta, \varphi, \chi$ parametrize~$S^3$. Note that
\eqn\wzmods{e^{2\beta_p x} |w|^2 + e^{2 \beta_q x} |z|^2 = 1~,}
so that~$x$ is always well-defined. (For fixed~$w, z$ the left-hand side is a monotonic function of~$x$.) While~$\theta$ is also always well-defined, the angles~$\varphi$ and~$\chi$ degenerate when~$w$ or~$z$ vanishes, respectively. They describe~$S^3$ as a torus fibered over the interval parametrized by~$\theta$, with one cycle of the torus shrinking to zero at each end. 

Hopf surfaces have non-zero Hodge numbers~$h^{0,0} = h^{0,1} = h^{2,1} = h^{2,2} = 1$, which shows that they are not K\"ahler. We can represent~$H^{0,1}(\CM_4^{p,q})$ by the~$(0,1)$-form
\eqn\zerooneform{
\omega^{0,1} = \b \d (-2x) = {e^{2 \beta_p x} w d \b w + e^{2 \beta_q x} z d\b z \over \beta_p \cos^2 {\theta \over 2} + \beta_q \sin^2{\theta \over 2}}~.}
Taking the tensor product with holomorphic vector fields whose coefficient functions are also holomorphic, we obtain elements of~$H^{0,1}(\CM_4^{p,q}, T^{1,0} \CM_4^{p,q})$. According to the discussion in section~2.1, these describe infinitesimal deformations of the complex structure. For generic~$p$ and~$q$, the only such holomorphic vector fields on~$\CM_4^{p,q}$ descend from the vector fields~$X^{( p)}  = w \d_w$ and~$X^{( q)} = z \d_z$ on~$\C^2 - (0,0)$. We can therefore take~$\Theta^{( p )} = X^{( p)} \otimes \omega^{0,1}$ and~$\Theta^{(q)} = X^{( q)} \otimes \omega^{0,1}$, which are independent elements of~$H^{0,1}(\CM_4^{p,q}, T^{1,0} \CM_4^{p,q})$ and describe infinitesimal deformations of~$p$ and~$q$, respectively. When~$p = q^n$ for some~$n \in \N$, there is a third holomorphic vector field~$X^{(\lambda)} = z^n \d_w$. The corresponding~$\Theta^{(\lambda)} = X^{(\lambda)} \otimes \omega^{0,1}$ describes a deformation from a surface of the first type with~$\lambda = 0$ to a surface of the second type with~$\lambda \neq 0$.\foot{When~$n=1$ there is a fourth holomorphic vector field, which is obtained from~$X^{(\lambda)}$ by exchanging~$w$ and~$z$. Not every linear combination of~$\Theta^{( p)}, \Theta^{(q)},\Theta^{(\lambda)}$ gives rise to an allowed finite deformation around~$p = q^n$. This is because we can use~$\Theta^{(p )}, \Theta^{(q)}$ to vary~$p,q$ independently, while deforming to a surface of the second type using~$\Theta^{(\lambda)}$ requires~$p = q^n$.} Note that studying infinitesimal deformations around the point~$\lambda = 0$ is not sufficient to conclude that all non-zero values of~$\lambda$ correspond to the same complex structure, although the fact that rescaling~$z$ also rescales~$\Theta^{(\lambda)}$ suggests this. If we work around a point where~$\lambda \neq 0$, the infinitesimal deformation describing changes in~$\lambda$ vanishes in cohomology.

\subsec{Hermitian Metrics on~$S^3 \times S^1$}

We can obtain Hermitian metrics on~$\CM^{p,q}_4$ by constructing Hermitian metrics on $\C^2 - (0,0)$ that are invariant under the identifications~\Hopfsurf. We will consider the metric\foot{A closely related family of Hermitian metrics on~$\CM^{p,q}_4$ was constructed in~\gaudorn.}
\eqn\metricst{ds^2= r^2 \left( \sqrt{\beta_q \over \beta_p}\,e^{2 \beta_p x} dw d \b w + \sqrt{\beta_p \over \beta_q}\,e^{2 \beta_q x} dz d \b z\right)~.}
For generic values of~$p$ and~$q$, it is not isometric to the usual round metric on~$S^3 \times S^1$. However, $\d_x$ is always a Killing vector, so that we can we can dimensionally reduce along~$S^1$. (We will return to this point in section~7.1.) The metric~\metricst\ always admits an anti-holomorphic Killing vector,
\eqn\holokilling{K=\beta_p \, \bar w \del_{\bar w}+\beta_q \, \bar z \del_{\bar z}~,}
which is nowhere vanishing and commutes with its complex conjugate, $[K, \b K] = 0$. As we reviewed in section~1.1, this allows us to preserve two supercharges of opposite~$R$-charge. 

We would like to understand when the metric~\metricst\ on~$\CM_4^{p,q}$ admits four supercharges. As was shown in~\refs{\FestucciaWS,\DumitrescuHA}, a necessary (but not sufficient) condition is that it be locally round. This happens when~$|p| = |q|$, i.e $\beta_p = \beta_q \equiv \beta$, so that~\metricst\ and~\holokilling\ reduce to
\eqn\metricsequalstbar{ds^2=r^2 {dw d\b w + dz d \b z \over |w|^2 + |z|^2}~, \qquad K = \beta \left(\b w \d_{\b w} + \b z \d_{\b z}\right)~.}
To see this explicitly, we define new angular variables,
\eqn\newangle{\t \varphi = \varphi + \vartheta_p x~, \qquad \t \chi = \chi + \vartheta_q x~,}
in which the metric and the Killing vector~$K$ take the following form,
\eqn\metnewangles{ds^2 = r^2 \left(\beta^2 dx^2 + d\Omega_3\right)~, \qquad K = - \half \left( \d_x - i\beta \big(\d_{\t \varphi} + \d_{\t \chi}\big)\right)~.}
Here~$d \Omega_3$ is the round metric on a unit three-sphere parametrized by~$\theta, \t \varphi, \t \chi$. Therefore~$r$ is the radius of the round~$S^3$, while~$\beta$ is the dimensionless ratio of the $S^1$ circumference to the~$S^3$ radius. When~$\vartheta_p = \vartheta_q = 0$, the angles~$\t \varphi, \t \chi$ in~\newangle\ are good coordinates on~$S^3$ and the space is globally isometric to the round~$S^3 \times S^1$. In general, this is not the case and we have the identifications~$(\t \varphi, \t \chi, x) \sim (\t \varphi + \vartheta_p, \t \chi + \vartheta_q, x + 1)$. This can be described by starting with the globally round~$S^3 \times S^1$ and rotating~$\t \varphi$, $\t \chi$ by~$\vartheta_p$, $\vartheta_q$ as~$x$ circles the~$S^1$ one full time. For the special choice~$\vartheta_p + \vartheta_q = 0$, i.e.~for~$p = \b q$, this leaves invariant the Hopf fiber of~$S^3$ that is acted on by the imaginary part of~$K$ in~\metnewangles. As we will see in section~4.4, this means that we can preserve four supercharges on~$\CM^{\b q, q}_4$.

In appendix~D, we show that Hopf surfaces of the second type, which correspond to the identification~\Hopfsurfii, do not admit Hermitian metrics with an anti-holomorphic Killing vector. Therefore, they only admit a single supercharge.

\subsec{Background Gauge Fields on~$S^3 \times S^1$}

Since~$H^2(S^3 \times S^1, \Z) = 0$, all complex line bundles over~$S^3\times S^1$ are topologically trivial. However, the Hopf surfaces~$\CM_4^{p,q}$ admit non-trivial holomorphic line bundles, which are classified by~$H^{0,1}(\CM_4^{p,q})$ (see section~2.3) and can be represented by~$\omega^{0,1}$ in~\zerooneform. Up to a globally defined complex gauge transformations, the~$(0,1)$-component of the corresponding background gauge field~$A_\mu$ is a complex multiple of~$\omega^{0,1}$, while its~$(1,0)$-component is arbitrary. Using this freedom, we can take~$A_\mu$ to be real, so that
\eqn\realahopf{\eqalign{A_\mu dx^\mu &= - \half (a_r - i a_i) \omega^{0,1} + ({\rm c.c.}) \cr
& = a_r dx + a_i {\left(\vartheta_p \cos^2 {\theta \over 2} + \vartheta_q \sin^2{\theta \over 2}\right) dx + \cos^2 {\theta \over 2} d \varphi + \sin^2 {\theta \over 2} d\chi \over \beta_p \cos^2 {\theta \over 2} + \beta_q \sin^2 {\theta \over 2} }~,}}
with~$a_{r, i} \in \R$. This configuration also satisfies the conditions~\twoscfdbundle\ for compatibility with the two supercharges associated with the Killing vector~\holokilling. Compatibility with four supercharges requires that~$A_\mu$ be a flat connection, so that~$a_i$ must vanish. (If we allow complex configurations for the gauge field, we can complexify~$a_r$ while keeping~$a_i = 0$.)

\subsec{$S^3 \times S^1$ and the Supersymmetric Index}

Any four-dimensional~$\CN=1$ theory with a~$U(1)_R$ symmetry can be placed on~$S^3 \times \R$ with the usual round metric while preserving all four supercharges~\refs{\SenPH, \RomelsbergerEG, \FestucciaWS}. The isometry group contains the~$SU(2) \times SU(2)'$ symmetry of the round~$S^3$, with generators~$J_i$ and~$J'_i \; (i = 1, 2, 3)$, and translations along~$\R$, generated by~$H$. The supercharges form two doublets of~$SU(2)$ with opposite~$R$-charge, while they are invariant under~$SU(2)'$ and~$H$. The supersymmetry algebra is~$SU(2|1)$, which contains~$SU(2)$. The supersymmetric theory on~$S^3 \times \R$ is unitary, so that it makes sense to talk about the adjoint of a supercharge. 

The supersymmetric index~$\CI(p,q,u)$ is defined by a trace over states on~$S^3$ in Hamiltonian quantization~\refs{\RomelsbergerEG, \KinneyEJ,\FestucciaWS},
\eqn\defIndex{\CI(p,q, u) = \Tr_{S^3} \left((-1)^F p^{J_3 + J'_3 - {R \over 2}} q^{J_3 - J'_3 - {R \over 2}} u^{Q_f}\right)~.}
Here~$F$ is the fermion number and~$Q_f$ is an Abelian flavor symmetry, which could reside in the Cartan subalgebra of a non-Abelian flavor symmetry group. The parameters~$p,q,u$ are independent fugacities for bosonic symmetries that commute with one of the supercharges and its adjoint, but in general they do not commute with all four supercharges. The index only receives contributions from states with $H = 2 J_3 - R$, so that it cannot depend on continuous parameters or renormalization group flow~\refs{\RomelsbergerEG, \FestucciaWS}. This explains its utility in probing supersymmetric dynamics or dualities (see for instance~\refs{\RomelsbergerEC\DolanQI \SpiridonovZA-\SpiridonovHF}). The definition~\defIndex\ naturally accommodates complex values of~$p,q,u$ and shows that~$\CI(p,q,u)$ should be holomorphic, as long as it is well defined. In practice, many computations of the index focus on the subspace
\eqn\chempotdef{p = e^{- \beta + i \vartheta} = \b q~, \qquad 0 < |p| = |q| < 1~, \qquad u = e^{i \alpha}~, \qquad |u| = 1~.} 
Substituting into~\defIndex\ and using the fact that only states with~$H = 2 J_3 -R$ contribute, 
\eqn\restrictedindex{\CI\left(e^{- \beta + i \vartheta}, e^{- \beta - i \vartheta}, e^{i\alpha}\right) = \Tr_{S^3} \left( (-1)^F e^{-\beta H+2 i \vartheta J'_3 + i \alpha Q_f}\right)~.}
Now the chemical potentials~$\beta, \vartheta, \alpha$ only couple to the operators~$H, J'_3, Q_f$, which commute with all four supercharges in~$SU(2|1)$. 

It is clear from~\defIndex\ that, up to local counterterms, the index is given by a partition function on~$S^3 \times S^1$. This is easiest to see on the subspace~\chempotdef, which preserves four supercharges. In this case it follows from~\restrictedindex\ that~$\beta$ is the ratio of the~$S^1$ circumference to the~$S^3$ radius~\FestucciaWS. As the~$S^3$ goes around~$S^1$ one full time, it is rotated by~$2 \vartheta$ along a Hopf fiber that is invariant under~$SU(2) \subset SU(2|1)$. Comparing with the discussion at the end of section~4.2, we see that this precisely describes a Hopf surface~$\CM_4^{\b q, q}$, where~$p = \b q$ is given by~\chempotdef. Similarly, the fugacity~$u$ for the global flavor charge~$Q_f$ corresponds to a flat connection along~$S^1$. In the notation of~\realahopf, we have~$a_r = \alpha$ and~$a_i = 0$. On the subspace~\chempotdef\ we have thus identified
\eqn\indpart{\CI(p, q, u) = Z_{\CM^{p, q}_4}~.}
Since we have argued that both sides are holomorphic functions, this equality must persist for all allowed values of~$p,q$ in~\Hopfsurf. The same argument applies to the analytic continuation of the fugacity~$u$, which corresponds to turning on a background gauge field~\realahopf\ with~$u = e^{a_i + i a_r}$ on~$\CM^{p,q}_4$.  Therefore, the supersymmetric partition function on the Hopf surface~$\CM^{p,q}_4$ geometrizes the index~$\CI(p,q,u)$ by identifying its fugacities as complex structure or holomorphic line bundle moduli.

Explicit computations of the index lead to a family of special functions that have recently been studied in the mathematical literature, see for instance~\refs{\Rains\Spiridonov-\Bult,\DolanQI\SpiridonovZA-\SpiridonovHF} and references therein.  Our results show that they can be interpreted as locally holomorphic functions on the moduli space of complex structures and holomorphic line bundles on the Hopf surfaces~$\CM_4^{p,q}$. It may be interesting to examine some of their properties from this point of view.

\subsec{$L(r,s) \times S^1$}

Complex manifolds obtained from primary Hopf surfaces by taking an additional quotient are known as secondary Hopf surfaces. Those with Abelian fundamental group are diffeomorphic to~$L(r,s) \times S^1$, where~$L(r,s)$ is a Lens space. The complex structures on these spaces have been classified in~\Nakagawa. On general grounds, they must admit at least one supercharge. Four-dimensional~$\CN=1$ theories on~$L(r,1) \times S^1$ were studied in~\refs{\BeniniNC,\RazamatOPA}. See also~\AldayAU\ for a related discussion in three dimensions.

We will restrict our discussion to the following class of identifications,
\eqn\lensid{(w, z) \sim (e^{2\pi i s \over r} w, e^{- {2 \pi i \over r}} z)~.}
Here~$r,s$ are relatively prime integers with~$1 \leq s < r$. If~$w,z$ lie on a three-sphere~$|w|^2 + |z|^2 = {\rm constant}$, the resulting quotient space~$S^3 / \Z_r$ is diffeomorphic to~$L(r,s)$.\foot{As discussed in~\Nakagawa, there are other quotients that are also diffeomorphic to~$L(r,s)$ but give rise to a different complex structure on~$L(r,s) \times S^1$.} We can use this quotient action to construct secondary Hopf surfaces of the form~$\CM^{p, q}_4/\Z_r$ by starting with~$\C^2 - (0,0)$ and imposing the identifications~\Hopfsurf\ and~\lensid. Since they leave the metric~\metricst\ and the Killing vector~\holokilling\ invariant, we can always preserve two supercharges of opposite~$R$-charge on the resulting quotient space. Repeating the arguments of the previous subsections, we see that four supercharges require~$p = \b q$ and~$s = 1$. In this case the~$\Z_r$ quotient only acts on~$\CM_4^{\b q, q}$ via an element of~$SU(2)'$, which does not affect the four supercharges in~$SU(2|1)$. We can therefore place any~$\CN=1$ theory with a~$U(1)_R$ symmetry on the resulting quotient space without modifying the background supergravity fields. 

For generic~$p,q$, we can preserve two supercharges on the quotient space. When~$s \neq 1$ this requires an additional Wilson line for the~$R$-symmetry gauge field~$A^{( R)}_\mu$. It must wrap the torsion one-cycle~$\Gamma$ that generates the fundamental group~$\pi_1\left(L(r,s)\right) = \Z_r$. As we will now show, this leads to a quantization condition for the allowed~$R$-charges. Consider the following~$(2,0)$-form on~$\C^2 - (0,0)$,
\eqn\holdtwoform{e^{\left(\beta_p + \beta_q - i (\vartheta_p + \vartheta_q)\right) x} dw \wedge dz~.}
It is nowhere vanishing and invariant under~\Hopfsurf, so that it descends to~$\CM^{p,q}_4$.\foot{This shows that the canonical bundle~$\Lambda^{2,0}$ of~$\CM^{p,q}_4$ is topologically, though not holomorphically, trivial. Hence there is no restriction on the allowed~$R$-charges for theories on~$\CM^{p,q}_4$.} Under the identification~\lensid\ leading to~$\CM^{p,q}_4/\Z_r$, it picks up a phase~$e^{2 \pi i (s-1) \over r}$. Therefore, the line bundle~$\Lambda^{2,0}$ on~$\CM^{p,q}_4/\Z_r$ has Chern class~$c_1(\Lambda^{2,0}) = s-1$ in~$H^2(L(r,s)\times S^1, \Z) = \Z_r$. As explained in~\DumitrescuHA, supersymmetry requires that the line bundle~$L$ corresponding to fields of~$R$-charge~$+1$ be chosen such that~$L^2 \otimes \Lambda^{2,0}$ is topologically trivial. Therefore, the bundle~$L^{2 \over s-1}$ is well-defined, and hence the allowed~$R$-charges for well-defined bosonic fields are integer multiples of~$2 \over s-1$.\foot{Alternatively, we can use the fact that~$\Gamma^r$ is a contractible cycle. A Wilson line wrapping~$\Gamma$ should not lead to a phase for a charged particle propagating~$r$ times around~$\Gamma$. This gives the same quantization condition for the~$R$-charges.} Since the supercharges~$Q$ and~$\t Q$ are well defined, so are the fermionic superpartners.

As in the previous subsection, the partition function on~$\CM^{p,q}_4/\Z_r$ will be a holomorphic function of the complex structure moduli~$p,q$. It is also straightforward to extend the discussion to holomorphic line bundles. Topologically these are classified by their first Chern class~$c_1 \in \Z_r$. Each bundle has a holomorphic modulus corresponding to~$\omega^{0,1}$ in~\zerooneform, which is invariant under~\lensid\ and thus descends to~$\CM^{p,q}_4/\Z_r$.

\subsec{$ S^2 \times T^2$}

Here we will briefly consider a family of complex manifolds~$\CM_4$ that are diffeomorphic to~$S^2 \times T^2$. (A more detailed discussion will appear in~\cs.) We will realize them as quotients of~$\C \times \C\P^1$ with metric
\eqn\metricprod{
ds^2 = d w d\b w + {4 r^2 \over (1 + |z|^2)^2} dz d \b z~.
}
Here~$w$ and~$z$ are coordinates on~$\C$ and~$\C\P^1$, respectively (we need to change coordinates to~$z' = 1/z$ to describe the point~$z = \infty$) and~$r$ is the radius of~$\C\P^1$. Note that~$K = \d_{\b w}$ is a nowhere vanishing, holomorphic Killing vector. We now impose the identifications
\eqn\complexstructTtwo{
(w, z) \sim (w+1~, e^{i \alpha}z) \sim (w + \tau, e^{i \beta} z)~,
}
where~$\tau$ is complex with~$\Im \tau > 0$ and~$\alpha, \beta$ are real angles with periodicity~$2\pi$. The identification of~$w$ results in a~$T^2$ with modular parameter~$\tau$, while the identification of~$z$ has the effect of rotating~$\C\P^1$ through~$\alpha$ and~$\beta$ as it goes around the two cycles of the torus. The resulting quotient space~$\CM_4$ is diffeomorphic to~$S^2 \times T^2$. Since the metric~\metricprod\ and the Killing vector~$K = \d_{\b w}$ are invariant under~\complexstructTtwo, they descend to the quotient, which therefore admits two supercharges of opposite~$R$-charge. As explained in~\DumitrescuHA, these two supercharges correspond to the usual BRST charges of the topological~$A$-model on~$\C\P^1$, while they generate a~$(2,0)$ supersymmetry algebra on~$T^2$. In particular, this background has a unit-flux monopole through~$\C\P^1$ for the~$R$-symmetry gauge field~$A^{( R)}_\mu$ (this also follows directly from~\fdbgfields), so that well-defined bosonic fields must carry integer~$R$-charge.

As in previous examples, the complex structure moduli are related to the parameters~$\tau, \alpha, \beta$ that appear in the identification~\complexstructTtwo. We can construct infinitesimal complex structure deformations by combining~$\omega^{0,1} = d\b w$, which represents~$H^{0,1}(\CM_4)$, with a holomorphic vector field~$X$, whose coefficient functions are also holomorphic. Choosing~$X^{(\tau)} = \d_w$ corresponds to deforming the modular parameter~$\tau$ of~$T^2$, while~$X^{(\sigma)}  = z \d_z$ corresponds to deformations of the complex parameter~$\sigma = \alpha \tau - \beta$, with fixed~$\tau$. Therefore the partition function~$Z_{\CM_4}$ will be a locally holomorphic function of two complex structure moduli~$\tau, \sigma$. From the point of view of the~$(2,0)$ theory on~$T^2$, this partition function computes the elliptic genus, which is known to have modular properties under~$SL(2, \Z)$ transformations of~$\tau$ (see the recent papers~\refs{\GaddeDDA\BeniniNDA-\BeniniXPA} and references therein). Similarly, there are global identifications in the moduli space parametrized by~$\tau$ and~$\sigma$,\foot{They correspond to a certain subgroup of~$SL(3,\Z)$, which contains the~$SL(2,\Z)$ transformations of~$\tau$ as a subgroup.} which should act on~$Z_{\CM_4}$ in a reasonable way. 

It is straightforward to include Abelian background gauge fields, which we take to be real for simplicity. According to the discussion around~\twoscfdbundle, the gauge field must be flat on~$T^2$ in order to preserve the two supercharges that are present on~$S^2 \times T^2$. The corresponding holomorphic line bundles are thus labeled by their first Chern class~$c_1 \in \Z$, which specifies the flux through~$\C\P^1$, and a single holomorphic modulus corresponding to~$\omega^{0,1} = d\b w$, whose real and imaginary parts correspond to Wilson lines wrapping the cycles of the torus.

\newsec{Deformation Theory in Three Dimensions}

We would like to repeat the preceding analysis to study the parameter dependence of the partition function~$Z_{\CM_3}$ on a three-manifold~$\CM_3$. In this section we set up the necessary mathematical machinery. Following~\ClossetRU, we review the geometric structure that is required to preserve a single supercharge on~$\CM_3$ and explain how it leads to a THF with a compatible transversely Hermitian metric. This allows us to define a three-dimensional notion of holomorphic forms and an analog, $\t \d$, of the~$\b \d$-operator on complex manifolds. These ingredients feature prominently in the deformation theory of THFs, which closely parallels the theory of complex structure deformations reviewed in section~2. We also discuss the three-dimensional version of holomorphic line bundles and their deformations. General references for THFs are~\refs{\DKi\GM-\Haef}. Three-manifolds admitting THFs are classified in~\refs{\BG\Brunella-\Ghys}. In the physics literature, THFs have previously arisen in the study of~D-branes in the topological~$A$-model~\KapustinIJ.

\subsec{Supersymmetry and Transversely Holomorphic Foliations}

In~\ClossetRU, the conditions for the existence of a supercharge on~$\CM_3$ were stated in terms of a vector~$\xi^\mu$, a one-form~$\eta_\mu$, and an endomorphism~${\Phi^\mu}_\nu$ that satisfy
\eqn\acsdef{{\Phi^\mu}_\nu {\Phi^\nu}_\rho = - {\delta^\mu}_\rho + \xi^\mu \eta_\rho~, \qquad \eta_\mu \xi^\mu =1~.}
This implies that~${\Phi^\mu}_\nu$ has rank two, while~$\eta_\mu$ and~$\xi^\mu$ are nowhere vanishing and span the left and right kernels of~${\Phi^\mu}_\nu$. In addition~$\xi^\mu$, $\eta_\mu$, ${\Phi^\mu}_\nu$  satisfy the integrability condition
\eqn\acsint{{\Phi^\mu}_\nu {\left(\CL_\xi \Phi\right)^\nu}_\rho = 0~,}
where~$\CL_\xi$ denotes the Lie derivative along~$\xi^\mu$. These conditions can be rephrased in terms of a well-studied mathematical structure: a transversely holomorphic foliation (THF). The orbits of the nowhere vanishing vector field~$\xi^\mu$ constitute the leaves of a one-dimensional oriented foliation of~$\CM_3$. The normal bundle~$\CD$ of the foliation consists of vector fields orthogonal to~$\eta_\mu$. Then~$\Phi$ induces an almost complex structure~$J = \Phi |_\CD$ on~$\CD$, i.e.~$J^2 = -1$. Since~$\CD$ is two-dimensional, $J$ is automatically integrable. As shown in appendix~B of~\ClossetRU, the integrability condition~\acsint\ ensures that~$J$ is constant along the orbits of~$\xi$, i.e.~the leaves of the foliation, so that~$\xi$ and~$J$ define a THF~\GM. 
 
It was shown in~\ClossetRU\ that we can cover~$\CM_3$ with patches of adapted coordinates,
\eqn\tdcoord{\tau = x^1~, \qquad z= x^2 + i x^3~, \qquad \b z = x^2 - i x^3~,}
in which~$\xi^\mu$, $\eta_\mu$, ${\Phi^\mu}_\nu$ are given by
\eqn\acsadcoord{\xi = \d_\tau~, \qquad \eta = d\tau + h dz + \b h d {\b z}~, \qquad {\Phi^\mu}_\nu = \pmatrix{0 & - i h  & i \b h \cr 0 & i & 0 \cr 0 & 0 & -i}~.}
Here~$\eta_z = h(\tau, z, \b z)$ is a complex function. Two overlapping adapted patches are related by
\eqn\adctrans{\tau' = \tau + t(z, \b z)~, \qquad z' = f(z)~, \qquad \d_{\b z} f(z) = 0~.}
The function~$t(z, \b z)$ is real and~$f(z)$ is holomorphic. Such a transformation preserves~$\xi$, while~$h$ transforms as follows,
\eqn\htrans{h'\left(\tau', z', \b z'\right) = {1 \over f'(z)} \left(h\left(\tau, z, \b z\right) - \d_z t\left(z, \b z\right)\right)~.}
The existence of adapted coordinates~$\tau, z, \b z$ transforming according to~\adctrans\ and such that~$\xi = \d_\tau$ directly follows from the fact that~$\xi$ generates a THF.

We will define various structures on~$\CM_3$ by working in adapted coordinates and ensuring good behavior under coordinate changes of the form~\adctrans. As an example, we can define a canonical orientation~$dx^1 \wedge dx^2 \wedge dx^3$ in each adapted patch. This defines an orientation on~$\CM_3$, because the corresponding transition functions~$|f'(z)|^2$ are positive.

\subsec{Holomorphic Forms and the~$\t \d$-Operator}

Define a complex projector
\eqn\projdef{{\Pi^\mu}_\nu = \half \left({\delta^\mu}_\nu - i {\Phi^\mu}_\nu - \xi^\mu \eta_\mu\right)~, \qquad {\Pi^\mu}_\nu {\Pi^\nu}_\rho = {\Pi^\mu}_\rho~.}
We can use~${\Pi^\mu}_\nu$ to split the complexified tangent and cotangent bundles, and we will call the corresponding invariant subspaces and their complements holomorphic and anti-holomorphic, respectively. A holomorphic vector~$X \in T^{1, 0} \CM_3$ and a holomorphic one-form~$\omega^{1,0} \in \Lambda^{1,0}$ satisfy
\eqn\holoformsdef{{\Pi^\mu}_\nu X^\nu = X^\mu~, \qquad \omega^{1,0}_\mu {\Pi^\mu}_\nu = \omega^{1,0}_\nu~.}
In adapted coordinates,
\eqn\holovec{X = X^z (\d_z - h \d_\tau) ~, \qquad \omega^{1,0} = \omega^{1,0}_z dz ~.}
Under an adapted coordinate change~\adctrans, both of them transform with holomorphic transition functions,
\eqn\holotransvecfm{\left(X'\right)^{z'} = f'(z) X^z~, \qquad \big(\omega'^{1,0}\big)_{z'} = {1 \over f'(z)} \omega^{1,0}_z~.}
Therefore, both~$T^{1,0} \CM_3$ and~$\Lambda^{1, 0}$ are complex line bundles with holomorphic transition functions. We will call such line bundles holomorphic. 

The complement of~$\Lambda^{1,0}$ defines the anti-holomorphic one-forms~$\omega^{0,1} \in \Lambda^{0,1}$, which satisfy~$\omega^{0,1}_\mu {\Pi^\mu}_\nu = 0$. In adapted coordinates,
\eqn\ahonefm{\omega^{0,1} =\omega^{0,1}_\tau \left(d\tau + h dz\right) +\omega^{0,1}_{\b z}d \b z~.}
These~$(0,1)$-forms span a two-dimensional subspace of the cotangent bundle that is not simply related to the line bundle of~$(1,0)$-forms by complex conjugation. We can split the space~$\Lambda^k$ of complex~$k$-forms into a direct sum over~$\Lambda^{p,q} = \wedge^p \Lambda^{1,0} \otimes \wedge^q \Lambda^{0,1}$. Explicitly, $\Lambda^2$ splits into~$(1,1)$-forms and~$(0,2)$-forms, while~$\Lambda^3$ consists of~$(1,2)$-forms. In adapted coordinates,
\eqn\higherforms{\eqalign{&  \omega^{1,1} = \omega^{1,1}_{\tau z} \, d \tau \wedge d z + \omega^{1,1}_{z \b z} \, d z \wedge d \b z  ~,\cr
& \omega^{0,2} = \omega^{0,2}_{\tau \b z} \, \left(d \tau + h dz\right) \wedge d \b z ~,\cr
& \omega^{1,2} = \omega^{1,2}_{\tau z \b z} \, d \tau \wedge dz \wedge d \b z ~.}}

We can now examine the action of the exterior derivative on~$\Lambda^{p,q}$. Given a holomorphic one-form~$\omega^{1,0}$, we compute in adapted coordinates,
\eqn\dholo{d\omega^{1,0} = \d_\tau \omega^{1,0} _z \, d \tau \wedge dz - \d_{\b z} \omega^{1,0} _z \,  dz\wedge d{\b z} \in \Lambda^{1,1}~.}
Therefore~$d\omega^{1,0} $ does not have a~$(0,2)$-component. As for complex manifolds, this implies that the exterior derivative~$d\omega^{p,q}$ of a~$(p,q)$-form is an element of~$\Lambda^{p+1, q} \oplus \Lambda^{p,q+1}$. We define an operator~$\t \d$ by projection onto the second summand,
\eqn\tdef{\t \d : \Lambda^{p,q} \rightarrow \Lambda^{p,q+1}~, \qquad \t \d \omega^{p,q} = d \omega^{p,q} \big|_{\Lambda^{p,q+1}}~.}
We have chosen the notation~$\t \d$ rather than~$\b \d$ to emphasize that the operator does not only involve differentiation with respect to~$\b z$. For future reference, we summarize the action of~$\t \d$ on~$(p,q)$-forms~$\omega^{p,q}$ in adapted coordinates:
\eqn\tdaction{\eqalign{& \t \d \omega^{0,0} = \d_\tau \omega^{0,0} \left(d\tau + h dz\right) + \d_{\b z} \omega^{0,0} \, d \b z~,\cr
& \t \d \left(\omega^{1,0}_z \, dz\right) =  \d_\tau \omega_z^{1,0} \, d\tau \wedge d z - \d_{\b z} \omega^{1,0}_z \, dz \wedge d\b z~\cr
& \t \d \left(\omega^{0,1}_\tau \left(d\tau + h dz\right) + \omega^{0,1}_{\b z} \, d\b z\right) = \left(\d_\tau \omega^{0,1}_{\b z} - \d_{\b z} \omega^{0,1}_{\tau} \right) \left(d\tau + h dz\right) \wedge d \b z~,\cr
& \t \d \left(\omega^{1,1}_{\tau z} \, d \tau \wedge d z + \omega^{1,1}_{z \b z} \, d z \wedge d \b z\right) = \left(\d_{\b z} \omega^{1,1}_{\tau z} + \d_\tau \omega^{1,1}_{z \b z}\right) \, d\tau \wedge d z \wedge d \b z~,\cr
& \t \d \omega^{0,2} = \t \d \omega^{1,2} = 0~.}}

It follows from~$d^2 = 0$ that~$\t \d$ satisfies 
\eqn\tdrels{\t \d^2 = 0~.}
We can therefore define its cohomology,
\eqn\tdcohodef{H^{p,q}(\CM_3) = { \{ \omega^{p,q} \in \Lambda^{p,q} | \t \d \omega^{p,q} = 0 \} \over \t \d \Lambda^{p, q-1}}~.}
As in the complex case, there is a Poincar\'e lemma for the~$\t \d$-operator: given a~$\t \d$-closed~$(p,q)$-form~$\omega^{p,q}$ on~$\R^3$ with~$q \geq 1$, there is a~$(p, q-1)$-form~$\varphi^{p,q-1}$ such that~$\omega^{p,q} = \t \d\varphi^{p,q-1}$. Again, it can be shown that~$H^{p,q}(\CM_3)$ is finite dimensional if~$\CM_3$ is compact. We will also need the cohomology~$H^{p,q}(\CM_3, T^{1,0}\CM_3)$ of forms with coefficients in the holomorphic tangent bundle~$T^{1, 0} \CM_3$. This is well defined, since the transition functions of~$T^{1,0}\CM_3$ in~\holotransvecfm\ are annihilated by~$\t \d$.

The cohomology~$H^{p,q}(\CM_3)$ has some unfamiliar features. For instance, we will see in section~$7.1$ that~$H^{0,1}(S^3)$ is non-trivial, even though~$S^3$ is simply connected. On the other hand, the~$(0,1)$-part~$\omega^{0,1}$ of a one-form~$\omega$ that is closed in the usual sense, $d \omega = 0$, satisfies~$\t \d \omega^{0,1} = 0$. Therefore, elements of the usual de Rham cohomology~$H^1(\CM_3)$ can give rise to non-trivial elements of~$H^{0,1}(\CM_3)$.

\subsec{Deformations of THFs}

We will now consider infinitesimal deformations~$\Delta \xi^\mu$, $\Delta \eta_\mu$, $\Delta {\Phi^\mu}_\nu$ of~$\xi^\mu$, $\eta_\mu$, ${\Phi^\mu}_\nu$. In coordinates adapted to the undeformed THF, the requirement that the deformation satisfies~\acsdef\ fixes all variations in terms of~$\Delta \xi^\mu$, $\Delta \eta_z$, $\Delta \eta_{\b z}$, $\Delta {\Phi^\tau}_\tau$, $\Delta {\Phi^z}_{\b z}$, $\Delta {\Phi^{\b z}}_z$. Explicitly,
\eqn\acsdeform{\eqalign{& \Delta \eta_\tau = - \eta_\mu \Delta \xi^\mu~,\cr
& \Delta {\Phi^z}_\tau = - i \Delta \xi^z~,\cr
& \Delta {\Phi^z}_z = - i h \Delta \xi^z~,\cr
& \Delta {\Phi^\tau}_z = - i \Delta \eta_z - i h \Delta \xi^\tau - i h^2 \Delta \xi^z + h \Delta {\Phi^\tau}_\tau - \b h \Delta {\Phi^{\b z}}_z~,}}
and their complex conjugates for~$\Delta {\Phi^{\b z}}_\tau$, $\Delta {\Phi^{\b z}}_{\b z}$, $\Delta {\Phi^\tau}_{\b z}$. Additionally requiring that the deformation preserve the integrability condition~\acsint\ leads to 
\eqn\tdefint{\eqalign{& \left(\Delta {\Phi^\tau}_\tau - i h \Delta \xi^z + i \b h \Delta \xi^{\b z}\right) \d_\tau h = 0~,\cr
& \d_\tau \left(\Delta{\Phi^z}_{\b z} - i \b h \Delta \xi^z\right) + 2 i \d_{\b z} \Delta \xi^z = 0~.}}
We can always satisfy the first condition by adjusting~$\Delta {\Phi^\tau}_\tau$ (see below). The second condition can be written in a more suggestive form by introducing a~$(0,1)$-form~$\Theta^z$ with coefficients in the holomorphic tangent bundle~$T^{1,0}\CM_3$,
\eqn\tdthetadef{\Theta^z =  - 2 i \Delta \xi^z \left(d\tau + h dz\right) + \left(\Delta {\Phi^z}_{\b z} - i \b h \Delta \xi^z\right) d{\b z}~.}
The second integrability condition in~\tdefint\ can now be written as
\eqn\thetareltd{\t \d \Theta^z = 0~.}
Trivial deformations induced by infinitesimal diffeomorphisms, which are parametrized by a real vector field~$\ep^\mu$, are given by 
\eqn\trivdef{\Delta \xi^z = - \d_\tau \ep^z~, \qquad \Delta {\Phi^z}_{\b z} = 2 i \d_{\b z} \ep^z - i \b h \d_\tau \ep^z~,}
or equivalently,
\eqn\trivtheta{\Theta^z = 2 i \t \d \ep^z~.}
We conclude that non-trivial deformations of the THF are parametrized by the cohomology class of~$\Theta^z$ in~$\t \d$-cohomology with coefficients in~$T^{1,0}\CM_3$,
\eqn\tdthetcoho{\left[\Theta^z\right] \in H^{0,1}\left(\CM_3, T^{1,0}\CM_3\right)~,}
in exact analogy to~\thcoho\ for complex manifolds. As was discussed there, not every cohomology class necessarily gives rise to an actual modulus, due to higher-order obstructions.

\subsec{Compatible Metrics}

Given~$\xi^\mu$, $\eta_\mu$, ${\Phi^\mu}_\nu$, it is always possible to find a compatible metric~$g_{\mu\nu}$, which satisfies
\eqn\tdcompmet{g_{\mu\nu} {\Phi^\mu}_\alpha {\Phi^\nu}_\beta  = g_{\alpha\beta} - \eta_\alpha \eta_\beta~.}
By multiplying with~$\xi^\beta$ we also find that~$\eta_\mu = g_{\mu\nu} \xi^\nu$. In adapted coordinates, such a metric takes the form
\eqn\tdadcompmet{ds^2 = \eta^2 + c\left(\tau, z, \b z\right)^2 dz d \b z = \left(d \tau + h\left(\tau, z, \b z\right) dz + \b h\left(\tau, z, \b z\right)d \b z\right)^2 + c\left(\tau, z, \b z\right)^2 dz d\b z~.}
Since the induced metric~$c\left(\tau, z, \b z\right)^2 dz d\b z$ on the normal bundle of the foliation is Hermitian for every~$\tau$, we refer to~\tdadcompmet\ as a transversely Hermitian metric compatible with the THF. It also follows from~\tdadcompmet\ that we can  express~${\Phi^\mu}_\nu = - {\ep^\mu}_{\nu\rho} \xi^\rho$, where we use the canonical orientation defined at the end of section~5.1. Therefore, $\eta_\mu$ and~${\Phi^\mu}_\nu$ can be viewed as auxiliary objects, which are derived from the orientable THF and the transversely Hermitian metric defined by~$\xi^\mu$ and~$g_{\mu\nu}$.

It is useful to note that we can obtain a THF with a transversely Hermitian metric by dimensionally reducing an integrable complex structure~${J^\mu}_\nu$ and a Hermitian metric~$G_{\mu\nu}$ in four dimensions. If~$G_{\mu\nu}$ possesses a Killing vector~$\d_x$ along which we can reduce, the nowhere vanishing vector field~$\xi^\mu = {1 \over \sqrt{G_{xx}}} {J^\mu}_x$ ($\mu\neq x$) and the three-dimensional metric~$g^{\mu\nu} = G^{\mu\nu}$ ($\mu, \nu \neq x$) define an orientable THF and a compatible transversely Hermitian metric. We will use this to generate various three-dimensional examples.

We can deform the THF and demand that the deformed metric~$g_{\mu\nu} + \Delta g_{\mu\nu}$ remain transversely Hermitian. At linear order, this leads to the constraint
\eqn\compcons{g_{\mu\nu} \Delta {\Phi^\mu}_\alpha {\Phi^\nu}_\beta + g_{\mu\nu} {\Phi^\mu}_\alpha \Delta {\Phi^\nu}_\beta + \Delta g_{\mu\nu} {\Phi^\mu}_\alpha {\Phi^\nu}_\beta = \Delta g_{\alpha\beta} - \Delta \eta_\alpha \eta_\beta - \eta_\alpha \Delta \eta_\beta~.}
In coordinates adapted to the undeformed case, we find that~$\Delta g_{z \b z}$ is unconstrained (we are always free to change~$c(\tau, z, \b z)$ in~\tdadcompmet), while the other components of~$\Delta g_{\mu\nu}$ are determined in terms of~$\Delta \xi^\mu$, $\Delta \eta_z$, $\Delta {\Phi^z}_{\b z}$,
\eqn\tdmetvarcoord{\eqalign{& \Delta g_{\tau \tau} = - 2 \eta_\mu \Delta \xi^\mu~,\cr
& \Delta g_{\tau z} = \Delta \eta_z - h \eta_\mu \Delta \xi^\mu - {c^2 \over 2} \Delta \xi^{\b z}~,\cr
& \Delta g_{z z } = {i c^2 \over 2} \Delta {\Phi^{\b z}}_{z} - {h c^2 \over 2} \Delta \xi^{\b z} + 2 h \Delta \eta_z~,}}
and their complex conjugates for~$\Delta g_{\tau \b z}$, $\Delta g_{\b z \b z}$. Note that~$\Delta g_{\mu\nu}$ does not explicitly depend on~$\Delta {\Phi^\tau}_\tau$. The supersymmetric theory on~$\CM_3$ is defined using the metric and~$\xi^\mu$, which is determined by the Killing spinor~$\zeta_\alpha$, while~${\Phi^\mu}_\nu = - {\ep^\mu}_{\nu\rho} \xi^\rho$ is an auxiliary object~\ClossetRU. Varying the metric and~$\xi^\mu$ in this formula, we find that~$\Delta {\Phi^\tau}_\tau = i h \Delta \xi^z - i \b h \Delta \xi^{\b z}$, which automatically satisfies the first integrability condition in~\tdefint.

\subsec{Abelian Gauge Fields and Holomorphic Line Bundles}

As we saw in section~1.5, supersymmetric configurations for three-dimensional Abelian background gauge fields correspond to a complex vector field~$\SA_\mu = A_\mu + i \sigma \eta_\mu$ on~$\CM_3$, whose curvature~$\SF_{\mu\nu} = \d_\mu \SA_\nu - \d_\nu \SA_\mu$ satisfies
\eqn\tdholcond{\SF_{\tau \b z} = 0~,}
where~$\tau, z, \b z$ are coordinates adapted to the THF on~$\CM_3$. This condition is reminiscent of the condition~\holcond, which defines a holomorphic line bundle over a complex manifold. We will now explain how~\tdholcond\ leads to an analogous construction in three dimensions. Locally, we can view~$\SA_\mu$ as a one-form and consider its anti-holomorphic part~$\SA^{0,1} = \SA_\tau \left(d\tau + h dz\right) +\SA_{\b z} d \b z$. Using~\tdaction, we see that~\tdholcond\ can be expressed as~$\t \d \SA^{0,1} = 0$. It follows from the Poincar\'e lemma for~$\t \d$ that we can locally express~$\SA^{0,1} = \t \d \lambda$ for some complex function~$\lambda(\tau, z, \b z)$. Therefore, we can locally set~$\SA^{0,1} = 0$ by a complex gauge transformation. The transition functions that preserve this gauge choice consist of~$GL(1,\C)$-valued functions that are annihilated by~$\t \d$, i.e.~they do not depend on~$\tau$ and are holomorphic in~$z$. This is the definition of a holomorphic line bundle over a three-manifold with a THF. 

We can change the structure of the holomorphic line bundle if we deform~$\SA^{0,1}$ by a well-defined~$(0,1)$-form~$\Delta \SA^{0,1}$ that preserves~\tdholcond,
\eqn\tddacond{\t \d  \big(\Delta \SA^{0,1}\big) = 0~.}
We will quotient by trivial deformations induced by globally defined gauge transformations,
\eqn\glaobalgttd{\Delta \SA^{0,1} = \t \d \ep~,}
where~$\ep(\tau, z, \b z)$ is a well-defined complex function on~$\CM_3$. Deformations of a holomorphic line bundle over~$\CM_3$ are thus parametrized by the~$\t \d$-cohomology class of~$\Delta \SA^{0,1}$,
\eqn\tdachoh{\left[\Delta  \SA^{0,1} \right] \in H^{0,1}\left(\CM_3\right)~.}

\newsec{Parameter Dependence of~$Z_{\CM_3}$}

In this section we study the dependence of the partition function~$Z_{\CM_3}$ on the THF and the transversely Hermitian metric on~$\CM_3$, as well as its dependence on the vector field~$U^\mu$ and the function~$\kappa$ in~\tdbgs\ and~\tdrgf. We also analyze the dependence on Abelian background gauge fields. As in four dimensions, we consider small deformations around flat space and rely on the~$Q$-cohomology of the~$\CR$-multiplet and a flavor current.

{\it Note: In this section, we raise and lower indices using the usual flat-space metric.}

\subsec{The $Q$-Cohomology of the~$\CR$-Multiplet}

The operators in the three-dimensional~$\CR$-multiplet reside in the following superfields (see for instance~\refs{\DumitrescuIU,\ClossetVG,\ClossetVP}),
\eqn\rcomp{\eqalign{ \CR_\mu =~& j_\mu^{(R)} - i \theta S_\mu - i \t \theta \t S_\mu - (\theta\gamma^\nu\t \theta) \left(2 T_{\mu\nu} + i \ep_{\mu\nu\rho} \d^\rho J^{(Z)}\right) \cr
& - i \theta \t \theta \left(2 j_\mu^{(Z)} + i \ep_{\mu\nu\rho} \d^\nu j^{(R)\rho} \right) + \half \theta^2 \t \theta \gamma^\nu \d_\nu S_\mu \cr
& + \half \t \theta^2 \theta \gamma^\nu \d_\nu \t S_\mu + {1 \over 4} \theta^2 \t \theta^2 \d^2 j_\mu^{( R)}~,\cr
\CJ^{(Z)} =~& J^{(Z)} - \half \theta \gamma^\mu S_\mu + \half \t \theta \gamma^\mu \t S_\mu + i \theta \t \theta T^\mu_\mu - (\theta \gamma^\mu \t \theta) j_\mu^{(Z)} \cr
& + {1 \over 4} \ep^{\mu\nu\rho}\theta^2 \t \theta \gamma_\mu \d_\nu S_\rho - {1 \over 4} \ep^{\mu\nu\rho}\t \theta^2 \theta \gamma_\mu \d_\nu \t S_\rho + {1 \over 4} \theta^2 \t \theta^2 \d^2 J^{(Z)}~.}}
Their~$Q$-transformations are given by
\eqn\tdrmulttrans{\eqalign{& \big\{Q, j_\mu^{( R )} \big\} = - i \zeta S_\mu~,\cr
&\big\{Q, S_{\alpha\mu} \big\} = 0~,\cr
& \big\{Q, \t S_{\alpha\mu} \big\} = \zeta_\alpha \left(2 j_\mu^{(Z)} + i \ep_{\mu\nu\rho} \d^\nu j^{( R )\rho} \right) + \left(\gamma^\nu\zeta\right)_\alpha \left(2 i T_{\mu\nu} + \d_\nu j_\mu^{( R  )} - \ep_{\mu\nu\rho} \d^\rho J^{(Z)} \right)~,\cr
&  \big\{Q, T_{\mu\nu} \big\} = {i \over 4} \ep_{\mu\rho\lambda} \zeta \gamma^\rho \d^\lambda S_\nu + {i \over4} \ep_{\nu\rho\lambda} \zeta \gamma^\rho \d^\lambda S_\mu~,\cr
& \big\{Q, j_\mu^{(Z)} \big\} = -{i \over 2} \zeta \gamma^\nu \d_\nu S_\mu - \half \ep_{\mu\nu\rho} \zeta \d^\nu S^\rho~,\cr
& \big\{Q, J^{(Z)} \big\} =  - \half \zeta \gamma^\mu S_\mu~.}}
There are six bosonic~$Q$-exact operators, $\big \{ Q, \t S_{\alpha\mu} \big\}$. Multiplying by~$\zeta$ or~$\zeta^\dagger$, we find that they are given by
\eqn\tdqexops{\eqalign{& \CT_{\tau\tau} = T_{\tau\tau} - i j_\tau^{(Z)} + 2 i \d_{\b z} j_z^{( R ) }~,\cr
& \CT_{\tau z} = T_{\tau z} - i j_z^{(Z)} + {i \over 2} \d_z j_\tau^{( R)} - i \d_\tau j_z^{( R )}+ \half \d_z J^{(Z)}~,\cr
& \CT_{\tau \b z} = T_{\tau \b z} - {i \over 2} \d_{\b z} j_\tau^{( R)} + \half \d_{\b z} J^{(Z)}~,\cr
& \CT_{z\b z} = T_{z \b z} - {i \over 2} \d_{\b z} j_z^{( R)} -{1 \over 4} \d_\tau J^{(Z)}~,\cr
& \CT_{\b z \b z} = T_{\b z \b z} -{i \over 2} \d_{\b z} j_{\b z}^{( R )}~,\cr
& \CJ^{(Z)}_{\b z} = j^{(Z)}_{\b z} - i \d_{\b z} J^{(Z)}~.}}
These are also the only~$Q$-closed bosonic operators in the~$\CR$-multiplet.

\subsec{Deformations around Flat Space}

The linearized couplings of the three-dimensional~$\CR$-multiplet to the corresponding bosonic supergravity fields is given by~\linearizedthreed,
\eqn\tdlinlag{\Delta \SL = - \half \Delta g^{\mu\nu} T_{\mu\nu} + A^{( R ) \mu} j_\mu^{( R )} + C^\mu j_\mu^{(Z)} + H J^{(Z)}~.}
Following the discussion in section~5, we deform the THF and the compatible transversely Hermitian metric around flat space. By linearizing~\tdbgs\ and~\tdrgf, we find that the supergravity background fields are given by
\eqn\tdlinbar{\eqalign{& A^{( R)\tau} = {i \over 2} \d_{\b z} \Delta \xi^{\b z} + i \d_z \Delta \eta_{\b z} - i \d_{\b z} \Delta \eta_z + {i \over 2} \d_\tau \Delta g_{z \b z}~,\cr
& A^{(R )z} = \half \d_z \Delta {\Phi^z}_{\b z} - i \d_{\b z} \Delta g_{z\b z} - 2i \d_{\b z} \Delta \xi^\tau - 2i \d_\tau \Delta \eta_{\b z}~,\cr
& A^{( R)\b z} = i \d_z \Delta g_{z \b z} + \half \d_{\b z} \Delta {\Phi^{\b z}}_z~,\cr
& C^\tau = - i \Delta \xi^\tau + \t C^\tau~, \qquad C^z = 2 i \Delta \eta_{\b z} + \t C^z~, \qquad C^{\b z} = 2 i \Delta \eta_z + \t C^{\b z}~,\cr
& H = - \d_\tau \Delta g_{z \b z} - \half \d_z \Delta \xi^z - \half \d_{\b z} \Delta \xi^{\b z} + \d_z \Delta \eta_{\b z} - \d_{\b z} \Delta \eta_z + i \kappa~.}}
Here~$\t C_\mu$ is such that its dual field strength is given by
\eqn\fsct{-i \ep^{\mu\nu\rho} \d_\nu \t C_\rho = U^\mu + \kappa \eta^\mu~,}
where~$U^\mu$ and~$\kappa$ are the ambiguities in~$V^\mu$ and~$H$, which satisfy the conditions in~\tdbgs,
\eqn\ukproprep{{\Phi^\mu}_\nu U^\nu = - iU^\mu~, \qquad \grad_\mu \left(U^\mu + \kappa \eta^\mu\right) = 0~.}
We can trade~$U^\mu$ and~$\kappa$ for a closed two-form,
\eqn\tutddef{W_{\mu\nu} = \half \ep_{\mu\nu\rho} \left(U^\rho + \kappa \eta^\rho\right)~, \qquad d W = 0~.}
It follows from~\ukproprep\ that~$U^\mu$ only has a~$\b z$-component, so that~$W_{\tau \b z} = 0$, and hence~$W$ is a~$(1,1)$-form. Since it is~$d$-closed, it is also~$\t \d$-closed,
\eqn\tuclosed{\t \d W = 0~.}
Just as in four dimensions, we will see that the partition function~$Z_{\CM_3}$ only depends on~$U^\mu$ and~$\kappa$ through the cohomology class of~$W$ in~$H^{1,1}(\CM_3)$.

We can now substitute the background fields~\tdlinbar\ into~\tdlinlag, use the conservation of the~$R$-current, and drop a total derivative to obtain the following Lagrangian,
\eqn\tddeflag{\eqalign{\Delta \SL =~& - 4 \Delta g_{z \b z} \CT_{z \b z} - 2 \Delta \eta_z \left(\CT_{\tau \b z} - i \CJ_{\b z}\right) - 2 \Delta \eta_{\b z} \CT_{\tau z}  \cr
& + \Delta \xi^\tau \CT_{\tau \tau} + \Delta \xi^{\b z} \CT_{\tau \b z} - i \Delta {\Phi^{\b z}}_z \CT_{\b z \b z} + \t C^\mu j_\mu^{(Z)} + i \kappa J^{(Z)} \cr
& + \Delta \xi^z \left(T_{\tau z} + \half \d_z J^{(Z)}\right) + i \Delta {\Phi^z}_{\b z} \left(T_{zz} + {i \over 2} \d_z j_z^{( R)}\right)~.}}
This Lagrangian is supersymmetric if~$\Delta \xi^z$ and~$\Delta {\Phi^z}_{\b z}$ satisfy the second integrability condition in~\tdefint\ with~$h = 0$, since we are deforming around flat space.

Recalling the~$Q$-exact operators in~\tdqexops, we conclude about the partition function:
\medskip
\item{$\bullet$} Given a fixed THF, $Z_{\CM_3}$ does not depend on the compatible transversely Hermitian metric, i.e.~it does not depend on the functions~$h(\tau, z, \b z)$ and~$c(\tau, z, \b z)$ in~\tdadcompmet. 
\item{$\bullet$} $Z_{\CM_3}$ does not depend on~$\Delta \xi^\tau$, $\Delta \xi^{\b z}$, $\Delta {\Phi^{\b z}}_z$. 
\medskip
\noindent 
As in four dimensions, the dependence on~$\Delta \xi^z$ and~$\Delta {\Phi^z}_{\b z}$ is cohomological. Recall from~\tdthetadef\ and~\thetareltd\ that the~$(0,1)$-form~$\Theta^z$ with coefficients in~$T^{1,0}\CM_3$, which was defined in terms of~$\Delta \xi^z$ and~$\Delta {\Phi^z}_{\b z}$, satisfies~$\t \d \Theta^z = 0$. According to~\trivtheta, a trivial deformation of the THF corresponds to~$\Theta^z = 2i \t \d \ep^z$. Since~$\ep^\mu$ parametrizes an infinitesimal diffeomorphism of~$\CM_3$, which cannot affect the partition function, we conclude:
\medskip
\item{$\bullet$} $Z_{\CM_3}$ only depends on deformations of the THF through the cohomology class of~$\Theta^z$ in~$H^{0,1}\left(\CM_3, T^{1,0}\CM_3\right)$. It is a locally holomorphic function of the corresponding moduli. 
\bigskip

To prove that~$Z_{\CM_3}$ only depends on~$\t C_\mu$ and~$\kappa$ through the~$\t \d$-cohomology class of the~$(1,1)$-form~$W$ in~\tutddef, we assume that~$W = \t \d \varphi^{1,0}$ for some well-defined~$(1,0)$-form~$\varphi^{1,0}$, so that~$\t C_\mu$ and~$\kappa$ are derivatives of~$\varphi^{1,0}_z$. Substituting into~\tddeflag\ and dropping a total derivative, we find that the terms containing~$\varphi^{1,0}_z$ are given by
\eqn\tdtriviallag{\Delta \SL = 4 i \varphi^{1,0}_z \CJ_{\b z}^{(Z)} + \cdots~.}
Since~$\CJ^{(Z)}_{\b z}$ is~$Q$-exact, we have shown:
\medskip
\item{$\bullet$} $Z_{\CM_3}$ only depends on~$W$ through its cohomology class in~$H^{1,1}(\CM_3)$. 
\medskip
\noindent As in four dimensions, the dependence on~$W$ drops out completely, if the~$\CR$-multiplet can be improved to an FZ-multiplet, which happens whenever the theory does not possess any parameters that can contribute to the central charge in the flat-space supersymmetry algebra, such as real masses or FI-terms~\DumitrescuIU. Certain choices of~$W$ may lead to non-trivial flux for the graviphoton~$C_\mu$. For instance, this happens for the~$S^2 \times S^1$ background that preserves four supercharges, which is discussed in section~7.2. In this case the central charge, which couples to the graviphoton, must be quantized in terms of these fluxes. This is similar to the quantization of FI-terms discussed at the end of section~3.2.

\subsec{Background Gauge Fields}

We now consider the dependence of~$Z_{\CM_3}$ on Abelian background gauge fields. A flavor current~$j_\mu$ resides in a real linear multiplet,
\eqn\tdlinmult{\CJ = J + i \theta j + i \t \theta \t j + i \theta \t \theta K - \big(\theta \gamma^\mu \t \theta\big) j_\mu  - \half \thetasq \t \theta \gamma^\mu \d_\mu j - \half {\t \theta}^2 \theta \gamma^\mu \d_\mu \t j + {1 \over 4} \thetasq {\t \theta}^2 \d^2 J~,}
which implies the following~$Q$-transformations:
\eqn\tdjvar{\eqalign{& \big\{Q, J\big\} = i \zeta j~,\cr
&  \big\{Q, j_\alpha \big\} = 0~,\cr
&  \big\{Q, \t j_\alpha \big\} = - i\left(\gamma^\mu\zeta\right)_\alpha \left(j_\mu - i \d_\mu J\right) + \zeta_\alpha K~,\cr
&  \big\{Q, j_\mu \big\} = i \ep_{\mu\nu\rho} \zeta \gamma^\rho \d^\nu j~,\cr
&  \big\{Q, K\big\} = - i \zeta \gamma^\mu \d_\mu j~.}}
Multiplying by~$\zeta$ or~$\zeta^\dagger$, we find that the bosonic~$Q$-exact operators~$\{Q, \t j_\alpha\}$ are given by
\eqn\tdcurrexop{\CJ_{\b z} = j_{\b z} - i \d_{\b z} J~, \qquad {\CK} = K - i j_\tau - \d_\tau J~.}
These are also the only~$Q$-closed operators.

At the linearized level, $\CJ$ couples to the bosonic fields in a background vector multiplet as follows,
\eqn\tdflavcoup{\Delta \SL = A^\mu j_\mu + \sigma K + DJ~.}
The background field~$D$ is given by the linearization of~\tdgfcons,
\eqn\flatdtd{D = \d_\tau \sigma - 2i \left(\d_z A_{\b z} - \d_{\b z} A_z\right)~.}
Substituting into~\tdflavcoup\ and dropping a total derivative, we find
\eqn\tdfllag{\Delta \SL = \SA_\tau j_\tau + 2 \SA_ z \CJ_{\b z} + 2 \SA_{\b z} \left(j_z + i \d_z J\right) + \sigma {\CK}~,}
written in terms of the gauge field~$\SA_\mu = A_\mu + i \sigma \eta_\mu$ defined in~\sadef. The Lagrangian~\tdfllag\ is supersymmetric provided the field strength~$\SF_{\mu\nu}$ of~$\SA_\mu$ satisfies~$\SF_{\tau \b z} = 0$. This implies that~$\SA_\mu$ determines a holomorphic line bundle, as explained in section~5.5.

Since~$\CJ_{\b z}$ and~$\CK$ are~$Q$-exact, the partition function only depends on~$\SA_\tau$ and~$\SA_{\b z}$, i.e.~the~$(0,1)$-part of~$\SA_\mu$. As before, invariance under complex gauge transformations renders this dependence cohomological, so that we conclude:
\medskip
\item{$\bullet$} $Z_{\CM_3}$ only depends on the cohomology class of~$\SA^{0,1}$ in~$H^{0,1}(\CM_3)$. Hence, it is a locally holomorphic function of the corresponding line bundle moduli.
\bigskip

\newsec{Examples in Three Dimensions}

In this section we explore three-manifolds~$\CM_3$ that are diffeomorphic to~$S^3$ and~$S^2 \times S^1$. In each case we analyze the moduli space of THFs and holomorphic line bundles. We discuss the implications for the partition functions~$Z_{\CM_3}$ on these manifolds, making contact with the literature on squashed spheres and supersymmetric indices in three dimensions.

\subsec{Squashed Spheres}

Much recent work has focused on the partition functions~$Z_{S^3}$ of three-dimensional $\CN=2$ theories with a~$U(1)_R$ symmetry on round and squashed~$S^3$ backgrounds. Here we will refer to any manifold that is diffeomorphic to~$S^3$, but not isometric to the round case, as a squashed sphere. The partition function on a round~$S^3$ with four supercharges residing in~$SU(2|1)$ was computed in~\refs{\KapustinKZ\JafferisUN-\HamaAV} using localization. This was subsequently generalized to various squashed spheres~\refs{\HamaEA\ImamuraUW\ImamuraWG\MartelliFU\NishiokaHAA\MartelliAQA\AldayLBA\NianQWA-\TanakaDCA}, all of which preserve at least two supercharges.\foot{The squashed spheres discussed in~\refs{\ImamuraUW,\ImamuraWG} preserve all four supercharges (see also~\ClossetRU).} By looking at the explicit expressions for the partition functions in these examples, one is lead to the following observations:
\medskip
\item{a)} $Z_{S^3}$ only depends on the squashing through a single complex parameter, usually called~$b$, where~$b=1$ corresponds to the round~$S^3$ of~\refs{\KapustinKZ\JafferisUN-\HamaAV}.
\item{b)} Some squashings do not affect $Z_{S^3}$, i.e. they give~$b=1$, even though they are not isometric to the round case.
\medskip
\noindent We will address these observations using our general framework. In order to explain~a), we show that all known examples belong to a one-parameter family of THFs on~$S^3$, while~b) follows from the observation that some squashings deform the transversely Hermitian metric without changing the THF. We will also describe THFs on~$S^3$ that do not belong to this one-parameter family.

A global description of all THFs on manifolds diffeomorphic to~$S^3$ is beyond the scope of this paper (see~\refs{\BG\Brunella-\Ghys} for a full classification). Instead, we will analyze small deformations of the THF that corresponds to the round spheres of~\refs{\KapustinKZ\JafferisUN-\HamaAV}. In adapted~$\tau, z, \b z$ coordinates, it takes the following form,
\eqn\cancon{\xi=\d_\tau~, \qquad \eta=d\tau+{i\over 2} {\b z dz - z d \b z \over 1+ |z|^2 }~,\qquad ds^2=\eta^2+{dz d\b z\over (1+ |z|^2)^2}~,} 
where~$\tau \sim \tau + 2\pi$ runs along a Hopf fiber over the~$\C\P^1$ base parametrized by~$z$. To cover the point~$z = \infty$, we must change adapted coordinates to~$z'={1\over z}$ and~$\tau'=\tau-{i\over 2}\log {\b z\over z}$. Since the radius~$r$ of the sphere does not affect the partition function, we have set~$r=1$. Note that~$\xi$ is a Killing vector, which is a special feature of this example.

We start with the simpler problem of computing the cohomology~$H^{0,1}(S^3)$, defined in~\tdcohodef. As discussed in section~5.5, it parametrizes moduli of holomorphic line bundles over~$S^3$. (Topologically, all complex line bundles over~$S^3$ are trivial.) According to~\tdaction, we must find~$(0,1)$-forms~$\omega^{0,1} = \omega_\tau^{0,1} \left(d\tau + h dz\right) + \omega_{\b z}^{0,1}$ that satisfy
\eqn\oneformsex{\d_\tau \omega_{\b z}^{0,1} = \d_{\b z} \omega_\tau^{0,1}~,}
modulo those~$(0,1)$-forms that can be written in terms of a well-defined complex function~$\ep(\tau, z, \b z)$ on~$S^3$, 
\eqn\trivialex{\omega^{0,1}_\tau = \d_\tau \ep~, \qquad \omega_{\b z}^{0,1} = \d_{\b z} \ep~.}
We can restrict ourselves to~$\d_\tau \omega^{0,1}_\tau = 0$, since terms with non-trivial~$\tau$-dependence can be removed by choosing a suitable~$\ep$ in~\trivialex. (To see this, Fourier-expand~$\omega_\tau^{0,1}$ in the periodic variable~$\tau$ and integrate term by term.) Therefore~$\omega_\tau^{0,1}(z,\b z)$ descends to a well-defined scalar on the~$\C\P^1$ base. Now, the only way to satisfy~\oneformsex\ with well-defined~$\omega_{\b z}^{0,1}$ is to separately set both sides of this equation to zero. First, it follows that~$\omega_\tau^{0,1}$ is a holomorphic function on~$\C\P^1$, and therefore constant, $\omega_\tau^{0,1} = \gamma \in \C$. Second, we find that~$\omega_{\b z}^{0,1}$ does not depend on~$\tau$. The transformation rules between adapted coordinates discussed in section~5.1 show that~$\omega_{\b z}^{0,1}$ does not descend to a~$(0,1)$-form on~$\C\P^1$, but the combination~$\omega_{\b z}^{0,1} - \b h \omega^{0,1}_\tau$ does, because~$h = \eta_{z}$ does not depend on~$\tau$. Every~$(0,1)$-form on~$\C\P^1$ can be written as~$\d_{\b z} \ep$ for a well-defined function~$\ep(z, \b z)$ on~$\C\P^1$ (the Dolbeault Cohomology~$H^{0,1}(\C\P^1)$ vanishes). We thus find that non-trivial elements of~$H^{0,1}(S^3)$ are complex multiples of the one-form~$\eta$ in~\cancon, i.e.~$\omega^{0,1} = \gamma \eta$ with~$\gamma \in \C$. This shows that the partition functions~$Z_{S^3}$ depend on a supersymmetry-preserving Abelian background gauge field through a single complex parameter, consistent with all known examples.

As in four-dimensional examples, we can determine~$H^{0,1}(S^3, T^{1,0} S^3)$ by multiplying~$\gamma \eta$, which represents~$H^{0,1}(S^3)$, with a holomorphic vector field~$X = X^z \left(\d_z - h \d_\tau\right)$. As long as~$X^z$ is annihilated by~$\t \d$, i.e.~it does not depend on~$\tau$ and is holomorphic in~$z$, we find that~$\Theta = \gamma X \otimes \eta$ is a non-trivial element of~$H^{0,1}(S^3, T^{1,0} S^3)$.\foot{It can be checked that this procedure generates all of~$H^{0,1}(S^3, T^{1,0}S^3)$.} Since~$X^z$ does not depend on~$\tau$, it reduces to a holomorphic vector field (with holomorphic coefficients) on~$\C\P^1$, of which there are only three. We divide the resulting THFs into two types:
\medskip
\item{1.)} The first type correspond to~$X = z \left(\d_z - h \d_\tau\right)$, so that~$\Theta^{( \gamma)} = \gamma X \otimes \eta$ defines a one-parameter family of deformations.
\item{2.)} The second type correspond to~$X = \left(\d_z - h \d_\tau\right)$ or~$X = z^2 \left(\d_z - h \d_\tau\right)$. The parameter~$\lambda$ in the infinitesimal deformation~$\Theta^{( \lambda)} = \lambda X \otimes \eta$ can be removed by rescaling~$z$. Like Hopf surfaces of the second type, discussed in section~4.1, these THFs should not possess any moduli. In appendix~D we construct such a rigid THF, which only preserves a single supercharge, by reducing a Hopf surface of second type to~$S^3$.   
\bigskip

Here we will focus on THFs of the first type. They can be realized by reducing the complex structure and the Hermitian metric~\metricst\ on primary Hopf surfaces~$\CM_4^{p,q}$ along~$\d_x$. (Various reductions from~$S^3 \times S^1$ to squashed three-spheres were discussed in~\refs{\DolanRP, \GaddeIA,\ImamuraUW,\ImamuraWG,\AharonyDHA}.) Since this family of THFs is one-dimensional, the corresponding partition functions~$Z_{S^3}$ should only depend on the geometry through a single complex parameter -- the squashing parameter~$b$. Given a particular squashing, we can expand the metric and the THF around~\cancon\ and ask what values of~$\gamma$ are realized by this squashing. (In the vicinity of~\cancon\ it is possible to establish a precise mapping between~$b$ and~$\gamma$, but we will not need it here.)

All known examples of squashed spheres realize THFs of the first type. The squashed spheres studied in~\MartelliAQA, which are reviewed in appendix E, realize deformations with complex~$\gamma$. In special cases, they are equivalent to the ellipsoid squashings with~$U(1) \times U(1)$ isometry studied in~\HamaEA, which only give rise to real~$\gamma$, and the squashings of~\refs{\ImamuraUW,\ImamuraWG}. Some of the squashings discussed in~\refs{\HamaEA,\MartelliFU} give~$\gamma = 0$, i.e.~they do not change the THF, even though they deform the transversely Hermitian metric, and this explains why they do not affect the partition function. On the other hand, the results of~\refs{\AldayLBA\NianQWA-\TanakaDCA} show that~$\gamma \neq 0$ can be realized on~$S^3$ with a round metric. Finally, the authors of~\NishiokaHAA\ considered a singular space with topology~$S^3$, which realizes a THF of the first type once the singularities are resolved.

\subsec{$S^2 \times S^1$ without Flux and the Supersymmetric Index}

Any~$\CN=2$ theory with a $U(1)_R$ symmetry can be placed on~$S^2\times \R $ with the usual round metric while preserving four supercharges residing in~$SU(2|1)$, which contains the~$SU(2)$ isometry of the sphere~\ImamuraSU. As in four dimensions, a suitable trace over states on~$S^2$ in Hamiltonian quantization defines a supersymmetric index~$\CI(y,u)$, where~$y$ and~$u$ are (generally complex) fugacities that couple to the Hamiltonian~$H$, which generates translations along~$\R$, and an Abelian flavor charge~$Q_f$ that commutes with the supercharges~\refs{\BhattacharyaZY,\ImamuraSU,\KapustinJM,\DimoftePY}. (As in four dimensions, generic values of the chemical potentials are only consistent with two supercharges.) Up to local counterterms, this index corresponds to a partition function on a space diffeomorphic to~$S^2\times S^1$. Here we describe the family of THFs that corresponds to the index and interpret the fugacities~$y, u$ as geometric moduli. A distinct family of THFs on~$S^2 \times S^1$ is discussed in the next subsection.

Our starting point is a round metric on~$S^2 \times S^1$, 
\eqn\metcylf{ds^2= dx^2+d\theta^2+ \sin^2\theta \, d\varphi^2~,}
where~$x \sim x +1$, $\varphi \sim \varphi + 2\pi$, $0 \leq \theta \leq \pi$, and we have set the radius of~$S^2$ to one. Up to rotations of~$S^2$, the THF that describes the case with four supercharges in~$SU(2|1)$ is given by
\eqn\chxi{\xi=\cos\theta \d_x+\sin \theta \d_\theta~.} 
Note that the flux of~$\xi$ through~$S^2$ vanishes. This is related to the fact that there is no flux of the~$R$-symmetry gauge field~$A_\mu^{( R)}$ through~$S^2$, which can be checked using~\tdrgf, and hence there is no restriction on the allowed~$R$-charges.

As long as~$\theta \neq \pi$, we can introduce adapted coordinates
\eqn\adcoor{\tau=x-2\log\Big(\cos{\theta\over 2}\Big)~,\qquad  z=e^{-x + i\varphi}\sin\theta~,} 
in which~$\xi = \d_\tau$. The periodicity of~$x$ implies the following identifications,
\eqn\idft{\left(\tau,z\right)\sim \left(\tau+1, e^{-1} z\right)~.}
In these coordinates, the metric~\metcylf\ takes the form
\eqn\metchc{ds^2=\eta^2+{16 e^{2\tau} dz d\b z\over (4+e^{2\tau} |z|^2 )^2}~, \qquad \eta= d\tau+{e^{2\tau}(\b z dz+ z d\b z)\over 4+e^{2\tau} |z|^2}~,}
so that~$\xi$ is not a Killing vector. The coordinate~$\tau$ becomes singular at~$\theta=\pi$. For~$\theta \neq 0$, we can instead use
\eqn\difptch{\tau'=-x+2\log\Big(\sin{\theta\over 2}\Big)=\tau+\log{|z|^2 \over 4}~.}

As before, we begin by examining~$H^{0,1}(S^2\times S^1)$, which parametrizes holomorphic line bundle moduli. This cohomology is computed in appendix~F, where we show that it is one-dimensional. The generator is given by~$\left(dx\right)^{0,1}$, the~$(0,1)$-component of the one-form~$dx$ representing the usual de Rham cohomology~$H^1(S^2 \times S^1)$. (Proving that this is the only non-trivial element of~$H^{0,1}(S^2 \times S^1)$ is complicated by the fact that~$\xi$ is not a Killing vector.) Thus, there is a single holomorphic line bundle modulus on this~$S^2 \times S^1$, which corresponds to the fugacity~$u$ for the Abelian flavor charge~$Q_f$ in the index~$\CI(y,u)$.

As in previous examples, the cohomology~$H^{0,1}(S^2\times S^1, T^{1, 0} S^2\times S^1)$ that parametrizes deformations of the THF can be obtained by tensoring the generator~$(dx)^{0,1}$ of~$H^{0,1}(S^2 \times S^1)$ with a holomorphic vector field~$X$, whose coefficients are holomorphic functions of~$z$. Due to the identifications~\idft, the only such vector fields are complex multiples of~$X = z \left(\d_z - h \d_\tau\right)$. Therefore, $H^{0,1}(S^2\times S^1, T^{1, 0} S^2\times S^1)$ is also one-dimensional, and hence the THF in~\chxi\ belongs to a one-parameter family. (This also follows from the classification of~\refs{\BG\Brunella-\Ghys}.) As in four dimensions, the corresponding modulus can be identified with the fugacity~$y$ in the index~$\CI(y,u)$. It is straightforward to write down explicit metrics and THFs on~$S^2 \times S^1$ that realize these deformations. For general complex~$y$, they can be chosen to preserve two supercharges of opposite~$R$-charge.

\subsec{$S^2 \times S^1$ with Flux}

There is another THF on~$S^2\times S^1$ with the round metric~\metcylf, which is obtained by taking~$\xi$ to be the Killing vector~$\xi=\d_x$. Note that this choice of~$\xi$ is topologically distinct from~\chxi, since it has flux through the~$S^2$. It follows from~\tdrgf\ that the~$R$-symmetry gauge field~$A^{(R )}$ has a unit-flux monopole through~$S^2$, so that well-defined bosonic fields must carry integer~$R$-charge. 

We can introduce adapted coordinates~$\tau=x$ and~$z=\tan{\theta \over 2} e^{i\varphi}$, which is a holomorphic coordinate on the~$\C\P^1$ base. Note that~$\eta = d \tau$, so that~$h = 0$. It is straightforward to repeat the arguments of section~7.1 to conclude that the cohomology~$H^{0,1}(S^2 \times S^1)$ is one-dimensional and generated by~$\eta$. Hence, there is a single holomorphic line bundle modulus on~$S^2 \times S^1$. Deformations of the THF are obtained by tensoring with a suitable holomorphic vector field~$X$ on~$\C\P^1$. As on~$S^3$, we find two types of deformations: 
\item{1.)} Choosing~$X = z \d_z$ gives rise to a one-parameter family of THFs. They can be obtained from four dimensions. For instance, we can reduce the~$S^2 \times T^2$ background discussed in section~4.6 along a cycle of the torus, or we can reduce the Hopf surfaces discussed in sections~4.1 and~4.2 along a suitable Hopf fiber. In both cases we loose one of the two complex structure moduli and end up with a single modulus in three dimensions. 
\item{2.)} Choosing~$X = \d_z$ or~$X = z^2 \d_z$ corresponds to a THF without any moduli, since the deformation parameter can be eliminated by rescaling~$z$.\foot{Just as for~$S^3$, an example of such a rigid THF arises by reducing the Hermitian metric on the Hopf surface of the second type described in appendix~D along the Killing vector~$\d_\varphi+\d_\chi$.}

\vskip 1cm

\noindent {\bf Acknowledgments:}
We would like to thank~D.~Cassani, C.~C\'ordova, S.~Gukov, G.~Gur-Ari, N.~Seiberg, I.~Shamir, A.~Strominger, S.-T.~Yau, and especially S.~Cecotti, M.~Kontsevich, and C.~Vafa for useful discussions. CC is a Feinberg postdoctoral fellow at the Weizmann Institute of Science. TD is supported by the Fundamental Laws Initiative of the Center for the Fundamental Laws of Nature at Harvard University, as well as DOE grant~DE-SC0007870 and NSF grants~PHY-0847457, PHY-1067976. The work of TD was also supported in part by NSF grant PHY-0756966 and a Centennial Fellowship from Princeton University. The work of GF was supported in part by NSF grant PHY-0969448 and a Marvin L. Goldberger Membership at the Institute for Advanced Study.  ZK was supported by NSF grant PHY-0969448, a research grant from Peter and Patricia Gruber Awards, a Rosa and Emilio Segre research award, a grant from the Robert Rees Fund for Applied Research, and by the Israel Science Foundation under grant number 884/11. ZK would also like to thank the United States-Israel Binational Science Foundation (BSF) for support under grant number 2010/629. The research of CC and ZK is supported by the I-CORE Program of the Planning and Budgeting Committee and by the Israel Science Foundation under grant number 1937/12. Any opinions, findings, and conclusions or recommendations expressed in this material are those of the authors and do not necessarily reflect the views of the funding agencies.

\appendix{A}{Conventions}

We follow the conventions of~\DumitrescuHA\ in four dimensions, and those of~\ClossetRU\ in three-dimensions, with the following exceptions:
\medskip
\item{$\bullet$} We denote the~$R$-symmetry gauge field in four and three dimensions by~$A_\mu^{( R)}$. It is related to the gauge field~$A_\mu$ used in~\refs{\DumitrescuHA, \ClossetRU} as follows:
\eqn\ararel{A^{( R)}_\mu = A_\mu - {3 \over 2} V_\mu~.}
\item{$\bullet$} In four dimensions, the complex structures~$J_{here}$ and~$\t J_{here}$ are determined in terms of the Killing spinors~$\zeta_\alpha$ and~$\t \zeta^\alphadot$ according to~\ccbil\ and~\tjdef. They differ from the complex structures used in~\DumitrescuHA\ by a sign, so that~$J_{here} = - J_{there}$ and~${\t J}_{here} = - {\t J}_{there}$. This preserves the orientation, but exchanges holomorphic and anti-holomorphic indices. For instance, the Killing vector~$K^\mu = \zeta \sigma^\mu \t \zeta$ is anti-holomorphic with respect to~$J_{here}$ and~$\t J_{here}$, but it was holomorphic in~\DumitrescuHA.
\item{$\bullet$} In four dimensions, we parametrize the ambiguity in the background fields through the $\b \d$-closed~$(2,1)$ form~$W_{ij \b k}$ in~\fdthreform, rather than a conserved, anti-holomorphic piece in~$V^\mu$, which was called~$U^\mu$ in~\DumitrescuHA.
\item{$\bullet$} In three dimensions, we defined~${\big(\Phi_{here}\big)^\mu}_\nu = - {\ep^\mu}_{\nu\rho} \xi^\rho$. This differs from the definition in~\ClossetRU\ by a sign, so that~$\Phi_{here} = - \Phi_{there}$. 
\item{$\bullet$} In~\ClossetRU, the mathematical structure required for the existence of a supercharge in three dimensions was characterized as an almost contact metric structure satisfying~\acsdef\ and~\acsint. Here we have stated this requirement in terms of an equivalent, well-studied mathematical structure: a THF with a compatible transversely Hermitian metric.
\medskip
\noindent Below, we collect several formulas that are useful for computations in four- and three-dimensional flat space. 

\subsec{Four Dimensions}

In flat Euclidean~$\R^4$, we define holomorphic coordinates,
\eqn\ccfour{w = x^1 + i x^2~, \qquad z = x^3 + i x^4~,}
so that the usual orientation~$\ep_{1234} = 1$ corresponds to~$\ep_{w\b w z \b z} = -{1 \over 4}$. In these coordinates, the sigma matrices~$\sigma^\mu_{\alpha\alphadot}, \t \sigma^{\mu\alphadot\alpha}$ are given by
\eqn\holsigmat{\eqalign{& \sigma^w = -\t \sigma^w =  \pmatrix{0 & 2 \cr 0 & 0}~, \quad \sigma^{\b w} = - \t \sigma^{\b w} = \pmatrix{0 & 0 \cr 2 & 0}~, \cr
& \sigma^z = - \t \sigma^{\b z} =  \pmatrix{2 & 0 \cr 0 & 0}~, \quad \sigma^{\b z} = -  \t \sigma^z = \pmatrix{0 & 0 \cr 0 & -2}~,}}
while the matrices~${\big(\sigma_{\mu\nu}\big)_\alpha}^\beta, {\big(\t \sigma_{\mu\nu}\big)^\alphadot}_\betadot$ take the following form:
\eqn\holsigtens{\eqalign{& \sigma_{w \b w} = \sigma_{z \b z} = \t \sigma_{w \b w} = - \t \sigma_{z \b z} = {1 \over 4} \pmatrix{1 & 0 \cr 0 & -1}~,\cr
& \sigma_{wz} = \t \sigma_{w \b z} = \half \pmatrix{0 & 0 \cr -1 & 0}~, \qquad \sigma_{\b w \b z} = - \t \sigma_{z \b w} = \half \pmatrix{0 & 1 \cr 0 & 0}~.}}
Given two supercharges~$Q, \t Q$, which are parametrized by constant spinors~$\zeta_\alpha, \t \zeta^\alphadot$, we define the supersymmetry variations~$\{Q, \CS\}$ and~$\{\t Q, \CS\}$ of a superfield~$\CS(x, \theta, \t \theta)$ as follows:
\eqn\svar{\eqalign{&\{Q, \CS\} = i \zeta^\alpha \CQ_\alpha \CS~, \qquad\{\t Q, \CS\} = i \t \zeta_\alphadot \t \CQ^\alphadot \CS~,\cr
& \CQ_\alpha = {\d \over \d \theta^\alpha} - i(\sigma^\mu \t \theta)_\alpha \d_\mu~, \qquad \t \CQ_{\alphadot} = - {\d \over \d \t \theta^\alphadot} + i(\theta \sigma^\mu)_\alphadot \d_\mu~.}}

\subsec{Three Dimensions}

In flat Euclidean~$\R^3$, we define adapted coordinates,
\eqn\tdacflat{\tau = x^1~, \qquad z = x^2 + i x^3~,}
so that the usual orientation~$\ep_{123} = 1$ corresponds to~$\ep_{\tau z \b z} = {i \over 2}$. In these coordinates, the gamma matrices~${\big(\gamma^\mu\big)_\alpha}^\beta$ are given by
\eqn\tdgammas{\gamma^\tau = \pmatrix{1 & 0 \cr 0 & -1}~, \qquad \gamma^z = \pmatrix{0 & -2 \cr 0 & 0}~, \qquad \gamma^{\b z} = \pmatrix{0 & 0 \cr -2 & 0}~.}
The supersymmetry variation of a superfield~$\CS(x, \theta, \t \theta)$ under the supercharge~$Q$ parametrized by~$\zeta_\alpha$ takes the form
\eqn\tdsfieldvar{\{Q, \CS\} = i \zeta^\alpha \CQ_\alpha \CS~, \qquad \CQ_\alpha = {\d \over \d \theta^\alpha} - i(\sigma^\mu \t \theta)_\alpha \d_\mu~.}

\appendix{B}{Basic Aspects of Complex Manifolds}

Here we review a few standard facts about complex manifolds, see for instance~\refs{\Kodairabook,\Huybrechts}. Given a differentiable manifold~$\CM$, an almost complex structure~${J^\mu}_\nu$ satisfies
\eqn\acplsdef{{J^\mu}_\nu {J^\nu}_\rho = -{\delta^\mu}_\rho~.}
We say that~${J^\mu}_\nu$ is an integrable complex structure if
\eqn\ntensor{{J^\lambda}_\nu \d_\lambda {J^\mu}_\rho - {J^\lambda}_\rho \d_\lambda {J^\mu}_\nu -{J^\mu}_\lambda \d_\nu {J^\lambda}_\rho + {J^\mu}_\lambda \d_\rho {J^\lambda}_\nu = 0~.}
The expression on the left-hand side of this equation is known as the Nijenhuis tensor of~${J^\mu}_\nu$. It follows from~\ntensor\ that we can cover~$\CM$ with adapted coordinate patches of holomorphic local coordinates
\eqn\holcdef{z^i = x^i + i y^i~, \qquad (i = 1, \ldots, n)~,}
and their complex conjugates~$\b z^{\b i}$, in which the non-vanishing components of~${J^\mu}_\nu$ are
\eqn\cscoord{{J^i}_j = i {\delta^i}_j~, \qquad {J^{\b i}}_{\b j} = - i {\delta^{\b i}}_{\b j}~.}
The~$x^i, y^i$ in~\holcdef\ are real local coordinates on the~$2n$-manifold~$\CM$, which is therefore even-dimensional. (The dimension of~$\CM$ as a complex manifold is~$n$.) In order to preserve~\cscoord, the transition functions~$z'^i = f^i(z)$ between two adapted patches must be holomorphic, $\d_{\bar j} f^i (z) = 0$. In each adapted patch we can define a canonical orientation $dx^1 \wedge dy^1 \wedge \cdots \wedge dx^n \wedge dy^n$. This consistently defines an orientation on~$\CM$, since the corresponding transition functions~$\big| \det (\d_j f^i ) \big|^2$ are positive. 

We can use~${J^\mu}_\nu$ to split the complexified tangent and cotangent bundles into holomorphic vectors~$X \in T^{1, 0}$ and holomorphic one-forms~$\omega^{1,0} \in \Lambda^{1,0}$,
\eqn\holvfdef{{J^\mu}_\nu X^\nu = i X^\mu~, \qquad \omega^{1,0}_\mu {J^\mu}_\nu = i \omega^{1,0}_\nu~,}
and their anti-holomorphic counterparts in~$T^{0,1}$ and~$\Lambda^{0,1}$, which satisfy~\holvfdef\ with~$i \rightarrow -i$. More generally, we can split the space~$\Lambda^k$ of complex~$k$-forms into a direct sum,
\eqn\formsplit{\Lambda^k = \bigoplus_{p + q = k} \Lambda^{p,q}~, \qquad \Lambda^{p,q} = {\bigwedge}^p \Lambda^{1,0} \bigotimes {\bigwedge}^q \Lambda^{0,1}~, \qquad 0 \leq p,q \leq k~.} 
In adapted coordinates, a holomorphic vector takes the form
\eqn\holvect{X = X^i (z, \b z) \, \d_i ~,}
while a general~$(p,q)$-form is given by
\eqn\pqform{\omega^{p,q} = {1 \over p ! \, q !} \omega^{p,q}_{i_1 \ldots i_p \b j_1 \ldots \b j_q} (z, \b z) \, dz^{i_1} \wedge \cdots \wedge dz^{i_p} \wedge d{\b z}^{\b j_1} \wedge \cdots \wedge d {\b z}^{\b j_q}~.}
It follows that the exterior derivative~$d \omega^{p,q}$ splits into~$\d \omega^{p,q} \in \Lambda^{p+1, q}$ and~$\b \d \omega^{p,q} \in \Lambda^{p, q+1}$, which we write as~$d = \d + \b \d$. Note in particular that the exterior derivative~$d \omega^{1,0}$ of a holomorphic one-form does not have a~$(0,2)$ component. This property is equivalent to the integrability condition~\ntensor. 

The fact that~$d^2 = 0$ implies that~$\d, \bar \d$ satisfy
\eqn\dprop{\d^2 = 0~, \qquad \b \d^2 = 0~, \qquad \d \b \d + \b \d \d = 0~.}
The Dolbeault cohomology~$H^{p,q}(\CM)$ of~$\CM$ is defined to be the cohomology of~$\b \d$,
\eqn\dchdef{H^{p,q}(\CM) = { \{ \omega^{p,q} \in \Lambda^{p,q} | \b \d \omega^{p,q} = 0 \} \over \b \d \Lambda^{p, q-1}}~.}
There is a Poincar\'e lemma for the the~$\b \d$-operator, which implies that the cohomology of~$\b \d$ is locally trivial: given a~$\b \d$-closed~$(p,q)$-form~$\omega^{p,q}$ with~$q \geq 1$ on~$\C^n$, there is a~$(p,q-1)$-form~$\varphi^{p,q-1}$ such that~$\omega^{p,q} = \b \d \varphi^{p,q-1}$. If~$\CM$ is compact, we can represent elements of~$H^{p,q}(\CM)$ by harmonic forms, which comprise a finite dimensional complex vector space. Hence, the Dolbeault cohomology is also finite dimensional.

\appendix{C}{Two Supercharges of Opposite~$R$-Charge in Four Dimensions}

In this appendix we establish the results outlined in section~3.4 by carrying out the linearized analysis around flat space in the presence of two supercharges~$Q$ and~$\t Q$ of opposite~$R$-charge. See section~1.1 and~\DumitrescuHA\ for additional details on the background geometry in in this case.

\subsec{The~$\t Q$-cohomology of  the~$\CR$-multiplet}

Like the~$Q$-transformations~\rmctrans\ of the operators in the~$\CR$-multiplet, their transformations under the second supercharge~$\t Q$, parametrized by~$\t \zeta^\alphadot$, follow from~\rmultcomp,
\eqn\rmctransii{\eqalign{
& \big\{\t Q, j_\mu^{( R )} \big\} =  i \t\zeta \t S_\mu~,\cr
& \big\{\t Q, S_{\alpha\mu}\big\} =  2 i( \sigma^\nu \t \zeta)_\alpha \t\CT_{\mu\nu}~,\cr
&\big \{\t Q, {\, \t S^\alphadot}_\mu \big\} =0~,\cr
& \big\{\t Q, T_{\mu\nu} \big\} = \half \t\zeta\, \t\sigma_{\mu\rho} \d^\rho \t S_\nu + \half \t\zeta\, \t\sigma_{\nu\rho} \d^\rho \t S_\mu~,\cr
& \big\{\t Q, \CF_{\mu\nu}\big\} = -{i \over 2} \t\zeta\, \t\sigma_\nu  \sigma_\rho \d_\mu \t S^\rho + {i \over 2} \t\zeta\, \t\sigma_\mu  \sigma_\rho \d_\nu \t S^\rho~.
}}
Here we have defined 
\eqn\cttildedef{\t\CT_{\mu\nu} = \CT_{\mu\nu} + i \d_\nu j_\mu^{( R)} = T_{\mu\nu} +{i \over 4} \ep_{\mu\nu\rho\lambda} \CF^{\rho\lambda} - {i \over 4} \ep_{\mu\nu\rho\lambda} \d^\rho j^{( R ) \lambda} + {i\over 2} \d_\nu j_\mu^{(R )}~.}
In particular, we will need
\eqn\compttilde{\eqalign{& \t \CT_{ww} = T_{ww} + {i \over 2} \d_w j_w^{( R )}~,\cr
& \t \CT_{wz} = T_{wz} + {i \over 2} \CF_{wz} -{i \over 4} \d_w j_z^{( R)} + {3 i \over 4} \d_z j_w^{( R)}~,}}
and two more with~$w\leftrightarrow z$. 

In order to extract the~$\t Q$-exact bosonic operators from~$\big \{ \t Q, S_{\alpha\mu}\big\}$, we follow the discussion around~\qst\ and multiply by~$\t \zeta^\dagger\, \t \sigma_\rho$ to obtain a projector proportional to~${\delta^\nu}_\rho + i {\, \t J^\nu}_\rho$. Here~${\, \t J^\mu}_\nu$ is the integrable complex structure determined by~$\t \zeta^\alphadot$ (see footnote~9). If we choose holomorphic coordinates~$w, z$ adapted to~${J^\mu}_\nu$ such that~$K = \d_{\b w}$, the coordinates that are holomorphic with respect to~${\, \t J^\mu}_\nu$ are~$w, \b z$. Since~${\delta^\nu}_\rho + i {\, \t J^\nu}_\rho$ projects onto coordinates that are anti-holomorphic with respect to~${\,\t J^\mu}_\nu$, we conclude that the eight bosonic~$\t Q$-exact operators are~$\t\CT_{\mu \bw}$ and~$\t \CT_{\mu z}$. As before, there are no other~$\t Q$-closed operators. Note that none of these operators are~$Q$-closed, since the two supercharges anti-commute to~$K = \d_{\b w}$, which does not annihilate any of the operators in the~$\CR$-multiplet.

\subsec{Deformations around Flat Space}

We now proceed as in section~3.2 and consider deformations~$\Delta {J^\mu}_\nu$ and~$\Delta g_{\mu\nu}$ of the complex structure and the metric around their flat space values, subject to the constraints~\acreq, \integrability, and~\defmet. The only new ingredient is the anti-holomorphic Killing vector~$K$, which must be deformed so that~$K^\mu + \Delta K^\mu$ is anti-holomorphic with respect to the deformed complex structure and a Killing vector of the deformed metric. This leads to further constraints on~$\Delta {J^\mu}_\nu$ and~$\Delta g_{\mu\nu}$, whose detailed form will not be needed here.

In terms of the operators~$\t \CT_{i j}$ in~\compttilde, the deformation Lagrangian~\deflag\ can be expressed as\foot{For the purpose of this discussion, we omit the term involving~$W_{i j \b k}$~.}
\eqn\deflagapp{\Delta {\scr L} =  - \Delta g^{i \b j} \CT_{i\b j}  - i \sum_{j = \b j} \left(\Delta {J^{\b i}}_j \CT_{\b j \b i} - \Delta {J^i}_{\b j} \t \CT_{i j}\right)~.}
Since~$\t \CT_{i z}$ is~$\t Q$-exact, we conclude that the partition function does not depend on~$\Delta {J^i}_{\b z}$. In section~3.2, we verified that the Lagrangian in~\deflagapp\ is~$Q$-closed up to a total derivative. If we want to check that it is also~$\t Q$-closed up to a total derivative, we must use the more stringent constraints on~$\Delta {J^\mu}_\nu$ and~$\Delta g_{\mu\nu}$ that follow from the presence of the anti-holomorphic Killing vector.

\subsec{Background Gauge Fields} 

The presence of a second supercharge~$\t Q$ also reduces the dependence of partition function on Abelian background gauge fields. The~$\t Q$-transformations of the flavor current multiplet~$\CJ$ follow from~\fdflavsf,
\eqn\jvarii{\eqalign{& \big\{\t  Q, J\big\} = -i \t\zeta \t j~,\cr
& \big\{\t Q, j_\alpha\big\} = - i (\sigma^\mu \t\zeta)_\alpha \t\CJ_\mu~,\cr
&\big \{\t Q,  \t j^\alphadot\big\} = 0~,\cr
&\big \{\t Q, j_\mu\big\} = - 2 \t\zeta \, \t\sigma_{\mu\nu} \d^\nu \t j~,}}
where~$\t\CJ_\mu = j_\mu + i \d_\mu J$. As before, we multiply~$\big\{ \t Q, j_\alpha \big\}$ by~$\t \zeta^\dagger \t \sigma_\rho$ to obtain a projector onto the anti-holomorphic indices of the complex structure~${\, \t J^\mu}_\nu$. This shows that~$\t\CJ_\bw$ and~$\t\CJ_z$ are~$\t Q$-exact. Comparing with the Lagrangian~\fldlag\ for the background gauge field, we conclude that the partition function does not depend on~$A_\bz$. 

\appendix{D}{$S^3 \times S^1$ and~$S^3$ Backgrounds Admitting One Supercharge}

In this appendix we will briefly discuss primary Hopf surface of the second type, which are obtained from~$\C^2 - (0,0)$ with coordinates~$w,z$ through the identification~\Hopfsurfii. For simplicity, we restrict ourselves to the case~$n=1$ in~\Hopfsurfii, 
\eqn\nisone{(w, z) \sim (q w + \lambda z, q z)~, \qquad 0 < |q| <1~, \qquad \lambda \in \C^*~.}
Recall that any~$\lambda \neq 0$ gives rise to the same complex structure (for instance, we could set~$\lambda = 1$ by rescaling~$z$), while~$q$ is a genuine complex structure modulus. Like all primary Hopf surfaces, the resulting quotient space is diffeomorphic to~$S^3 \times S^1$. However, we will show that it does not admit a Hermitian metric with a suitable holomorphic Killing vector, and hence it only preserves one supercharge. We will then construct an explicit example of a Hermitian metric with two real Killing vectors, which admits a reduction to a metric on~$S^3$. This leads to a THF of the second type, whose existence was inferred from an infinitesimal deformation analysis in section~7.1.

\subsec{Absence of Holomorphic Killing Vectors}

As explained in~\DumitrescuHA\ and reviewed in section~1.1, the presence of two supercharges of opposite~$R$-charge requires a Hermitian metric with a holomorphic Killing vector, whose coefficient functions are also holomorphic.\foot{This requirement was previously expressed in terms of an anti-holomorphic Killing vector, but since the metric is real its complex conjugate is also a Killing vector.} Here we will show that the Hopf surfaces corresponding to~\nisone\ do not admit such a Killing vector. The most general holomorphic vector field with holomorphic coefficients is given by (see for instance example~2.15 in~\Kodairabook)
\eqn\holvec{a \left(w \d_w + z \d_z\right) + b z \d_w~.}
We will argue that it is not possible to construct a well-behaved Hermitian metric on $\C^2 - (0,0)$ that is invariant under the identifications~\nisone\ and admits~\holvec\ as a Killing vector. In order to simplify the formulas, we will set~$\lambda = q$ in this subsection. This does not affect the argument. 

It is convenient to introduce the coordinates
\eqn\defwsh{v^1={w\over z}~,\qquad v^2=-\log z~,}
which are subject to the following identifications:
\eqn\idnwc{\left(v^1,v^2\right) \sim \left(v^1,v^2+2\pi i\right) \sim \left(v^1+1 ,v^2-\log q\right)~.}
In these coordinates, the holomorphic vector field~\holvec\ takes the simple form
\eqn\holf{a \d_{v^1}+b \d_{v^2}~.}
Since the one-forms~$dv^{1,2}$ are invariant under~\idnwc, the most general Hermitian metric consistent with the identifications is given by~$ds^2 = g_{i \b j} dv^i d{\b v}^{\b j}$, where the coefficient functions~$g_{i \b j}$ must also be invariant under~\idnwc. It is clear from~\defwsh\ that~$v^i \rightarrow \infty$ as~$z \rightarrow 0$ for fixed~$w \neq 0$. (The precise way in which they approach infinity depends on the phase of~$z$, as well as on~$w$.) Requiring the metric to be smooth and positive definite in the original~$w, z$ coordinates leads to various constraints, such as
\eqn\metlim{g_{1\b 1} \rightarrow C |z|^2 \; \,{\rm as} \; \,z \rightarrow 0 \; \,{\rm with} \,\; w \neq 0 \, \; {\rm fixed}\,  \; {\rm and}\, \; C>0~,}
where~$C$ may depend on~$w$. We will use this establish a contradiction.

If~$a = 0$ in~\holf, the metric components~$g_{i \b j}$ do not depend on~$v^2$. It follows from~\metlim\ that~$g_{1\b1}$ is proportional to~$1 \over |v^1|^2$ as~$v^1 \rightarrow \infty$, but this is inconsistent with the identification~$v^1 \sim v^1 + 1$. Therefore~$a \neq 0$ and we can set~$a =1$ without loss of generality. Now~$g_{i\b j}$ can only depend on~$y = b v^1 - v^2$ (and its complex conjugate), which is subject to the identifications~$y \sim y + 2 \pi i \sim y + b + \log q$. For generic~$b, q$, this describes a compact two-torus, and hence~$g_{1\b 1}$ cannot have the correct limiting behavior~\metlim. The exceptional case occurs when~$b + \log q$ is purely imaginary, so that the real part~$\Re y$ is not identified. Since~$|q| < 1$, this case requires~$b  \neq 0$. As~$z \rightarrow 0$, we then have~$y = b v^1 + \CO(\log v^1)$. Again, \metlim\ is not consistent with the identifications of~$y$.

\subsec{Hermitian Metrics and Reduction to Three Dimensions}

Here we will write down a class of metrics on the Hopf surfaces discussed above. They possess two commuting Killing vectors that enable a reduction to three dimensions.\foot{See~\Belgun\ for a discussion of metrics that have various other desirable properties.} As before, we introduce real variables~$x, \theta, \varphi, \chi$, subject to the same identifications as in~\zoztworealangles, so that~$w, z$ are given by
\eqn\ztoan{w = q^x \cos{\theta \over 2} e^{i \varphi} + {\lambda x \over q} z~, \qquad z = q^x \sin{\theta \over 2} e^{i\chi}~.}
Now~$x$ is only well defined if~$|\lambda|$ is sufficiently small. (The precise bound, which depends on~$q$, will not be important.) We can construct two holomorphic one-forms that are invariant under the identification~\nisone,
\eqn\forms{e^1=q^{-x} \Big(dw - {\lambda x \over q} dz\Big)~,\qquad e^2 = q^{-x} dz~.}
Given a constant~$2 \times 2$ Hermitian matrix~$A_{i\b j}$, which does not depend on~$w, z$ and satisfies a suitable positivity requirement, $ds^2= A_{i\b j} e^i {\b e}^{\b j}$ is a well-defined, smooth Hermitian metric on the Hopf surface. By expressing it in terms of~$x, \theta, \varphi, \chi$, it can be checked that~$\d_x$ and~$\d_{\varphi}+\d_{\chi}$ are Killing vectors. 

We can now reduce to three dimensions along the Killing vector~$\d_x$ to obtain a THF with a compatible transversely Hermitian metric on a squashed~$S^3$. In order to simplify the formulas, we will choose~$q$ real and positive, i.e.~$0 < q < 1$, and take
\eqn\choices{\lambda= -\ep q \log q~,\qquad A_{i \b j} = \pmatrix{1 & \ep \cr \ep & 1 + \ep^2}~, \qquad \ep >0~,}
with~$\ep$ sufficiently small to ensure that~$x$ is a good coordinate. We also switch to coordinates~$\theta,\psi,\chi$ with~$\psi = \varphi - \chi$. The one-form~$\eta$ is then given by
\eqn\etared{\eta= -{\ep\over 2}  \cos^2 {\theta\over 2} \sin \psi d\theta + \cos^2 {\theta\over 2} d\psi+\left( 1+{\ep \over 2} \sin \theta \cos \psi \right)d\chi~,}
while the compatible metric takes the form~$g_{\mu\nu} = \eta_\mu \eta_\nu + \t g_{\mu\nu}$ with
\eqn\mettdn{\eqalign{&\t g_{\theta \theta}= {1\over 4}\left(1+{\ep^2 \over 4} \sin^2 \theta-\ep \cos \psi \sin \theta\right)~,\cr
& \t g_{\theta \psi}=-{\ep\over 8}\sin^2 \theta \sin\psi~,\cr
&\t g_{\theta \chi}=-{\ep \over 2} \sin^2 {\theta\over 2} \sin \psi~,\cr
&  \t g_{\psi\psi}= {1\over 4} \sin^2 \theta~,\cr
& \t g_{\psi \chi}={\ep\over 2} \cos \psi \sin\theta \sin^2{\theta\over 2}~,\cr 
& \t g_{\chi \chi}=\ep^2 \sin^4{\theta\over 2}~.}}
It can be checked explicitly that~$\eta_\mu$ and~$g_{\mu\nu}$ satisfy the integrability condition~\acsint, and hence this squashed sphere admits one supercharge. The metric only depends on~$\theta$ and~$\psi $, so that~$\d_\chi$ is a Killing vector.\foot{When we change variables from~$\varphi, \chi$ to~$\psi, \chi$, we have~$\d_\varphi |_\chi + \d_\chi |_\varphi = \d_\chi |_\psi$.} However, the Killing vector is not proportional to~$\eta^\mu$, and hence this space does not admit a second supercharge of opposite~$R$-charge~\ClossetRU. For small~$\ep$, the expressions in~\etared\ and~\mettdn\ approach the THF on a round~$S^3$ of unit radius, which was discussed in section~7.1. If we switch to coordinates adapted to the round case and expand to first order in~$\ep$, we find that~\etared\ and~\mettdn\ realize an infinitesimal deformation of the second type, with~$\Theta = - 2 \ep z^2 \left(\d_z - h \d_\tau\right) \otimes \eta$. Thus, primary Hopf surfaces of the second type give rise to THFs of the second type on~$S^3$.

\appendix{E}{Squashed Three-Spheres of the First Type}

In this appendix we briefly review various known squashings of~$S^3$ that preserve at least two supercharges. We expand them around~\cancon\ and show that they correspond to deformations of the first type.

Our starting point is the metric studied in~\MartelliAQA, which depends on two real parameters~$\gamma_r,\gamma_i$,
\eqn\mep{\eqalign{ds^2=~&{d\theta^2\over  f\left(\theta\right)}+{f\left(\theta\right)}\sin^2 \theta \left({d\varphi\over \left(2+\gamma_i\right)^2+\gamma_r^2}+{d\chi\over \left(2-\gamma_i\right)^2+\gamma_r^2}\right)^{2}\cr
 &+ \left ({2 + 2\cos\theta+\gamma_i \sin^2 \theta \over \left(2+\gamma_i\right)^2+\gamma_r^2}d\varphi-{2-2\cos\theta-\gamma_i \sin^2 \theta \over \left(2-\gamma_i\right)^2+\gamma_r^2} d\chi\right)^{2}~,\cr f\left(\theta\right)= &\left(2-\gamma_i \cos\theta\right)^2+\gamma_r^2~.}} 
As before, $0 \leq \theta \leq \pi$ and~$\varphi, \chi$ and angles with periodicity~$2\pi$. This metric has two commuting Killing vectors~$\d_\varphi, \d_\chi$. The THF corresponding to the supercharge considered in~\MartelliAQA\ is defined by
\eqn\acmsmp{\eqalign{\xi=~&\gamma_r \sin\theta \d_\theta-{\left(2+\gamma_i\right)^2+\gamma_r^2\over 4 f\left(\theta\right)} \left(4-2\gamma_i-\left(2\gamma_i-\gamma_i^2-\gamma_r^2\right)\cos\theta \right)\d_\varphi\cr
&- {\left(2-\gamma_i\right)^2+\gamma_r^2\over 4 f\left(\theta\right)} \left(4+2\gamma_i-\left(2\gamma_i+\gamma_i^2+\gamma_r^2\right)\cos\theta \right)\d_\chi~.}}
As~$\gamma_r, \gamma_i \rightarrow 0$, \mep\ and~\acmsmp\ approach the THF on the round sphere in~\cancon. Switching to the coordinates adapted to the round case, we find that~$\xi$ in~\acmsmp\ corresponds to a deformation of the first type, i.e.~$\Theta = \gamma z \left(\d_z - h \d_\tau\right) \otimes \eta$,  with complex deformation parameter~$\gamma = - 2 i\left(\gamma_r + i \gamma_i\right)$. 

The metric~\mep\ is also viable for purely imaginary~$\gamma_r=-i \t \gamma_r$ with real~$\t \gamma_r$, but now~$\xi$ in~\acmsmp\ is complex, which is not allowed. In this case~\MartelliAQA\ choose
\eqn\acmsmpt{\xi=-\left(2-\gamma_i+\t \gamma_r\right){\left(2+\gamma_i\right)^2-{\t \gamma_r}^2\over 8+4\left(\t \gamma_r-\gamma_i\right)\cos \theta}\d_\varphi-\left(2+\gamma_i-\t \gamma_r\right){\left(2-\gamma_i\right)^2-{\t \gamma_r}^2\over 8+4\left(\t \gamma_r-\gamma_i\right)\cos \theta}\d_\chi~.}
As~$\gamma_i, \t\gamma_r \rightarrow 0$ we again find a deformation of the first type, with~$\gamma = 2 \left(\t \gamma_r + \gamma_i\right)$, so that the deformation parameter is real. 

For some choices of~$\gamma_{r, i}$, we recover other cases studied in the literature:
\medskip
\item{1.)} If~$\gamma_r = -i \t \gamma_r$ is purely imaginary and~$\t\gamma_r= \gamma_i$, then~\mep\ is isometric to the ellipsoid with~$U(1) \times U(1)$ isometry studied in~\HamaEA. According to the preceding discussion, this is a deformation of the first type with real deformation parameter~$\gamma = 4 \t \gamma_r$. 

\item{2.)} If~$\gamma_i=0$, then~\mep\ is isometric to the squashed sphere with~$SU(2)\times U(1)$ isometry studied in~\refs{\ImamuraUW,\ImamuraWG}, which preserves four supercharges. This corresponds to a deformation of the first type with~$\gamma = - 2 i \gamma_r$, which may be real or purely imaginary.

\appendix{F}{Computing~$H^{0,1}(S^2 \times S^1)$}

In this appendix we compute the cohomology~$H^{0,1}(S^2 \times S^1)$ for the THF in~\chxi\ on~$S^2\times S^1$. Below, we will consider a path on~$S^2 \times S^1$ that approaches the poles of~$S^2$ at~$\theta = 0, \pi$ along a meridian with constant~$\varphi$. In the adapted coordinates~\adcoor, it takes the form
\eqn\intlines{\theta(\tau)=2 \arctan e^{\tau}~,\qquad x(\tau)=-\log\left(2 \cosh\tau \right)~,} 
with fixed~$z$. As~$\tau \rightarrow -\infty$, it follows that~$\theta \rightarrow 0$ and~$x \rightarrow -\infty$, while~$\tau \rightarrow \infty$ corresponds to~$\theta \rightarrow \pi$ and~$x \rightarrow \infty$. Therefore, the path approaches the poles on~$S^2$ while rapidly circling the~$S^1$. 

\subsec{Constraining the Cohomology}

According to~\tdaction, non-trivial elements of~$H^{0,1}(S^2\times S^1)$ are~$\t\d$-closed $(0,1)$-forms,
\eqn\eqnfrm{\omega^{0,1}=\omega_\tau^{0,1}(d\tau+h dz)+\omega_{\b z}^{0,1} d\b z~,\quad \d_{\tau}\omega_{\b z}^{0,1}= \d_{\b z} \omega_\tau^{0,1}~,} 
modulo those that can be expressed in terms of a well-defined function~$\ep(\tau, z, \b z)$ on~$S^2\times S^1$,
\eqn\exactcnd{\omega_\tau^{0,1}=\d_\tau \ep~,\qquad \omega_{\b z}^{0,1}=\d_{\b z}\ep~.}
The identifications~\idft, i.e.~$(\tau, z) \sim (\tau + 1, e^{-1} z)$, and~\eqnfrm\ imply that
\eqn\seromt{\eqalign{& \omega^{0,1}_\tau= \sum_{k \in \Z} c_k e^{2 \pi i k \tau}+\sum_{n\geq 1} \sum_{m,k \in \Z} d_{n m k} z^{n+{m\over 2}} \b z^{- {m\over 2}} e^{(n+2 \pi i k)\tau}~,\cr &\omega^{0,1}_{\b z}= - \half \sum_{n\geq 1} \sum_{m,k \in \Z} {m\over  (n+2 \pi i k)} d_{n m k} z^{n+{m\over 2}} \b z^{- {m\over 2}-1} e^{(n+2 \pi i k)\tau}~,}} 
as long as~$\theta \neq \pi$. Note that~$\omega^{0,1}_{\b z}$ and the derivatives~$\d_{\b z}\omega^{0,1}_\tau$, $\d_{ z}\omega^{0,1}_\tau$ are exponentially damped as~$\tau\rightarrow -\infty$ at fixed~$z$, i.e.~as we approach~$\theta=0$ along the path~\intlines, while~$\omega^{0,1}_{\tau}$ approaches the constant~$c_0$. We can repeat this analysis in~$\tau', z, \b z$ coordinates, which describe the region~$\theta \neq 0$, where we find analogous results for the behavior of~$(\omega')^{0,1}_{\tau'}$ and~$(\omega')^{0,1}_{\b z}$ near~$\theta=\pi$. In order to understand which terms in~\seromt\ can be written in terms of a well-defined function~$\ep$ as in~\exactcnd, we use the identifications~\idft\ to restrict the form of~$\ep$ as follows,
\eqn\serexp{\ep(\tau, z, \b z) =  \sum_{k\in \Z}  \ep_k e^{2 \pi i k \tau}+\sum_{n\geq 1} \sum_{m,k \in \Z} \ep_{n m k} z^{n+{m\over 2}} \b z^{- {m\over 2}} e^{(n+2 \pi i k)\tau}~.} 
Its derivatives can generate all terms in~\seromt, except the constant term~$c_0$. In~$\tau', z , \b z$ coordinates we find that no well-defined~$\ep$ can give rise to a constant term~$c_0'$ in~$(\omega')^{0,1}_{\tau'}$. 

In summary, nontrivial elements of~$H^{0,1}(S^2 \times S^1)$ must behave as 
\eqn\smh{\eqalign{\omega^{0,1}_\tau \rightarrow  c_0~,~\omega^{0,1}_{\b z}\rightarrow 0 \quad &{\rm as} \quad \theta \rightarrow 0 \;\; {\rm along \; \;  (G.1)}~,\cr
(\omega')^{0,1}_{\tau'} \rightarrow  c'_0~,~(\omega')^{0,1}_{\b z}\rightarrow 0 \quad &{\rm as} \quad \theta \rightarrow \pi \;\; {\rm along \; \;  (G.1)}~.} }
Changing adapted coordinates from~$\tau$ to~$\tau' = \tau + \log {|z|^2 \over 4}$ implies that~$(\omega')_{\tau'} = \omega_\tau$ while $(\omega')^{0, 1}_{\b z}=\omega^{0,1}_{\b z} - {1\over \b z} \omega^{0,1}_\tau $. Combining this with~\smh, we find that 
\eqn\limt{\eqalign{\omega^{0,1}_\tau \rightarrow  c_0'~,~\omega^{0,1}_{\b z}\rightarrow {c_0'\over \b z} \quad &{\rm as} \quad \theta \rightarrow \pi \;\; {\rm along \; \;  (G.1)}~,\cr
(\omega')^{0,1}_{\tau'} \rightarrow  c_0~,~(\omega')^{0,1}_{\b z}\rightarrow -{c_0\over \b z} \quad &{\rm as} \quad \theta \rightarrow 0 \;\; {\rm along \; \;  (G.1)}~.} }
In section~7.2, we have exhibited a non-trivial element of~$H^{0,1}(S^2 \times S^1)$,
\eqn\derx{(dx)^{0,1}= {4-e^{2\tau} |z|^2\over 4+ e^{2\tau} |z|^2}(d\tau+h dz) -{e^{2\tau} z d\b z \over 4+e^{2\tau} |z|^2}~,\qquad h= {e^{2 \tau} \b z \over 4+e^{2\tau} |z|^2}~.}
By examining the limits~$\tau \rightarrow \pm \infty$ at fixed~$z$, we find that this satisfies the conditions in~\smh\ and~\limt\ with~$c_0=-c'_0=1$. We will now show that linearly independent solutions with~$c_0 \neq - c_0'$ do not exist. 

\subsec{Proof that~$c_0 = - c_0'$}

It suffices to show that there is no solution with~$c_0=0$ and~$c'_0\neq 0$. We will argue by contradiction, making use of an auxiliary one-form~$\omega$, which is obtained from~$\omega^{0,1}$ by adding a well-defined~$(1,0)$-form~$\omega^{1,0}=\omega^{1,0}_z dz $,
\eqn\auxform{\omega=\omega_\tau^{0,1}(d\tau+h dz)+\omega^{1,0}_z dz+\omega_{\b z}^{0,1} d\b z~.}
Since~$\t \d \omega^{0,1} = 0$, it follows that~$d \omega$ does not have a~$d\tau\wedge d\bz$-component. If it is possible to choose~$\omega^{1,0}$ so that the~$d\tau\wedge d z$-component also vanishes, we can integrate~$d \omega$ along a surface of constant~$\varphi$, which is bounded by two circles at~$\theta=0,\pi$. Since~$z$ is real on this surface, $dz\wedge d\bar z=0$ and~$d\omega$ integrates to zero. However, the integral of~$\omega$ over the boundary circles evaluates to~$c_0'\neq 0$, which contradicts Stokes' theorem.  

To complete the argument we must find a well-defined~$\omega^{1,0}$, such that the~$d\tau\wedge dz$-component of~$d\omega$ vanishes, i.e.~we must solve 
\eqn\constr{\d_\tau\big(\omega^{0,1}_\tau h+\omega_z^{1,0}\big)=\d_z \, \omega_\tau^{0,1}~.}
Consider the integral
\eqn\intepdef{\ep(\tau,z,\b z)=\int_{-\infty}^{\tau} ds \, \omega_\tau^{0,1}(s,z,\b z)~.}
This integral is finite, since~$\omega_\tau^{0,1}$ behaves as in~\smh\ with~$c_0=0$ when~$\tau \rightarrow - \infty$. While~$\ep$ is a well-defined function on the covering space~$S^2 \times \R$, it is not well-defined on~$S^2 \times S^1$ when~$c_0'\neq 0$, because it diverges as~$\tau \rightarrow \infty$ (see below). By contrast, $\d_{\b z} \ep$, $\d_{\b z} \ep$ are well-defined on~$S^2\times S^1$. In fact, since~$\d_{\b z} \omega^{0,1}_\tau=\d_{\tau}\omega^{0,1}_{\b z}$, we have~$\d_{\b z} \ep = \omega_{\b z}^{0,1}$. When~$\tau \rightarrow +\infty$ at constant~$z$, it follows from~\limt\ that~$\omega^{0,1}_\tau\rightarrow c_0'$ and~$\omega^{0,1}_{\b z}\rightarrow {c_0'\over \b z}$. Therefore,
\eqn\limfor{\ep(\tau,z,\b z)\rightarrow c_0'\left(\tau+ \log|z|^2\right) + ({\rm constant}) \quad {\rm as} \quad \tau \rightarrow +\infty~.}
Here~$\log z$ and~$\log \b z$ must arise together, since~$\ep$ is well defined on the covering space~$S^2\times \R$. 
Finally, define
\eqn\solex{\omega_z^{1,0}(\tau,z,\b z)=\d_z \ep(\tau,z, \b z)-\omega^{0,1}_\tau(\tau,z,\b z) h(\tau,z,\b z)~,}
which satisfies~\constr. It is clearly well-defined as~$\tau\rightarrow -\infty$. When~$\tau \rightarrow + \infty$, it follows from~\limfor\ that~$\d_z \ep \rightarrow {c_0'\over z}$, while we know that~$\omega_\tau^{0,1} \rightarrow c_0'$ and~$h \rightarrow {1 \over z}$. Therefore, all divergent contributions proportional to~$1 \over z$ cancel and~$\omega^{1,0}$ is well defined on~$S^2 \times S^1$.

\listrefs

\end